\documentclass[preprint,11pt]{elsarticle}

\usepackage{float}
\usepackage{url}
\usepackage[hidelinks]{hyperref}
\usepackage[margin=1in]{geometry}
\usepackage{bm}
\usepackage{listings}
\usepackage{parcolumns}
\usepackage{inconsolata}
\usepackage{algorithm}
\usepackage{algpseudocodex}
\usepackage{comment}
\usepackage{amssymb}
\usepackage{amsmath}
\newcommand{\sci}[2]{#1 \times 10^{#2}}
\newcommand{\ds}              { \displaystyle }
\newcommand{\bfm}[1]             { \mathbf{#1}     }
\newcommand{\uu}              { \bfm{u} }
\newcommand{\xx}              { \bfm{x} }

\newcommand{\FF}              { \bfm{F} }
\newcommand{\Nabla}       { \boldsymbol{\nabla}  }
\newcommand\dt[1]{\frac{\partial #1}{\partial t}}
\newcommand\Dt[1]{\frac{d #1}{dt}}
\newcommand\dx[1]{\frac{\partial #1}{\partial x}}
\newcommand\dy[1]{\frac{\partial #1}{\partial y}}

\newcommand\areaa{\int_{\Omega}}
\newcommand{\elev}{\bm{\zeta}}
\newcommand{\modal}{\hat{\bm{\zeta}}}
\newcommand{\mean}[1]{\langle #1 \rangle}
\newcommand{\lvx}[1]{\alpha_{\text{#1}}}
\newcommand{\lvy}[1]{\beta_{\text{#1}}}
\newcommand{\lvz}[1]{\gamma_{\text{#1}}}

\newcommand{\mm}{\text{ m}}

\usepackage{lineno}
\usepackage{graphicx}
\usepackage{subfigure}
\usepackage{notoccite}

\usepackage{booktabs}
\usepackage{float}
\usepackage{tikz}
\usepackage{tabularx}
\usetikzlibrary{shapes.geometric, arrows, calc}
\tikzstyle{process} = [rectangle, minimum width=3cm, minimum height=1cm, text centered, draw=black] %fill=orange!30]

\begin{document}
\begin{frontmatter}

\title{Coupled Continuous-Discontinuous Galerkin Finite Element Solver for Compound Flood Simulations}
 \author[inst1]{Chayanon Wichitrnithed\corref{cor1}}
 \ead{namo@utexas.edu}
 \cortext[cor1]{Corresponding author}

\affiliation[inst1]{organization={The Oden Institute for Computational Engineering and Sciences, The University of Texas at Austin},%Department and Organization
            addressline={201 E. 24th St. Stop C0200},
            city={Austin},
            postcode={78712},
            state={Texas},
            country={United States of America}}

\author[inst2,inst3]{Shintaro Bunya}
\author[inst4]{Ethan J. Kubatko}
\author[inst1]{Clint Dawson}
\author[inst5,inst6]{Eirik Valseth}

\affiliation[inst2]{organization={Coastal Resilience Center, University of North Carolina at Chapel Hill},%Department and Organization
            %addressline={2070 Neil Ave},
            city={Chapel Hill},
            postcode={27517},
            state={North Carolina},
            country={United States of America}}
\affiliation[inst3]{organization={Institute of Marine Sciences, University of North Carolina at Chapel Hill},%Department and Organization
            %addressline={2070 Neil Ave},
            city={Morehead City},
            postcode={28557},
            state={North Carolina},
            country={United States of America}}
\affiliation[inst4]{organization={The Department of Civil, Environmental, and Geodetic Engineering, The Ohio State University},%Department and Organization
            addressline={2070 Neil Ave},
            city={Columbus},
            postcode={43210},
            state={Ohio},
            country={United States of America}}
\affiliation[inst5]{organization={Department of Mechanical Engineering and Technology Management, The Norwegian University of Life Science},%Department and Organization
            addressline={Drøbakveien 31},
            city={Ås},
            postcode={1433},
            country={Norway}}
\affiliation[inst6]{organization={Department of   Numerical Analysis and Scientific Computing, Simula Research Laboratory},%Department and Organization
            addressline={Kristian Augusts gate 23},
            city={Oslo},
            postcode={0164},
            country={Norway}}

\titlepage
\begin{abstract}

Several recent tropical cyclones, e.g., Hurricane Harvey (2017), have lead to significant rainfall
and resulting runoff. When the runoff interacts with storm
surge, the resulting floods can be greatly amplified and lead to effects that cannot be correctly
modeled by simple superposition of its distinctive sources. In an effort to develop accurate numerical simulations of runoff, surge, and compounding floods, we develop a locally conservative coupled DG-CG discretization of the shallow water equations and integrate it into the ADvanced CIRCulation Model (ADCIRC).
We also modify the continuity equation to include spatially and temporally variable rainfall into the model using parametric rainfall models. We demonstrate the capabilities of the scheme though a sequence of physically relevant numerical tests, including small scale test cases based on laboratory measurements and large scale experiments with Hurricane Harvey in the Gulf of Mexico. The results highlight the conservation properties and robustness of the developed method and show the potential of compound flood modeling using our approach.
\end{abstract}
\end{frontmatter}

%% main text
\section{Introduction} \label{sec:introduction}
Numerical models based on finite element (FE) discretizations to simulate hydrodynamics in coasts, estuaries, and the ocean have widespread use in engineering, marine, and other scientific disciplines. The model that is the focus of our current investigation are the depth-averaged two dimensional Shallow Water Equations (SWE)~\cite{vreugdenhil1994numerical}. In this particular case, we use the SWE to model flows as a result of the interaction between rivers, rainfall runoff, and storm surge, i.e., compound floods~\cite{wahl2015increasing}. 
Famous examples from the last decade includes Hurricane Harvey (2017)~\cite{watson2018characterization} Hurricane Irma (2017)~\cite{cangialosi2018tropical}, and Hurricane Florence (2018)~\cite{callaghan2020extreme}. The flow resulting from the interaction of these processes requires careful treatment to ensure accurate modeling and has therefore been frequently studied in recent literature, see, e.g.,~\cite{loveland2021developing,santiago2019comprehensive,orton2020flood,kumbier2018investigating}. The aim of the present paper is to introduce a new numerical model in which we use  a modified form of the SWE to develop to simulate compound flooding during tropical cyclones~\cite{wahl2015increasing}. 

A widely used existing FE based numerical model for the SWE is the ADvanced CIRCulation (ADCIRC) model~\cite{luettich1992adcirc,Pringle2020} which has wide usage in both academia, industry, and government, see, e.g.,~\cite{bunya2010high,blanton2012urgent,funakoshi2012development}.  ADCIRC solves a surrogate to the SWE in which the continuity equation is replaced by a generalized wave continuity equation~\cite{Lynch1979-ux}. To discretize in space, ADCIRC uses a FE formulation based on the Bubnov-Galerkin, or continuous Galerkin (CG) method~\cite{carey1983finite,ern2004theory} and thus continuous polynomial basis functions. As a result of this CG method, element-wise conservation of mass is not guaranteed in ADCIRC simulations. To discretize in time, ADCIRC offers several options, including fully explicit and implicit-explicit flavors of finite differences~\cite{Pringle2020}. ADCIRC has successfully been applied to model compound flooding during Hurricane Harvey (2017)~\cite{loveland2021developing,pachev2023one} by incorporating rainfall runoff through riverine boundary conditions. 

We also mention two other FE based numerical models from recent literature: the Adaptive Hydraulics (AdH) \cite{savant2020theory} and Semi-implicit Cross-scale Hydro-science Integrated System Model (SCHISM) \cite{zhang2014schism} which both use CG methods in their spatial FE discretizations. While these FE models have the potential to model compound floods as demonstrated in e.g., \cite{zhang2020simulating}, the conservation properties of the underlying CG method are not exact.

A model that is related to ADCIRC is the discontinuous Galerkin Shallow Water Equation Model (DG-SWEM)~\cite{kubatko2006hp,bunya2009wetting,dawson2011discontinuous}, which solves the conservative SWE using a local discontinuous Galerkin (DG) method with an explicit strong stability preserving Runge Kutta time stepping scheme. In the recent paper \cite{wichitrnithed2024discontinuous}, we extended DG-SWEM to incorporate rainfall onto the FE mesh in an effort to model compound floods from rainfall and storm surge.  These results indicated that a exact locally mass conserving FE method can accommodate rainfall onto the computational mesh without inducing notable computational issues. The approach in \cite{wichitrnithed2024discontinuous}  was to modify the right hand side (RHS) of the continuity equation in the SWE with a source. This source term was defined, e.g., from  parametric rainfall models from literature such as R-CLIPER. 

In this paper, we present a new numerical model based on mixed continuous and discontinuous approximation spaces for the  SWE. In particular, we develop a model in which the continuity equation is modified with a nonzero RHS  is discretized using a DG scheme and the momentum equation using a CG scheme. Hence, the proposed method inherits the exact local conservation properties of the DG method and the efficiency of the CG method for the momentum equation keeps the solution cost as low as possible. This will allow us to incorporate rainfall onto the FE mesh in a fashion analogous to \cite{wichitrnithed2024discontinuous} and thus model compound flooding from runoff and storm surge in a single computational model. Such solution schemes have been investigated in past works as well, e.g., \cite{dawson2002discontinuous}. 
However, the present work differs in the use of the conservative SWE, as well as its application to large scale computational domains. Last, we also mention the open-source code SWEMniCS~\cite{dawson2024swemnics}, where an implicit approach was taken to solve the SWE using a mix of continuous and discontinuous function spaces. The SWEMniCS framework is currently in development but does not presently support large-scale inundation studies with the mixed function space approach.

In the following, we introduce, verify, and validate the proposed numerical model for compound flooding simulations. In section \ref{sec:swe}, we introduce the governing model, i.e., the modified SWE. Section \ref{sec:CD-FEM} details the FE discretization of the SWE and  present an overview of the implementation details.
Sections \ref{sec:experiments} - \ref{sec:performance} present a comprehensive numerical study of our methodology including small benchmark tests and large scale hurricanes with compound flood effects. Finally, in Section~\ref{sec:conclusions}, we draw conclusions and discuss potential future research directions.

\section{Governing Equations} \label{sec:swe}
\begin{figure}[h!]
    \centering
    \begin{tikzpicture}[scale=1.5]

    % Define colors for water and land
    \definecolor{waterblue}{RGB}{173,216,230}
    \definecolor{sand}{RGB}{194,178,128}
    
    % Draw the horizontal datum
    \draw[black, very thick] (-4,0) -- (4,0);
    %\node[right] at (4.5,0) {Horizontal Datum};
    
    % Draw the seabed (variable bathymetry)
    \draw[brown, thick] plot[smooth, tension=.7] coordinates {(-4,-2.5) (-2,-1.8) (0,-2.3) (2,-2.0) (4,-2.4)};
    \fill[sand] plot[smooth, tension=.7] coordinates {(-4,-3) (-4,-2.5) (-2,-1.8) (0,-2.3) (2,-2.0) (4,-2.4) (4,-3)} -- cycle;

    % Draw the water surface (variable)
    \draw[blue, thick] plot[smooth, tension=.7] coordinates {(-4,1.0) (-2,0.8) (0,1.2) (2,1.1) (4,1.3)};
    \fill[waterblue] plot[smooth, tension=.7] coordinates {(-4,1.0) (-2,0.8) (0,1.2) (2,1.1) (4,1.3)} -- plot[smooth, tension=.7] coordinates { (4,-2.4) (2,-2.0) (0,-2.3) (-2,-1.8) (-4,-2.5) } -- cycle;
    
    % Depth line and label at a specific point
    %\draw[dashed] (1, 1.1) -- (1, -2.2);
    \node[right] at (1, -0.6) {$\quad H = \zeta + h_b$};
    
    % Add labels
    \node[right] at (4.5, 0.450) {$\quad \zeta $};
    \node[right] at (4.5, -1.2) {$\quad h_b$};

    % Depth arrow
    \draw[<->] (1.3, 1.1) -- (1.3, -2.1);

    % Arrows and labels for water surface elevation and bathymetry
    \draw[<->] (4.5, 1.3) -- (4.5, 0);
    \node[above] at (4.5, 1.3) { };

    \draw[<->] (4.5, 0) -- (4.5, -2.4);
    \node[below] at (4.5, -2.4) { };
    
    \draw[black , thick] (-4,0) -- (4,0);

\end{tikzpicture}
    \caption{Definition of shallow water elevations. The horizontal line denotes the geoid, where $\zeta = h_b = 0$.}
    \label{fig:elevation_def}
\end{figure}
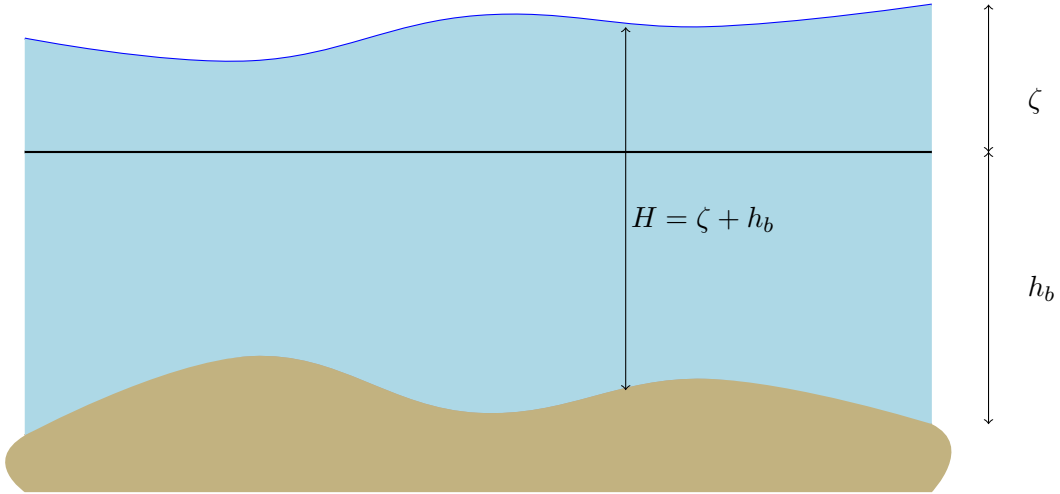

The governing two-dimensional shallow water equations (SWE) which consist of the conservative depth-averaged equations of mass conservation as well as $x$ and $y$ momentum conservation \citep{tan1992shallow} are the model used in this work. This set of equations is stated as follows (see Figure \ref{fig:elevation_def} for a schematic of the shallow water depths):

{Find }  $(\zeta, \uu)$  { such that:}  
\begin{align}   
\ds \frac{\partial  \zeta}{\partial t} + \nabla \cdot (H{\uu})  = R, \text{ in } \Omega,  \label{eq:SWEa}\tag{1a}\\
\ds\frac{\partial (Hu_x)}{\partial t} + \nabla \cdot \left( Hu_x^2 + \frac{g}{2}(H^2-h_b^2), Hu_xu_y \right) - g\zeta \frac{\partial h_b}{\partial x} + \kappa u_x = F_x, \text{ in } \Omega, \label{eq:SWEb}\tag{1b}\ \\ 
\ds\frac{\partial (Hu_y)}{\partial t} + \nabla \cdot \left( Hu_xu_y, Hu_y^2 + \frac{g}{2}(H^2-h_b^2) \right) - g\zeta \frac{\partial h_b}{\partial y} + \kappa u_y = F_y, \text{ in } \Omega, \label{eq:SWEc}\tag{1c}\
\end{align}
where $\zeta$ is the free surface elevation (positive upwards from the geoid),  $h_b$ the bathymetry (positive downwards from the geoid), $H$ is the total water column,  $\uu = \{ u_x,u_y\}^{\text{T}}$ is the depth-averaged velocity field, $\kappa$ is the bottom friction factor, $R$ is the mass source term, and the source terms $F_x,F_y$ represent other forces. These can include Coriolis force, tidal potential forces, wind stresses, and wave radiation stresses, and vertically integrated lateral stresses~\citep{luettich2004formulation}. 
The bottom friction terms can assume a linear or nonlinear form, or a mix of both. For the linear formulation, $\kappa$ is constant; in the nonlinear formulation $\kappa = C_d\sqrt{u_x^2+v_x^2}$ where $C_d$ is the drag coefficient. Commonly used examples of the nonlinear formulation includes Manning's $n$~\cite{Manning1891}, and Chezy~\cite{Prony1790} friction laws.  Coriolis force is expressed as $fHu_y$ and $fHu_x$, where $f = 2 \times 7.29212 \times 10^{-5} \times \sin{\phi}$ and $\phi$ is the latitude. When wind stress is included, it contributes through the pressure gradient terms $\dx{P}$ and $\dy{P}$, acting on the fluid surface.
The computational domain is denoted by $\Omega$ and its boundary $\Gamma$ is typically specified by three distinctive sections $\Gamma = \Gamma_{ocean}\cup\Gamma_{land}\cup\Gamma_{river}$. On these sections, the boundary conditions include specified elevation conditions, zero normal flow, and specified normal flow, respectively.

These equations are effectively used to model flows in the coastal regions, estuaries, and rivers, where the vertical dimension is much smaller than the horizontal scale. This justifies the use of depth-averaging.

\subsection{The Generalized Wave Continuity Equation} \label{sec:GWCE}
In ADCIRC, the continuity equation is formulated differently. Since it is well known that the CG method creates instabilities when used to solve hyperbolic systems, a different formulation called the Generalized Wave Continuity Equation (GWCE) has been used. 
In the original formulation, we first consider  (\ref{eq:SWEa}) without the source term:
\begin{align}
    \dt{\zeta} + \Nabla \cdot (H\uu) = 0
\end{align}
and differentiate in time to obtain:
\begin{align} \label{eq:cont2}
    \frac{\partial^2 \zeta}{\partial t^2} + \Nabla \cdot \left(\dt{H \uu}\right) = 0.
\end{align}
Substituting the momentum equation (in vector form): 
\begin{align}
    \dt{(H\uu)} + \Nabla \cdot (H\uu\uu) + gH\Nabla\zeta + \kappa H \uu = 0
\end{align}
for $\partial H\uu/\partial t$ and (\ref{eq:cont2}) in $\Nabla \cdot (H\uu)$ that arises yields: 
\begin{align} \label{eq:gwce}
    \frac{\partial^2 \zeta}{\partial t^2} + \kappa \dt{\zeta} - \Nabla \cdot [\Nabla \cdot (H\uu\uu) + gH\Nabla\zeta] - H \uu \cdot \Nabla \kappa = 0.
\end{align}
Note that we have ignored the forcing terms except bottom friction for simplicity.
This form of the continuity equation, discretized spatially with finite elements and in time with finite difference, yields better numerical properties by suppressing short waves. 

An additional parameter $\tau_0$ is used in ADCIRC to add weight to the contribution of the primitive continuity equation \ref{eq:SWEa}. This is done by multiplying $\tau_0$ to (\ref{eq:cont2}) and adding it to (\ref{eq:gwce}). This parameter $\tau_0$ is typically set to vary spatially with the depth $H$. 
Still, in practice we sometimes encounter issues when the flow is highly advective flows (either blows up or damps the solution) and difficulties in adding a source term, i.e. precipitation. With the understanding that the continuity equation plays a key role in storm surge simulations, it is worthwhile to explore the viability of using the original continuity equation (\ref{eq:SWEa}). This will be detailed in the following section.

\section{Discontinuous and Continuous Galerkin Discretization} \label{sec:CD-FEM}
\setcounter{equation}{0}

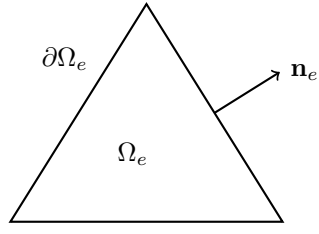
\begin{figure}[h!]
    \centering
\begin{tikzpicture}[scale=3.6, every node/.style={font=\small}]

  % Triangle vertices
  \coordinate (P1) at (0,0);
  \coordinate (P2) at (1,0);
  \coordinate (P3) at (0.5,0.8);
  \coordinate (P4) at (1.5,0.8);

  % Triangle
  \draw[thick] (P1) -- (P2) -- (P3) -- cycle; 
  \node[black] at (0.2,0.6) {$\partial \Omega_e$};
  % Labels
  \node at (0.45,0.25) {$\Omega_e$};

  \draw[->, black, thick] ($(P2)!0.5!(P3)$) -- ++(0.24,0.15) node[right] {$\mathbf{n}_e$};  

\end{tikzpicture}
    \caption{Definition of a triangular element $\Omega_e$}
    \label{fig:elem_def}
\end{figure}

To find an approximate solution for the shallow water equations above, we discretize them in space using finite elements. All solution variables $(\zeta, u_x, u_y)$ and the water column $H$ referenced in this and the following sections are to be understood in the discrete sense and should be referred to by $ \zeta^h, u_x^h, u_y^h,$ and $H^h$, but we will drop the $h$ superscript for ease of notation.
Given the computational domain $\Omega \subset \mathbb{R}^2$, we define the FE mesh partition $\mathcal{T}^h$ of $\Omega$ into a set of non-overlapping elements $\Omega_e \in \mathcal{T}^h$. For each element we also define its boundary $\partial \Omega_e$ and outwards unit normal vector $\mathbf{n}_e$, see Figure~\ref{fig:elem_def}.
With these definitions in hand, we also define two function spaces of linear polynomials: 
\begin{equation} \label{eq:discontspace}
   V_h = \Big\{ v \in L^2(\Omega) \, : \, v|_{\Omega_e} \in P^1(\Omega_e) \, \forall \Omega_e \in  \mathcal{T}^h\Big\},
\end{equation}
\begin{equation} \label{eq:contspace}
   \mathbf{W}_h(\Omega) =   P^1(\Omega) \times P^1(\Omega).
\end{equation}
Hence, $V_h$ is a space of \emph{discontinuous} piecewise linear polynomials, i.e., a discontinuous Galerkin (DG) space and  $\mathbf{W}_h$ a vector valued space of    \emph{continuous}  piecewise linear polynomials,  i.e., a continuous Galerkin (CG) space. 
The elevation solution $\zeta$ will be found in $V_h$ and the velocity solutions $(u_x,v_x)$ in $\mathbf{W}_h$. 
At each timestep $s$, ADCIRC solves for each state variable at the future timestep $s+1$ sequentially:
\begin{enumerate}
    \item Solve for $\zeta^{s+1} $ in the GWCE using $\zeta^s,u_x^s,u_y^s$
    \item Solve for $(u_x^{s+1},u_y^{s+1}) \in \bfm{W}_h$ through the CG method using $\zeta^s,\zeta^{s+1},u_x^s,u_y^s$.
\end{enumerate}
In this work, we modify step (1) to solve the original continuity equation (\ref{eq:SWEa}) using the DG method. The exact interaction and implementation of this coupling will be detailed in the following sections. 
We will also make use of the following variables to simplify notation for the rest of the section:
\begin{itemize}
    \item $N_p =$ number of nodes in the mesh
    \item $N_e = $ number of elements in the mesh
    \item $N_{k,i} = $ node number of the $i^{th}$ local node of element $k$, $1 \leqslant i \leqslant 3$.
    \item $N^{-1}_{k,j} = $ local node number (1, 2, or 3) of node $j$ in element $k$. Returns $-1$ if node $j$ is not connected to element $k$.
    \item $E_j$ = set of elements connected to node $j$
    \item $A_n$ = area of element number $n$.
    \item $\Lambda_j$ = total area of elements connected to node $j$
\end{itemize}

\subsection{Continuity Equation} \label{sec:continuity}
As one of our key goals is a locally mass conservative method, we will discretize the weak form corresponding to the continuity equation~\eqref{eq:SWEa} using a DG scheme. This means, for each element $\Omega_e$, we seek a local solution $\zeta^e$ where $\zeta(\bfm{x},t) \in P^1(\Omega_e)$ and $\zeta(\bfm{x},t) = 0 \; \forall \bfm{x} \notin \Omega_e$ such that:
\begin{align} \label{eq:weak_cont}
\displaystyle   \int_{\Omega_e} \dt{\zeta^e}{v} - \int_{\Omega_e} v\nabla \cdot \{Hu_x,Hu_y\} -  \int_{\Omega_e} Rv = 0,
   \end{align}
for all test functions $v \in V_h$. Next, integrating the spatial divergence term by parts yields the requirement:
\begin{align} \label{eq:weak2}
 \displaystyle  \int_{\Omega_e} \dt{\zeta^e}{v} - \int_{\Omega_e} \nabla v \cdot \FF + \int_{\partial\Omega_e} (\hat{\FF} \cdot \bfm{n}) v - \int_{\Omega_e} Rv= 0, 
\end{align}
where $\hat{\FF}$ is a numerical flux which represents the continuity flux $\FF = [u_xH, \; u_yH]^T$ on $\partial \Omega_e$. As in finite volume schemes, $\hat{\FF}$ accepts values at each side of the element (cell) boundary and connects the otherwise independent elements together. Local mass conservation is satisfied by letting $v=1$ which yields:
\begin{align}
    \int_{\Omega_e} \dt{\zeta^e} - \int_{\partial\Omega_e} (\hat{\FF} \cdot \bfm{n}) - \int_{\Omega_e} R = 0. 
\end{align}
This is simply the integral form of conservation of mass in $\Omega_e$, but now with respect to the numerical flux $\hat{\FF}$.
The global solution is formally defined to be the direct sum of the local solutions:
\begin{align}
    \zeta(\bfm{x},t) = \bigoplus_{e=1}^{N} \zeta^e(\bfm{x},t).
\end{align}
In the case of linear polynomials, each local approximation is represented by 3 local polynomial basis functions $\psi_n(\bfm{x})$:
\begin{align} \label{eq:modal}
    \zeta^e(\bfm{x},t) = \sum_{n=1}^3 \hat{\zeta}_n^e(t)\psi_n(\bfm{x})_, \quad \bfm{x} \in \Omega_e.
\end{align}
Our solver uses the Dubiner basis functions \citep{Dubiner1991-ts} for $\psi_n$ which are orthogonal and therefore yield a diagonal mass matrix which is trivially inverted. Note that we use the notation $\psi_n, \; n=1,2,3$ to represent the linear Dubiner bases $\phi_{00},\phi_{01},\phi_{10}$.
We will also frequently refer directly to the vector of modal coefficients:
\begin{align}
    \hat{\elev}^e \equiv [\hat{\zeta}^e_1, \hat{\zeta}^e_2, \hat{\zeta}^e_3].
\end{align}
Substituting the modal expansion (\ref{eq:modal}) into the weak form (\ref{eq:weak2}) and using  $v = \psi_j, \; j=1,2,3$ as test functions, we obtain: 
\begin{align}
  \displaystyle  \int_{\Omega_e} \left( \frac{\partial}{\partial t} \sum_{i=1}^3  \hat{\zeta}^e_i  \psi_i \right) \psi_j & = \notag \\  -\int_{\partial\Omega_e} & \hat{\FF} \cdot \bfm{n} \psi_j + \int_{\Omega_e} \left( R \psi_j + \frac{d \psi_j}{dx} H u_x  + \frac{d \psi_j}{dy} H u_y \right), \quad j=1,2,3
\end{align}
where we have also grouped the RHS into integrals over the area and edges of element $e$. Factoring out the left-hand side sum yields:
\begin{align}
    \sum_{i=1}^3 \left( \frac{d}{dt}\hat{\zeta}^e_i \int_{\Omega_e} \psi_i \psi_j \right) &= \notag \\ -\int_{\partial\Omega_e} & \hat{\FF} \cdot \bfm{n} \psi_j + \int_{\Omega_e} \left( R \psi_j + \frac{d \psi_j}{dx} H u_x  + \frac{d \psi_j}{dy} H u_y \right), \quad j=1,2,3
\end{align}
which can be written as a $3\times 3$ linear system on each element:
\begin{align}
    \bfm{M}^e\Dt{\hat{\elev}^e} = \bfm{L}^e(\zeta,u_x,u_y),
\end{align}
where: 
\begin{align}
    \bfm{M}^e_{ij} = \int_{\Omega_e} \psi_i \psi_j,
\end{align}
denotes the local mass matrix and $\bfm{L}^e_j$ contains the right hand terms.
As stated, the local mass matrix $\bfm{M}^e$ is trivially inverted and we obtain a system of ODEs for each element $e$:
\begin{align}
    \frac{d}{dt} \modal^e = (\bfm{M}^e)^{-1}\bfm{L}^e(\zeta, u_x, u_y) \equiv \bfm{\widetilde{L}}^e(\zeta,u_x,u_y). 
\end{align}
This is then explicitly solved using the forward Euler method:
\begin{align}
    (\modal^e)^{n+1} := (\modal^e)^{n} + \Delta t \; \bfm{\widetilde{L}}^e(\zeta^n,u_x^n,u_y^n)
\end{align}
In using the DG method to solve the full SWE, it is standard to use a higher-order total-variation bounded Runge-Kutta scheme like the Strong Stability-Preserving RK scheme (SSPRK) \citep{Shu1987-ws}. In this coupled case, we have not encountered stability issues with only using forward Euler, and this significantly reduces the computational cost.

\subsection{Numerical Flux and Boundary Conditions}
To compute the numerical flux between adjacent elements, we use the Local Lax-Friedrichs (LLF) flux. Because we are assuming continuous velocities, this leads to some amount of simplification from the standard form. To start, the LLF flux assumes the form:
\begin{equation}
    \hat{\FF}_n  = \frac{1}{2}(\FF^- + \FF^+) \cdot \bfm{n} + \lambda_{\max} (\zeta^+ - \zeta^-)
\end{equation}
where quantities with superscripts $\pm$ indicate each side of the edge. This flux can be viewed as the average of the two sides of the flux that is damped with an extra viscous term. 
The term $\lambda_{\max}$ refers to the maximum eigenvalue of the Jacobian, which, when we require $u_x = u_x^+ = u_x^-$ and $u_y = u_y^+ = u_y^-$, simplify to:
\begin{align}
    \lambda_1^{\pm} &= u_xn_x + u_yn_y - \sqrt{gH^{\pm}} \\
    \lambda_2 &= u_x n_x + u_yn_y \\
    \lambda_3^{\pm} &= u_xn_x + u_yn_y + \sqrt{gH^{\pm}}.
\end{align}
The resulting LLF flux is then:
\begin{align}
    \hat{\FF}_n(\zeta^+,\zeta^-,u_x,u_y) &= \frac{1}{2}[(u_xH)^+n_x + (u_yH)^+n_y + (u_xH)^-n_x + (u_yH)^-n_y] \notag \\ \qquad &-  \frac{1}{2}\lambda_{\max}(\zeta^+ -  \zeta^-) \notag \\
    &= \frac{1}{2}(H^+ + H^-)(\bfm{u} \cdot \bfm{n}) - \frac{1}{2}\lambda_{\max}(\zeta^+ - \zeta^-).
\end{align}
In the full DG scheme, apart from representing the flux at the Riemann interface, the numerical flux is used to weakly impose boundary conditions. In the DG-CG scheme, some modifications were made. First, integration at zero-normal flow boundaries (land) is skipped completely. This is because we set $\zeta^+ = \zeta^-$ and $\bfm{u} \cdot \bfm{n}$ is set to zero in the CG section. Second, for the specified elevation (ocean) boundaries we do continue to use the weak enforcement by using $\hat{\FF}(\zeta^{BC}, \zeta^-)$. However, we do use the exactly specified elevation values when computing the spatial pressure terms in the momentum equations, described in the next section.

\subsection{Momentum Equation}
To solve the momentum equations, we use the existing CG discretization in ADCIRC as described in the theory report \citep{luettich2004formulation}. 
Specifically, we use the (default) nonconservative formulation which transforms Eqs. (\ref{eq:SWEb}) and (\ref{eq:SWEc}) into
\begin{align} \label{eq:NC_momentum}
     \dt{u_x} &= - {u_x \dx{u_x}} - {u_y \dy{u_x}} - {g \dx{(\zeta +P/g\rho)}} + fu_y - \frac{\kappa u_x}{H\rho} \\
     \dt{u_y} &= - {u_x \dx{u_y}} - {u_y \dy{u_y}} - {g \dy{(\zeta + P/g\rho)}} + fu_x - \frac{\kappa u_y}{H\rho} 
\end{align}
where we have limited the forcing terms to be bottom friction and wind forcing for simplicity. Here, $f$ represents the Coriolis term, $\rho$ the water density, and $P$ the atmospheric pressure at sea surface. We typically use a quadratic bottom friction $\kappa = C_d \sqrt{u_x^2+u_y^2}$ where $C_d$ is the drag coefficient.
We apply the nodal CG  spatial discretization, such that  each global finite element node $j$ defines a global nodal basis function $\phi_j$ which we multiply with~\eqref{eq:NC_momentum} and integrate over the whole domain $\Omega$. The goal is then to find $(u_x,v_x) \in \bfm{W}_h$ such that:
\begin{align}
    \areaa \dt{u_x}\phi_j &= - \areaa{u_x \dx{u_x}}{\phi_j} - \areaa{u_y \dy{u_x}}{\phi_j} - \areaa{g \dx{\zeta+P/g\rho}}{\phi_j}  \notag \\ &+ \areaa fu_y\phi_j - \areaa{\frac{\kappa u_x}{H\rho}}\phi_j, \\
    \areaa{\dt{u_y}}{\phi_j} &= - \areaa{u_x \dx{u_y}}{\phi_j} - \areaa{u_y \dy{u_y}}{\phi_j} - \areaa{g \dy{\zeta+P/g\rho}}{\phi_j} \notag \\  &+ \areaa fu_x\phi_j - \areaa{\frac{\kappa u_y}{H\rho}}\phi_j,
\end{align}
for all $\phi_i \in W_h$.
Applying quadrature rules for linear triangular elements on each term then yields the semidiscrete equations for $\partial u_x/\partial t$ and $\partial u_y / \partial t$. The time derivative term is approximated using a lumped  approach to produce a diagonal mass matrix on the LHS (in the following we omit the formulation for the $y-$momentum equation for brevity):
\begin{align}
    \int_{\Omega} \dt{u_x} \phi_j &\approx \frac{\Lambda_j}{3} \frac{d}{dt}u_{x}(\bfm{x}_j). \\
\end{align}
The advection terms are integrated using a linear integration rule and averaging of the nodal values,
\begin{align}
    \int_{\Omega} u_x \dx{u_x} \phi_j \approx \sum_{n\in E_j} \frac{A_n}{3} \langle{u}_{x}\rangle_n \left(\dx{u_x} \right)_n, \\
    \int_{\Omega} u_y \dy{u_x} \phi_j \approx \sum_{n\in E_j} \frac{A_{n}}{3} \langle{u}_{y} \rangle_n \left(\dy{u_y} \right)_n, \\
\end{align}
where  $\langle u_x \rangle_n$ the average of $u_x$ over element $n$.
Similarly for the surface elevation gradient term:
\begin{align}
    \int_{\Omega} g\dx{\zeta} \phi_j &\approx \frac{1}{3}g\sum_{n\in E_j}A_{n} \left( \dx{\zeta + P/g\rho} \right)_n. \\
\end{align}
Finally, we again use lumped integration for Coriolis and bottom friction terms:
\begin{align}
\areaa fu_x\phi_j &\approx \frac{\Lambda_j}{3} f u_x, \\
    \areaa \frac{\kappa u_x}{H\rho_0}\phi_j &\approx \frac{\Lambda_j}{3} \frac{\kappa u_x}{H\rho_0}. \\
\end{align}
The resulting semidiscrete equations for $u_x$ at each node $j$ are:
\begin{align}
    \Dt{u_x}(\xx_j) &= -\frac{1}{\Lambda_j} \sum_{n\in E_j} A_n \left[ \mean{u_x}_n \left(\dx{u_x} \right)_n + \mean{u_y}_n \left(\dy{u_x} \right)_n \right] \notag \\
   & -\frac{g}{\Lambda_j} \sum_{n\in E_j} A_n \left( \dx{[\zeta + P/g\rho]}\right)_n + \frac{\Lambda_j}{3} f u_y(\xx_j) - \frac{\kappa u_x(\xx_j)}{H\rho},\\
\end{align}
and similarly for $u_y$.
ADCIRC utilizes a two-level time discretization, whereby some terms are averaged between current and future time steps $s$ and $s+1$. The LHS terms use a forward difference:
\begin{align}
    \frac{du_{x}}{dt}(\xx_j) &\approx \frac{u^{s+1}_{x}(\xx_j) - u^{s}_{x}(\xx_j)}{\Delta t}, \\
\end{align}
the advection terms use values at step $n$,
while the barotropic pressure, Coriolis, and bottom friction terms are replaced with their temporal average:
\begin{align}
&\frac{g}{\Lambda_j} \sum_{n\in E_j} A_n \frac{1}{2} \left( \dx{[\zeta + P/g\rho]^s} + \dx{[\zeta + P/g\rho]^{s+1}} \right)_n, \\
   & \frac{1}{2}f(u_x^s+u_x^{s+1}), \\
    &\frac{1}{2}\left(\frac{\kappa u_x^s}{H^s\rho} + \frac{\kappa u_x^{s+1}}{H^{s+1}\rho_0}\right). \\
\end{align}
We then group the future terms to the LHS and current terms to the RHS. The final discretization for $u_x$ is shown below.
\begin{align}
    \left(1+\frac{\Delta t \; \kappa}{2H(\xx_j)} \right) &u_x^{s+1}(\xx_j) - \frac{f\Delta t }{2} u_y^{s+1}(\xx_j) = \notag \\
    &\left(1+\frac{\Delta t \; \kappa}{2H(\xx_j)} \right) u_x^{s}(\xx_j) \notag  \\
    &-\frac{1}{\Lambda_j} \sum_{n \in E_j} A_n \left[ \mean{u_x^s}_n \left(\dx{u_x^s} \right)_n + \mean{u_y^s}_n \left(\dy{u_x^s} \right)_n \right] \notag \\
    &- \frac{g}{\Lambda_j} \sum_{n\in E_j} A_n \frac{1}{2} \left( \dx{[\zeta + P/g\rho]^s} + \dx{[\zeta + P/g\rho]^{s+1}} \right)_n.
\end{align}
When coupled with the discretization of $u_y$, this yields a $2\times2$ linear system for $u_x^{s+1},u_y^{s+1}$ at each node $j$. After abstracting away the known terms, this can be viewed more concisely as
\begin{align}
    \gamma_{\text{fric}}(j) \;u_x^{s+1}(\xx_j) + \lvz{Coriolis}(j) \; u_y^{s+1}(\xx_j) &= 
    \lvx{fric}(j) + \lvx{Coriolis}(j) \notag \\ &+ \sum_{e\in E_j} [\lvx{advect}(e) + \lvx{barotropic}(e)] \notag \\
    &\equiv \bfm{X}(\zeta^s,\zeta^{s+1},u_x^s,u_y^{s}),  \\
    \gamma_{\text{fric}}(j) \;u_y^{s+1}(\xx_j) + \lvz{Coriolis}(j) \; u_x^{s+1}(\xx_j) &= 
    \lvy{fric}(j) + \lvy{Coriolis}(j) + \notag \\ &\sum_{e\in E_j} [\lvy{advect}(e) + \lvy{barotropic}(e)] \notag  \\
    &\equiv \bfm{Y}(\zeta^s,\zeta^{s+1},u_x^s,u_y^{s}).
\end{align}
where the nodal terms are denoted with $(j)$ and elemental terms with $(e)$. The $\bfm{X},\bfm{Y} \in R^{N_p}$ are the load vectors.
During the assembly stage, we first loop through each element and add its contribution to the appropriate locations of the load vectors, and finally loop through each node and solve for $u_x^{s+1},u_y^{s+1}$ using Cramer's rule (see Alg. \ref{alg:cg}). 

\subsection{Coupling between different solution spaces}
This section details the implementation issue of mixing solutions from different solution spaces $V_h$ and $W_h$. 
In a previous work \cite{Dawson2006-ve}, the coupling is done through the projection of discontinuous $\zeta$ into the continuous space $W_h$, defined by finding $\eta \in W_h$ satisfying:
\begin{align}
    \areaa \eta w = \areaa \zeta w, \quad \forall w \in W_h.
\end{align}
Which is equivalent to the constraint:
\begin{align}
    \areaa \eta \phi_j = \areaa \zeta \phi_j, \quad j=1,..,N_p.
\end{align}
where $\phi_j$ are, as before, the linear basis functions in the CG discretization.
This implies that the continuous solution still conserves mass globally:
\begin{align}
     \areaa \eta  = \areaa \zeta .
\end{align}
Note that while local mass conservation is not guaranteed in $\eta$, it is still guaranteed during the computation of $\zeta$ at every timestep.
Similar to the CG momentum formulation, the LHS integral is approximated using mass lumping. For the RHS, if we follow the CG integration choice of using the elemental average, we obtain:
\begin{align}
    \frac{\Lambda_j}{3} \eta_j &= \sum_{n\in E_j} \int_{\Omega_n} \zeta^n\phi_j \notag \\
    &\approx \sum_{n\in E_j} \frac{A_n}{3} \mean{\zeta}_n,
\end{align}
i.e.,
\begin{align}
    \eta_j \approx  \frac{1}{\Lambda_j} \sum_{n\in E_j} A_n \mean{\zeta}_n.
\end{align}
Note that mass lumping is used to approximate the LHS integral to avoid the need to solve a global linear system which would be too costly in practice, on top of the extra computations in the DG scheme. The accuracy of the RHS approximation can, however, be improved with other integration schemes. For example, we can first express the modal $\zeta$ as a nodal expansion of the local shape functions $\Phi^e_k$:
\begin{align}
    \zeta^e(\xx) = \sum_k \zeta^e(\xx_k)\Phi^e_k(\xx),
\end{align}
where each $\Phi^e_k = 1$ at local node $k$ and zero at all other nodes.
Then each summand in the RHS integral is:
\begin{align}
    \int_{\Omega_e} \zeta^e\phi_j &= \int_{\Omega_e} \sum_k \zeta^e(\xx_k)\Phi^e_k(\xx)\phi_j(\xx) \; d\xx \notag \\ 
    &= \sum_k \zeta^e(\xx_k)\int_{\Omega_e} \Phi^e_k(\xx)\phi_j(\xx) \; d\xx \notag \\
    &= \sum_k \zeta^e(\xx_k)\int_{\Omega_e} \Phi^e_k(\xx)\Phi^e_{N^{-1}_{e,j}}(\xx) \; d\xx.
\end{align}
This yields:
\begin{align}
    \eta(\xx_j) = \frac{3}{\Lambda_j}\sum_{e\in E_j}\sum_{k=1}^3 \zeta^e(\xx_k)\int_{\Omega_e} \Phi^e_k(\xx)\Phi^e_{N^{-1}_{e,j}}(\xx) \; d\xx.
\end{align}

Alternatively, it is useful to consider this issue from a computational standpoint. We have "elemental" and "nodal" variables which need to communicate with each other.
On the DG side, we need to perform Gaussian quadrature involving the nodal values $u_x,u_y$ when calculating the edge and area integrals in (\ref{eq:weak2}). This is straightforward, since we can simply project to the DG modal representation and reuse the quadrature weights.
This follows from the choice of the Dubiner basis $\psi_n$ (for $p=1$). For a function $u \in W_h$,  the modal expansion of $u^e$ on element $\Omega_e$ is:
\begin{align}
    u^e(\bfm{x}) = \sum_{n=1}^3 \hat{u}^e_n \psi_n(\bfm{x}), \quad \bfm{x} \in \Omega_e
\end{align}
where:
\begin{align}
    \hat{u}^e_1 &= \frac{1}{3}(u(\xx^e_1)+u(\xx^e_2)+u({\xx^e_3})) \\
    \hat{u}^e_2 &= -\frac{1}{6}(u(\xx_1^e)+u_2(\xx_2^e)) + \frac{1}{3}u(\xx_3^e) \\
    \hat{u}^e_3 &= -\frac{1}{2}(u(\xx_1^e) - u(\xx_2^e)).
\end{align}
This projection is then applied to $u_x$ and $u_y$ for each time step in the DG computation. 

Conversely, the conversion from modal values to nodal values shows up in several places and is less well-defined, as we are effectively reducing degrees of freedom, and there are several options. 
We require the gradient values $\partial\zeta/\partial x$ and $ \partial\zeta/\partial y$ on each element for the barotropic pressure terms, and the actual nodal values of $\zeta$ for the rest of the terms like bottom friction and Coriolis. Thus there are two approaches in doing this:
\begin{itemize}
\item Directly compute the gradient from the DG representation on each element to use in the pressure terms, and simultaneously define nodal values for $\zeta$ for the other terms.
    \item Only define nodal values for $\zeta$, then compute everything using these.
    
\end{itemize}
In practice, we found that the first approach is less accurate, possibly due to the inconsistency between using the elemental and nodal values. Moreover, there is no way to avoid computing nodal values of $\zeta$, as are eventually required in the CG wetting and drying scheme (see next section). The second method is more straightforward and robust, since it obeys the strict separation between continuous and discontinuous representation. In this case, we follow the standard procedure used in finite volume schemes which performs a {weighted average} of each cell (element) surrounding a node based on its area:
\begin{align}
    \eta(\xx_i) \equiv \frac{1}{\Lambda_i }{\sum_{n \in E(i)} A_n\zeta^n(\bfm{x}_i)  }.
\end{align}
Here, each element connected to node $i$ computes its local DG solution at that node, and this is averaged among all elements. This is essentially what we arrived at by defining the projection from the discontinuous space to the continuous space, except that the RHS integral is approximated by using the nodal value as the average.

\subsection{Summary of DG and CG discretization}
We have so far described the DG formulation in abstract terms. For use and clarity in subsequent sections, we list in Table~\ref{tab:basic_routines} the basic routines that we will reference.
\begin{table}[h!]
    \centering
    \begin{tabular}{l c p{0.4\textwidth}}
    \hline
        Routine & Output Dimension & Description \\
        \hline
        Nodal$\rightarrow$Modal($u$) & $3 \times N_e$ & Returns the elemental DG representation of nodal function $u$ \\
        Modal$\rightarrow$Nodal($\hat{u}$) & $N_p$ & Returns the nodal averages of modal $\hat{u}$ \\ 
        Modal$\rightarrow$Edge($\hat{u},l,i$) & $1$ & Returns the value of modal $\hat{u}$ at the $i^{th}$ quadrature point on edge number $l$ \\
        Modal$\rightarrow$Quad($\hat{u},e,i$) & $1$ & Returns the value of modal $\hat{u}$ at the $i^{th}$ quadrature point on element $e$ \\
         \hline
    \end{tabular}
    \caption{Basic routines for DG computations.}
    \label{tab:basic_routines}
\end{table}
The pseudocode for the DG continuity solver is shown in Algorithm \ref{alg:dg}
and for the CG momentum solver in Algorithm \ref{alg:cg}, where $\eta$ is now used to denote nodal water elevation.  
\begin{algorithm}[h!]
    \caption{DG solver at timestep $s$.}
\label{alg:dg}
    \begin{algorithmic}[1]
        \Procedure{DG\_Timestep}{}
        \State $\hat{u}_x,\hat{u}_y$ := \Call{Nodal$\rightarrow$Modal}{$u_x^s,u_y^s$}
        \For{each edge $l$}
        \State $e^+$ := right element of $l$
        \State $e^-$ := left element of $l$
        \For{each quadrature point $i$}
        \State $U$ := \Call{Modal$\rightarrow$Edge}{$\hat{u}_x,l,i$}
        \State $V$ := \Call{Modal$\rightarrow$Edge}{$\hat{u}_y,l,i$}
        \State $\eta^+$ := \Call{Modal$\rightarrow$Edge}{$\zeta^{e^+},l,i$}
        \State $\eta^-$ := \Call{Modal$\rightarrow$Edge}{$\zeta^{e^-},l,i$}

        \State Compute $\widehat{\FF} := \widehat{\FF}(U,V,\eta^+,\eta^-,h_b)$
        \For{each mode $k=1,2,3$}
        \State Accumulate the quadrature $\bfm{L}_k^{e^+}$ using $\widehat{\FF} \cdot \bfm{n} \psi_k$
        \State Accumulate the quadrature $\bfm{L}_k^{e^-}$ using $\widehat{\FF} \cdot \bfm{n} \psi_k$
        \EndFor
        \EndFor
        \EndFor

        \For{each element $e$}
        \For{each quadrature point $i$}
        \State $U := $ \Call{Modal$\rightarrow$Quad}{$\hat{u}_x, e, i$}
        \State $V := $ \Call{Modal$\rightarrow$Quad}{$\hat{u}_y, e, i$}
        \State $\eta := $ \Call{Modal$\rightarrow$Quad}{$\zeta, e, i$}
        \State $h := $ \Call{Modal$\rightarrow$Quad}{$h_b, e, i$}

        \State Compute $\FF := [U(\eta+h), V(\eta+h)]^T$
        \For{each modal basis $k=1,2,3$}
        \State Accumulate $\bfm{L}^e_k$ with $R(e)\psi_k$ and $\nabla \psi_k \cdot \FF$
        \EndFor
        \EndFor
        \EndFor
        \For{each element $e$}
        \For{each mode $k$}
        \State $(\zeta^e)^{s+1} := (\zeta^e)^{s} + \Delta t \; \bfm{L}^e$
        \EndFor
        \EndFor
        \State $\eta^s := \eta^{s+1}$
        \State $\eta^{s+1}$ := \Call{Modal$\rightarrow$Nodal}{$\zeta^{s+1}$}
        \EndProcedure
    \end{algorithmic}

\end{algorithm}

\begin{algorithm}[h!]
    \caption{CG momentum solver at timestep $s$.}
\label{alg:cg}
    \begin{algorithmic}[1]
        \Procedure{CG\_Momentum\_Timestep}{$s$}
        \For{each element $e$}
        
        \State $n_1 := N(e,1)$
        \State $n_2 := N(e,2)$
        \State $n_3 := N(e,3)$
       
        \For{$i=1,2,3$}
        \State $\eta^s_{i} := \eta^s(\xx_{n_i})$
        \State $\eta^{s+1}_{i} := \eta^{s+1}(\xx_{n_i})$
        \EndFor
  
        \For{$i=1,2,3$}
          \State $H^s_{i} := \eta^s_{i} + h(\xx_{n_i})$
        \State $H^{s+1}_{i} := \eta^{s+1}_{i} + h(\xx_{n_i})$

          \EndFor
        \State Compute elemental $\alpha,\beta_{\text{barotropic}}$ from $\eta^s_i, \eta^{s+1}_i, P^s(\xx_i),P^{s+1}(\xx_i), \;i=1,2,3$
       
        \State Compute elemental $\alpha,\beta_{\text{advect}}$ from $u_x^n,u_y^n$
        \For{$i = 1,2,3$} \Comment{Global assembly}
        \State $\bfm{X}_{n_i} := \bfm{X}_{n_i} + \alpha_{\text{barotropic}} + \alpha_{\text{advect}}$
         \State $\bfm{Y}_{n_i} := \bfm{Y}_{n_i} +\beta_{\text{barotropic}} + \beta_{\text{advect}}$
        \EndFor
        \EndFor

\Comment{ Assembly of the load vectors }
    \For{each node $i$} 
    \State Compute nodal $\alpha,\beta_{\text{fric}}$ using $H^s_i, H^{s+1}_i,u_x^s,u_x^{s+1},u_y^s,u_y^{s+1}$
    \State Compute nodal $\alpha,\beta_{\text{coriolis}}$ using $u_x^s,u_x^{s+1},u_y^s,u_y^{s+1}$
    \State $\bfm{X}_{i} := \bfm{X}_{i} +\alpha_{\text{fric}} + \alpha_{\text{coriolis}}$
    \State $\bfm{Y}_{i} := \bfm{Y}_{i} +\beta_{\text{fric}} + \beta_{\text{coriolis}}$
    \EndFor
    \For{each node $i$}
    \State Compute $\lvz{fric}, \lvz{coriolis}$
    \State Solve the system for $u_x^{s+1}(\xx_i),u_y^{s+1}(\xx_i)$
    \EndFor
    \EndProcedure
    \end{algorithmic}

\end{algorithm}

\section{Wetting and Drying} \label{sec:wetdry}
The previous discretization alone is not sufficient for the general case. In the context of storm surge simulation, the domain often includes the coastline which contains an interface between wet ($H > 0$) and dry ($H = 0$) regions. This presents a unique difficulty since the SWE and its discretization is only defined for $H > 0$. One approach is to treat this as a moving boundary problem, where the domain changes with time. Another approach, used by both ADCIRC and DG-SWEM in different ways, applies an artificial, thin layer of water over the whole domain. This effectively models dry regions as regions with very little flow. A node an an element is marked as "wet" or "dry" through comparison with a specified minimum  water depth $H_0$. This comparison is repeated at every time step, and dry nodes/elements are processed in different ways.

\subsection{CG approach}
In ADCIRC, we maintain two wet/dry state arrays, one for the nodes and one for the elements. We denote them here as $\Theta_i\; (1 \leqslant i \leqslant N_p)$ and $\omega_i \; (1 \leqslant i \leqslant N_e)$, respectively. The entries are set to 0 (dry) or 1 (wet) based on five main criteria, evaluated sequentially:
\begin{enumerate}
    \item Any node with depth less than $H_0$ is flagged as dry, as long as it has been wet for some amount of time steps
    \item Any element that has one dry node is checked to see if the velocities from the remaining wet nodes are sufficient to wet that node
    \item Any element with flows coming from "barely wet" nodes (i.e. $H < cH_0$ for some $c > 1$) is flagged as dry
    \item Any node lying on an incoming flux boundary is wet
    \item Any node connected to only dry elements is flagged as dry, even if $H > H_0$.
\end{enumerate}
During the assembly of momentum load vectors, an element $k$ contributes to the load vectors if and only if $\Theta_{N(k,1)}\Theta_{N(k,2)}\Theta_{N(k,3)}\omega_k = 1$. That is, partially dry elements are ignored in the computation.

\subsection{DG approach}
Wetting and drying in DG-SWEM is considerably simpler as we only have to deal with elements and not nodes \citep{bunya2009wetting}. It is constructed such that elemental mass is always conserved, and momentum whenever possible. For each element we apply the \textsc{PositiveDepth()} procedure which performs the following:
\begin{enumerate}
    \item If the water depth at each vertex is greater than $H_0$, do nothing
    \item If the mean water depth is less than $H_0$, set the depth at each vertex to be the mean value (and recompute the modal representation). Velocity at each vertex is set to zero.
    \item Otherwise, redistribute the water such that the depth at each node is greater than $H_0$, but the relative ordering is still preserved. This also maintains mass conservation. Momentum values at the vertices are also redistributed accordingly.
    \item If the element was previously wet and case (2) was encountered, set $\omega_k := 0$, otherwise set $\omega_k := 1$. 
    \item If the element was previously dry, compute $|\zeta_k|_{Hmax}$, the surface elevation at the point of greatest depth, and $|h_k|_{min}$, the shallowest point of bathymetry. If $|\zeta_k|_{Hmax} > H_0 - |h_k|_{min}$, set $\omega_k := 1$, otherwise set $\omega_k := 0$.
\end{enumerate}
Computations of integral terms are then modified for dry elements ($\omega = 0$). For example, edge integrals where both elements are dry are skipped.
\subsection{Combined approach}
Using each approach above separately requires that we maintain three separate wet/dry states: one nodal and one elemental for CG, and one elemental for DG. In practice, we found that this adds a lot of computational overhead and leads to instability; moreover the conversion between CG-DG states is not well-defined. In the current approach, we instead maintain only the CG wet/dry states $\Theta, \omega$, and treat DG conditions as adjustments to those states as sparingly as possible. To achieve this, we modify the positive depth operator and formally summarize it in Algorithm \ref{alg:pd}. Like the DG approach, mass in each element is always conserved. We also explicitly mark the element as wet $(\omega = 1)$ in the first case.

\begin{algorithm}[h!]
\caption{Modified positive depth operator}
\label{alg:pd}
    \begin{algorithmic}
    \Procedure{PositiveDepth}{}
    \For{each element $e$}
    \State Compute the water elevation $\zeta_1,\zeta_2,\zeta_3$ at the vertices from $\elev^e$
       \State Compute vertex depths $H_i := h_{N_{e,i}} + \zeta_i$ for $i=1,2,3$
       \State $\overline{H} := (H_1+H_2+H_3)/3$
       \If{$H_i > H_0, \; \forall i=1,2,3$}
       \State $\omega_e := 1$
       \For{$i=1,2,3$}
       \State $\Theta_i := 1$
        \EndFor
       \State continue
       \ElsIf{$\overline{H} < H_0$}
       \For{$i=1,2,3$}
       \State $\zeta_i := \overline{H} - h_{N_{e,i}}$ 
           \EndFor
       \State $\omega_e := 0$ 

       \Else
       \State Redistribute mass among $\zeta_i$ such that $\zeta_i + h_{(N(e,i)} \geqslant H_0$
        \For{$i=1,2,3$}
         \State $(u_x)_{N_{e,i}} := 0$
       \State $(u_y)_{N_{e,i}} := 0$
       \EndFor
       \EndIf
    \State Reproject nodal $\zeta_1,\zeta_2,\zeta_3$ onto $\elev^e$
    \EndFor

    \EndProcedure
    \end{algorithmic}
\end{algorithm}
Additionally in the CG portion, we augment the 5 criteria above with one simple rule:
\begin{itemize}
    \item Any node with depth greater than $H_0$ is flagged as wet
\end{itemize}
This is necessary because it allows for a node to become wet \textit{without} incoming flux, e.g. from rainfall or source term. The existing criteria only wets nodes by assuming some sort of momentum, and will fail to wet a mesh that starts out completely dry and without flux boundary conditions. We also aggressively reset $u_x^s,u_y^s$ to zero when the element is partially dry. This was found to be necessary to keep some cases stable. However, this is a limitation as it does reduce flux across some wet/dry interfaces. Unlike the DG approach, it is not possible to simply redistribute the corresponding momentum values due to $u_x,u_y$ being continuous.

\section{Incorporating Rainfall Data} \label{sec:rain_models}

Having the capability of  adding rainfall as a spatially and temporally variable source to the SWE and our simulations is one of the main goals of the DG-CG coupling. 
\begin{figure}[h!]
    \centering
    \includegraphics[width=0.995\linewidth]{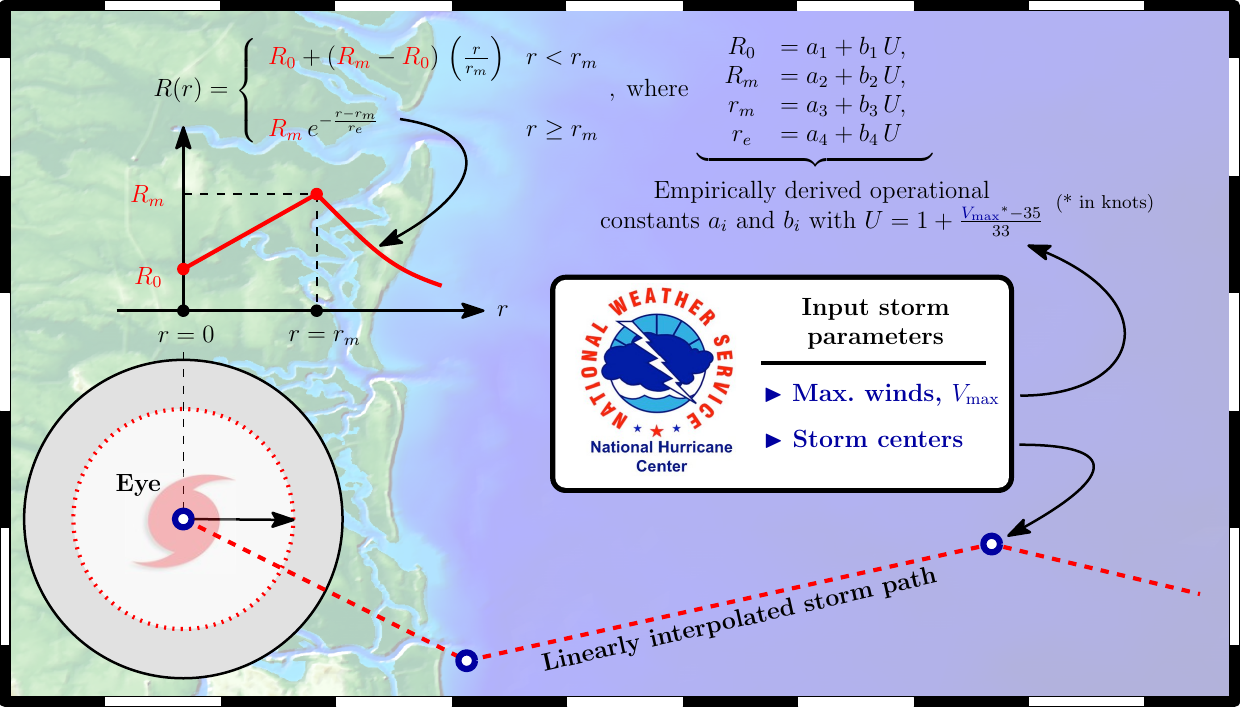}
    \caption{Usage of the R-CLIPER parametric model.}
    \label{fig:Usage of the R-CLIPER parametric model.}
\end{figure}
Specifically, we want to satisfy three requirements:
\begin{itemize}
\item Rain should not introduce instability to the solution
\item The source of rain should be easily extensible, e.g. constant rain, parametric rain, observed rain, etc.
\item Rain should work on both wet and dry elements, and dry elements should be able to become wet from rain
\end{itemize}
The first requirement is satisfied from the fact that DG uses the primitive continuity equation, and rain is simply the source term $R$ which can be produced through any method. The second requirement is satisfied from the fact that we compute rain on the area integral \textit{before} wet/dry check is performed. 

We essentially have two options: forecast rainfall data from atmospheric models and observed and spatially interpolated rainfall data available post rain or storm event. As an overarching goal of the present work is to develop numerical compound flood models that are capable of forecasting, post event observed data is less relevant. However, for validation and hindcasting purposes, high quality post-event rainfall data is critical.

During tropical cyclones, rainfall patterns in the vicinity of the storm exhibit defined structures that can be exploited by so-called parametric rainfall models. The basic idea behind this approach is to construct rainfall fields using simple analytic expressions based on a small number of (predicted or observed) storm parameters that are made publicly available by the National Hurricane Center (NHC) pre and post storm events. This type of approach has been implemented in the code based on the so-called R-CLIPER (Rainfall CLImatology and PERsistence) and the Interagency Performance Evaluation Task Force Rainfall Analysis (IPET) model, which computes rainfall rates $R(r)$ at any point in the grid, where $r$ is the distance from that point to the storm center. 
As an example, the R-CLIPER datapath is illustrated in Figure \ref{fig:Usage of the R-CLIPER parametric model.}; see \citep{Tuleya:2007,Brackins2020-xp} for more details on general parametric schemes. Note that the primary inputs of the model are the reported storm centers and maximum wind velocities, which are generally provided at 6-hour intervals by the NHC. Within DG-SWEM, linear interpolation is used to obtain the storm parameters at the model time step, and the rainfall rate at each finite element mesh node is computed based on its radial distance from the storm center.

Alternatively, the program should also accept observed or reanalysis rainfall data in hindcasting scenarios in the GRIB2 format \citep{grib2} which is read during each timestep. As this type of data is given with spatially varying distributions over the ocean and land, this data is readily interpolated onto finite element meshes. This option has been implemented in DG-SWEM and  will be incorporated into  ADCIRC in the future. 

\section{Numerical Experiments} \label{sec:experiments}
In this section, we thoroughly evaluate the developed solver by performing extensive numerical experiments that are designed to validate the solver. 
 First, we consider simplified test cases where analytic solutions exist in simple rectangular and annular geometries. 
This is followed by the classical Shinnecock Inlet test case, a modified lake-at-rest~\cite{leveque1998balancing} test case, and a relatively new compound flood modeling benchmark of the Neches River. Last, we perform large-scale validation test cases for Hurricanes Ike (2008) and Harvey (2017) where we simulate the entire western North Atlantic Ocean, Caribbean Sea, and Gulf of Mexico.

\subsection{Lynch and Gray test case}
\begin{figure}[h!]
    \centering
    \includegraphics[width=0.8\linewidth]{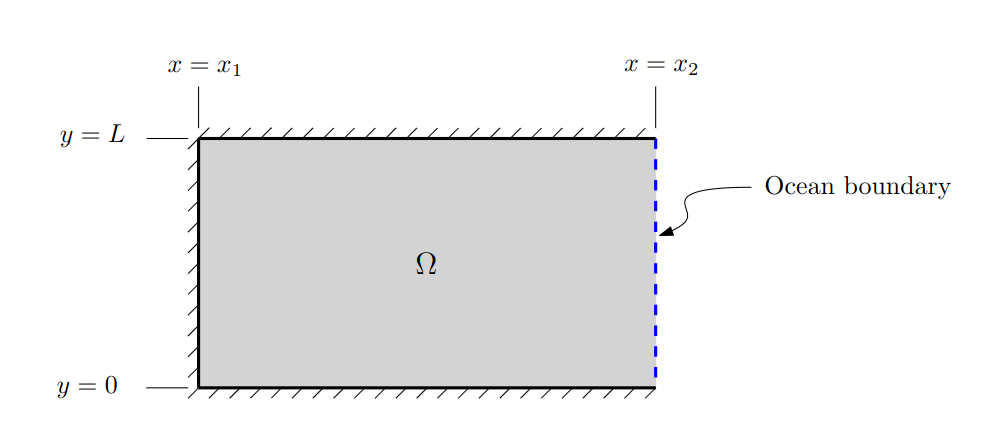}
    \caption{Domain and boundary conditions of the Lynch and Gray test case.}
    \label{fig:lg_domain}
\end{figure}

To verify the numerical accuracy of the scheme, we investigate the convergence of the error as we refine the mesh spacing $h$. This test case was first described in \citep{Lynch1978-ih} which also presented an analytical steady-state solution to be used in the verification. This is for the linearized version of the SWE, obtained by neglecting the advection terms and linearizing the friction term:
\begin{align}
    \dt{\zeta} + \nabla \cdot (H\bfm{u}) &= 0 \\
    \dt{u_x} + g \dx{\zeta} + \tau u_x &= 0 \\
    \dt{u_y} + g \dy{\zeta} + \tau u_y &= 0.
\end{align}
The domain consists of wall boundaries (zero normal-flow) and a periodic elevation (ocean) boundary (Figure \ref{fig:lg_domain}). The corresponding analytical solution, uniform in the y-direction, when the bathymetry is flat is given by 
\begin{align}
    \zeta(x,t) = \text{Re}\left\{ \zeta_0 \cdot e^{i\omega t}\frac{\cos(\beta (x-x_1))}{\cos(\beta (x_2-x_1))} \right\},
\end{align}
where $\zeta_0$ is the forcing amplitude, $\omega$ the angular frequency of the forcing, $\tau$ the bottom friction, $h_0$ the bathymetry, and $\beta = \sqrt{(\omega^2 - i \omega \tau)/gh_0}$.

\begin{figure}[h!]
    \centering
    \includegraphics[width=0.49\linewidth]{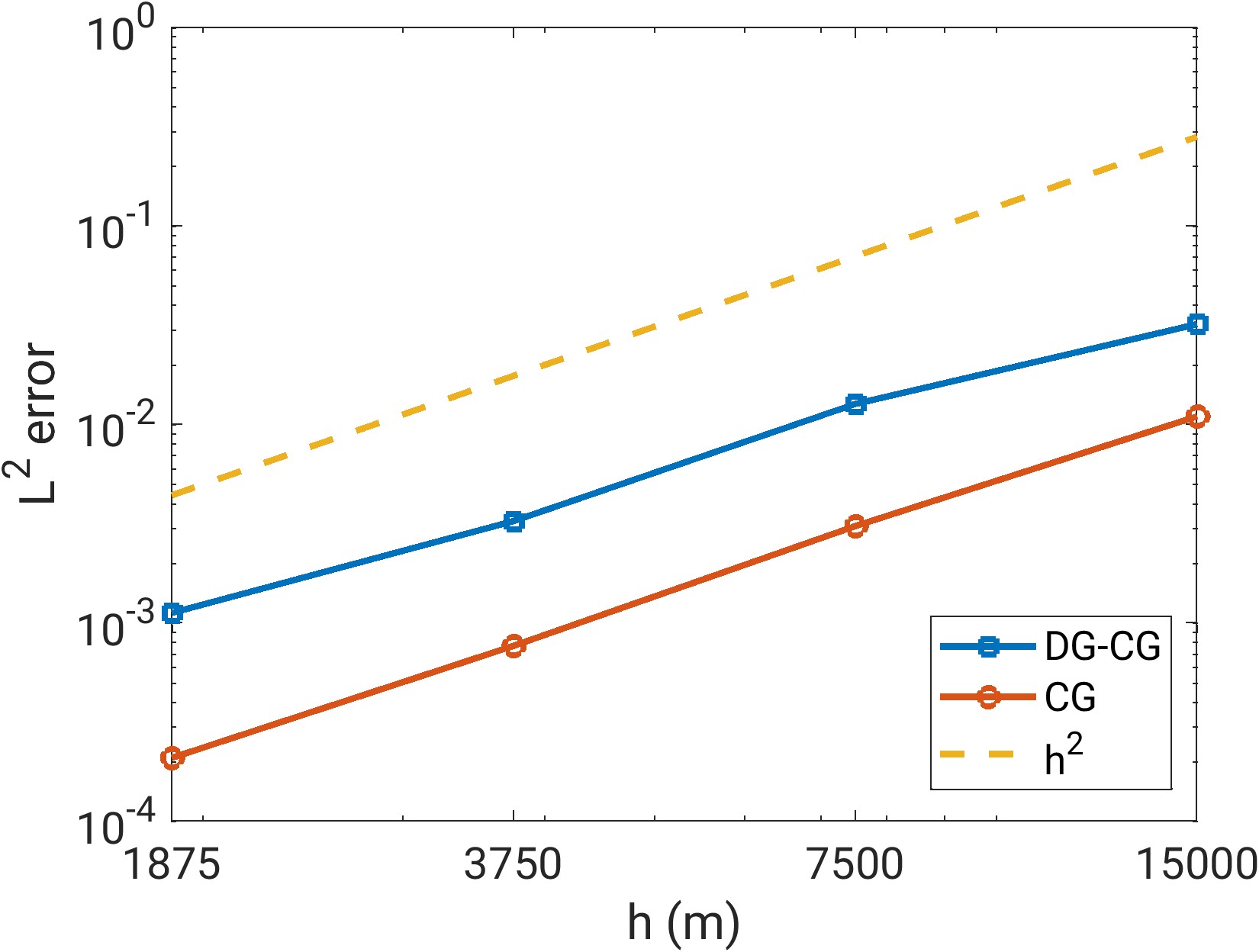}
    \includegraphics[width=0.49\linewidth]{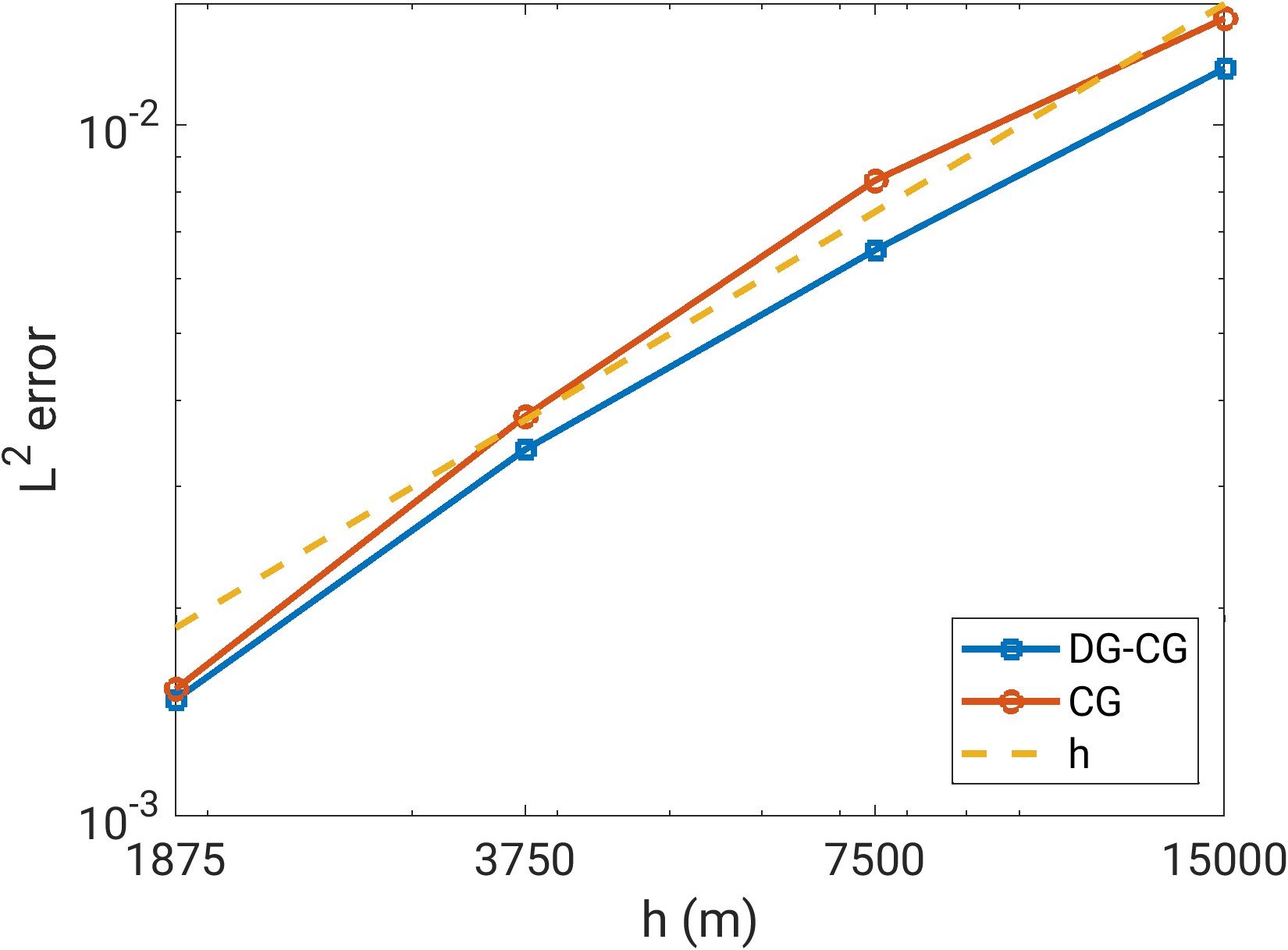}
    \caption{$h-$convergence  in surface elevation (left) and $u_x$ (right) for the Lynch and Gray test case.}
    \label{fig:error}
\end{figure}

In our versions of the mesh, we set $x_1 = 60,000$ m, $x_2 = 150,000$ m, $\zeta_0 = 0.3$, $\omega = 0.0001407$, $\tau = 0.005$, and $h_0 = 3$ m. 
The element spacings are 1,875, 3,750, 7,500, and 15,000 meters. We run the test case with $\Delta t = 1$ s for 5 days, at which the nodal outputs are compared. For evaluation, we use the nodal $L^2$ error, defined as
\begin{align}
    \text{error}_{L^2} = \sqrt{\frac{1}{N} \sum_i(s(\bfm{x}_i) - s_h(\bfm{x}_i))^2}
\end{align}
where $i$ ranges over all the nodes.
The convergence plot of the error is shown in Figure \ref{fig:error} and the values are presented in Table \ref{tab:error}. In the pure CG case, we do observe an optimal convergence rate of 2, while in the DG-CG case, the rate is close to optimal after reaching the asymptotic range of convergence. This is likely an effect of nodal averaging. Strictly speaking, the $L^2$ error should take into account the linear interpolation of each element, but we choose the nodal definition here because that is what matters in practice.
\begin{table}[h!]
    \centering
    \begin{tabular}{c c c c c}
       $h$ (m) & $L^2$ error, CG  &  Rate, CG & $L^2$ error, CG-DG & Rate, DG-CG \\
       \hline 

       15000 & $\sci{1.1}{-2}$ & - & $\sci{3.2}{-2}$ &  - \\
        7500 & $\sci{3.1}{-3}$ & 1.836 & $\sci{1.3}{-2}$ &  1.336  \\
       3750 & $\sci{8.0}{-4}$ & 2.004   & $\sci{3.3}{-3}$ &  1.959  \\
        1875 & $\sci{2.0}{-4}$ & 1.876 & $\sci{1.0}{-3}$ & 1.538 \\
         \hline 
    \end{tabular}
    \caption{Error and convergence rates on $h$ for the Lynch and Gray test case. }
    \label{tab:error}
\end{table}
\begin{table}[h!]
    \centering
    \begin{tabular}{c c c c c}
       $h$ (m) & $L^2$ error, CG  & Rate, CG & $L^2$ error, CG-DG & Rate, DG-CG \\
       \hline 
       15000 & $\sci{1.43}{-2}$ &  -  & $\sci{1.2}{-2}$ &  -  \\
        7500 & $\sci{8.3}{-3}$ &  0.7795  & $\sci{6.6}{-3}$ & 0.8741   \\
       3750 & $\sci{3.8}{-3}$ &  1.133  & $\sci{3.4}{-3}$ &  0.9583  \\
       1875 & $\sci{1.5}{-3}$ & 1.313 & $\sci{1.5}{-3}$ & 1.203 \\       
         \hline 
    \end{tabular}
    \caption{Error and convergence rates of $u_x$ on $h$ for the Lynch and Gray test case. }
    \label{tab:u-error}
\end{table}

\subsection{Quarter Annular Harbor}
\begin{figure}[h!]
    \centering
    \includegraphics[width=0.40\linewidth]{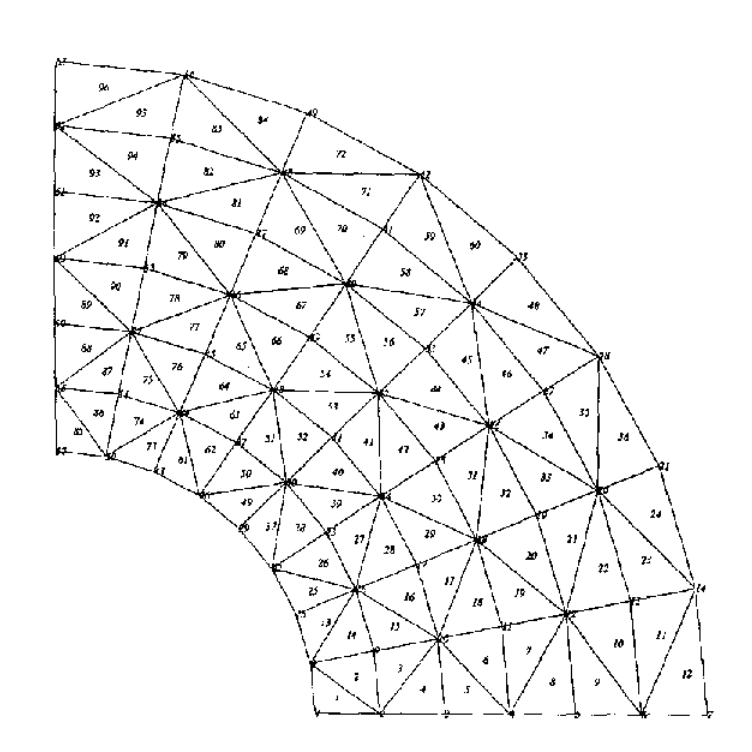}
    \includegraphics[width=0.40\linewidth]{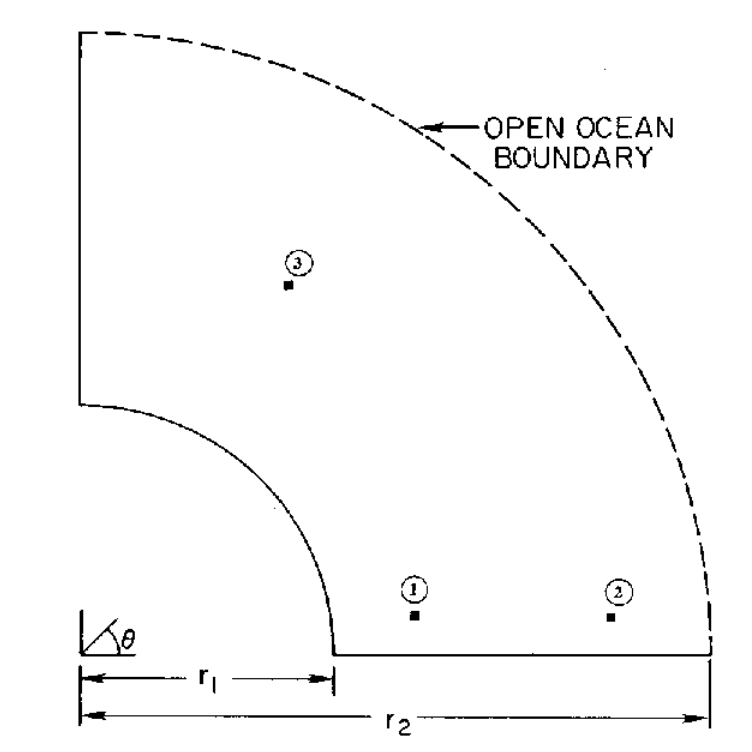} 

    \caption{Left: the mesh used in the Quarter Annular test case. Right: the boundary condition and locations (1-3) used to compare outputs.}
    \label{fig:qa_domain}
\end{figure}

\begin{figure}
    \centering
    \includegraphics[width=0.40\linewidth]{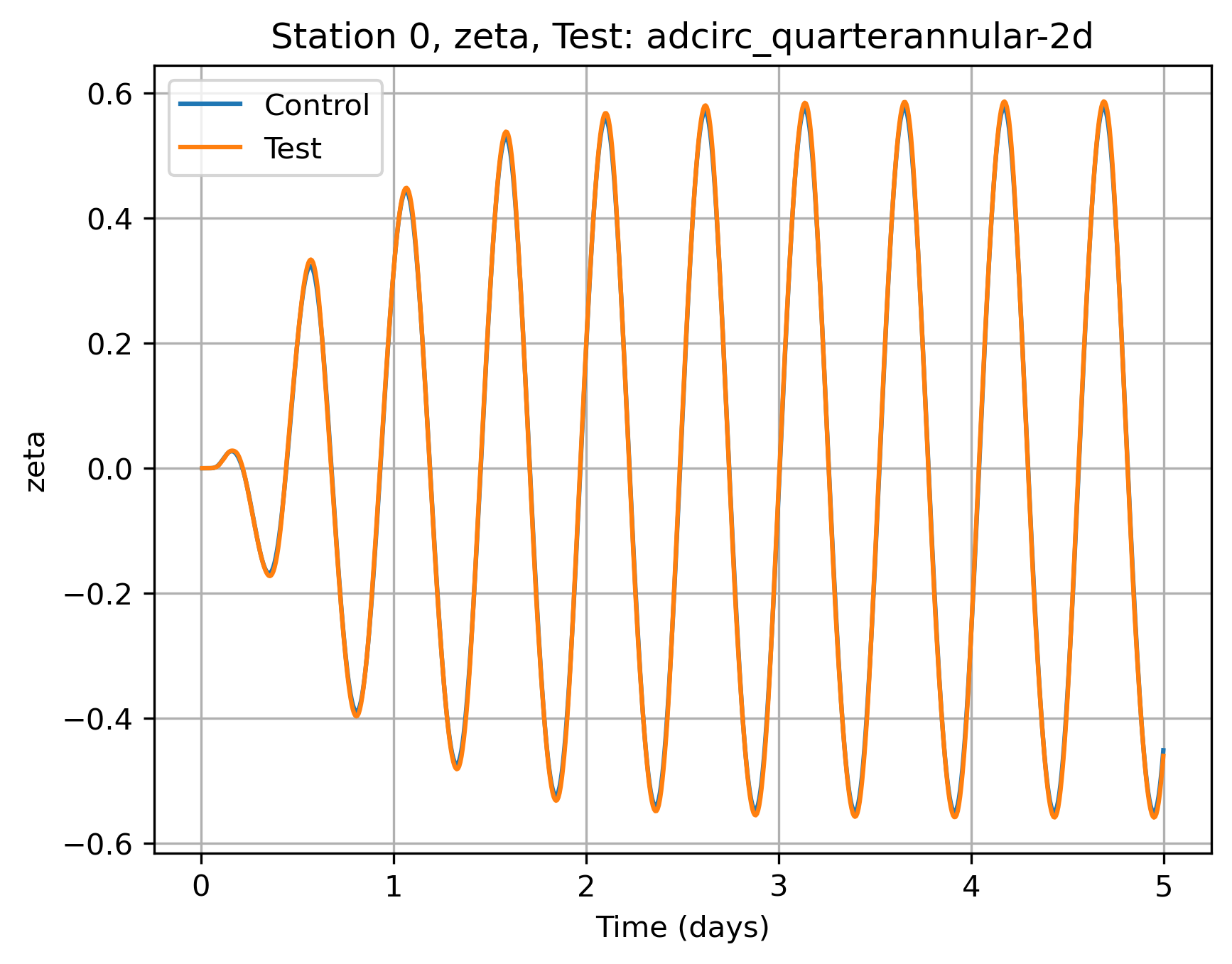}
    \includegraphics[width=0.40\linewidth]{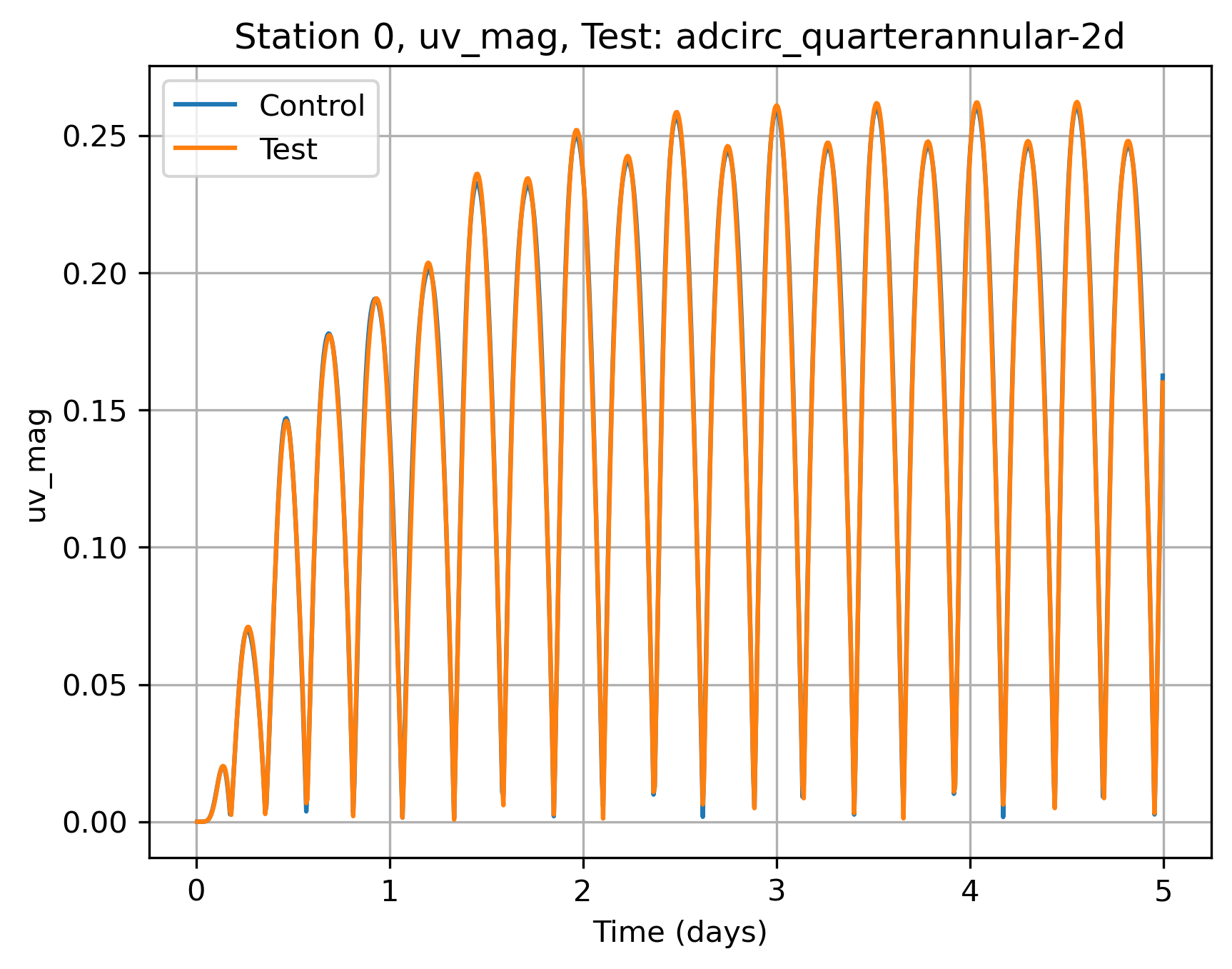} \\
      \includegraphics[width=0.40\linewidth]{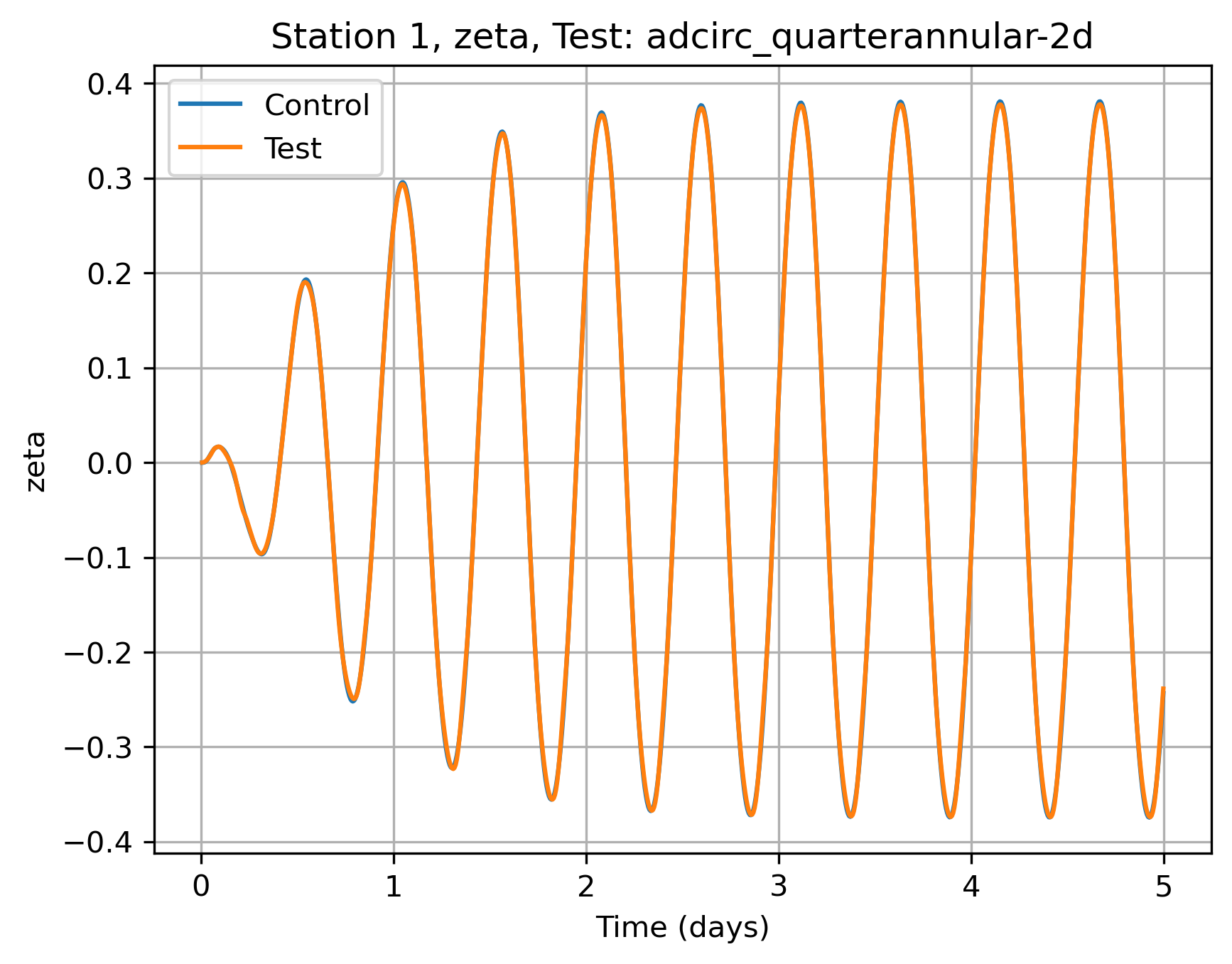}
    \includegraphics[width=0.40\linewidth]{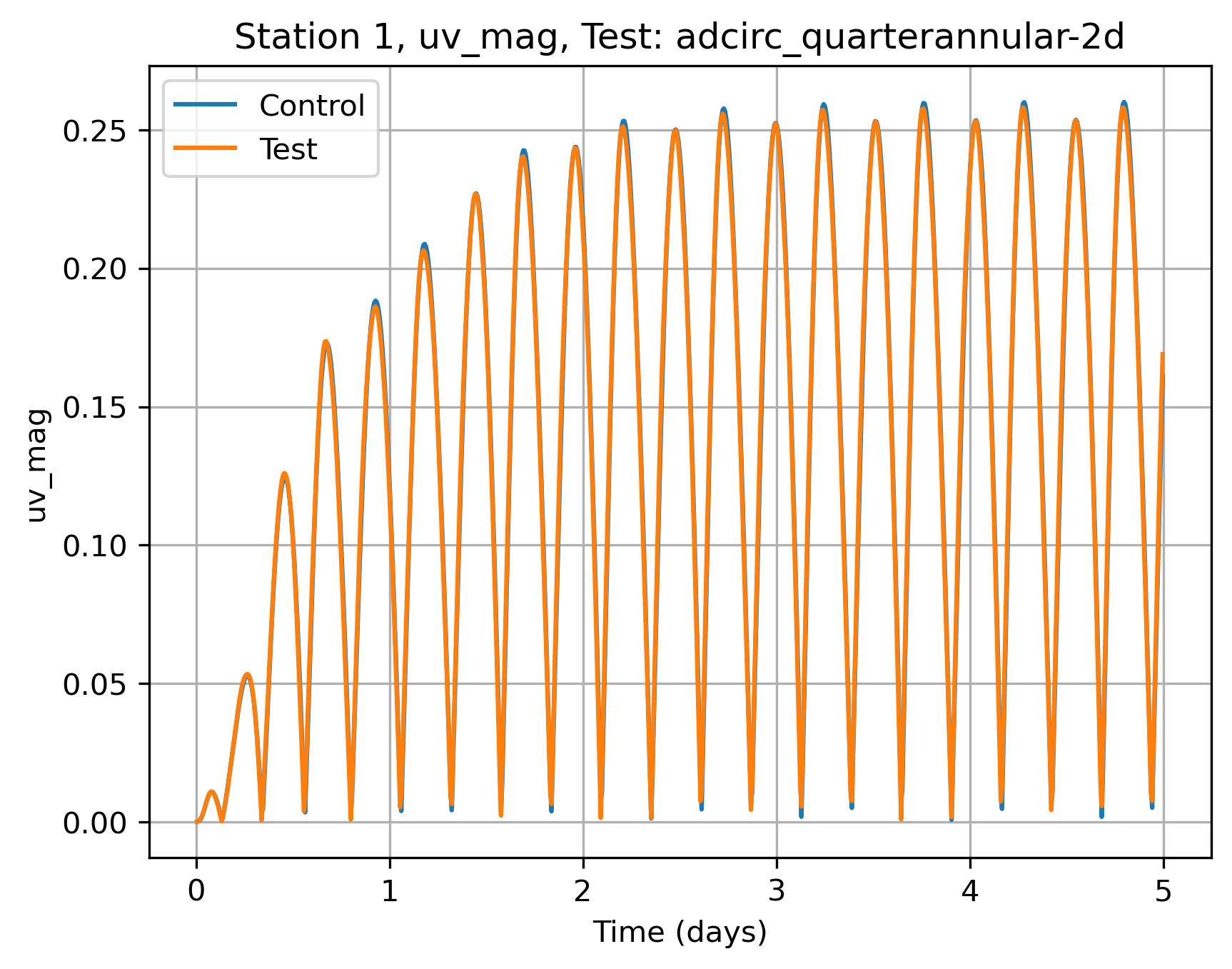}\\
      \includegraphics[width=0.40\linewidth]{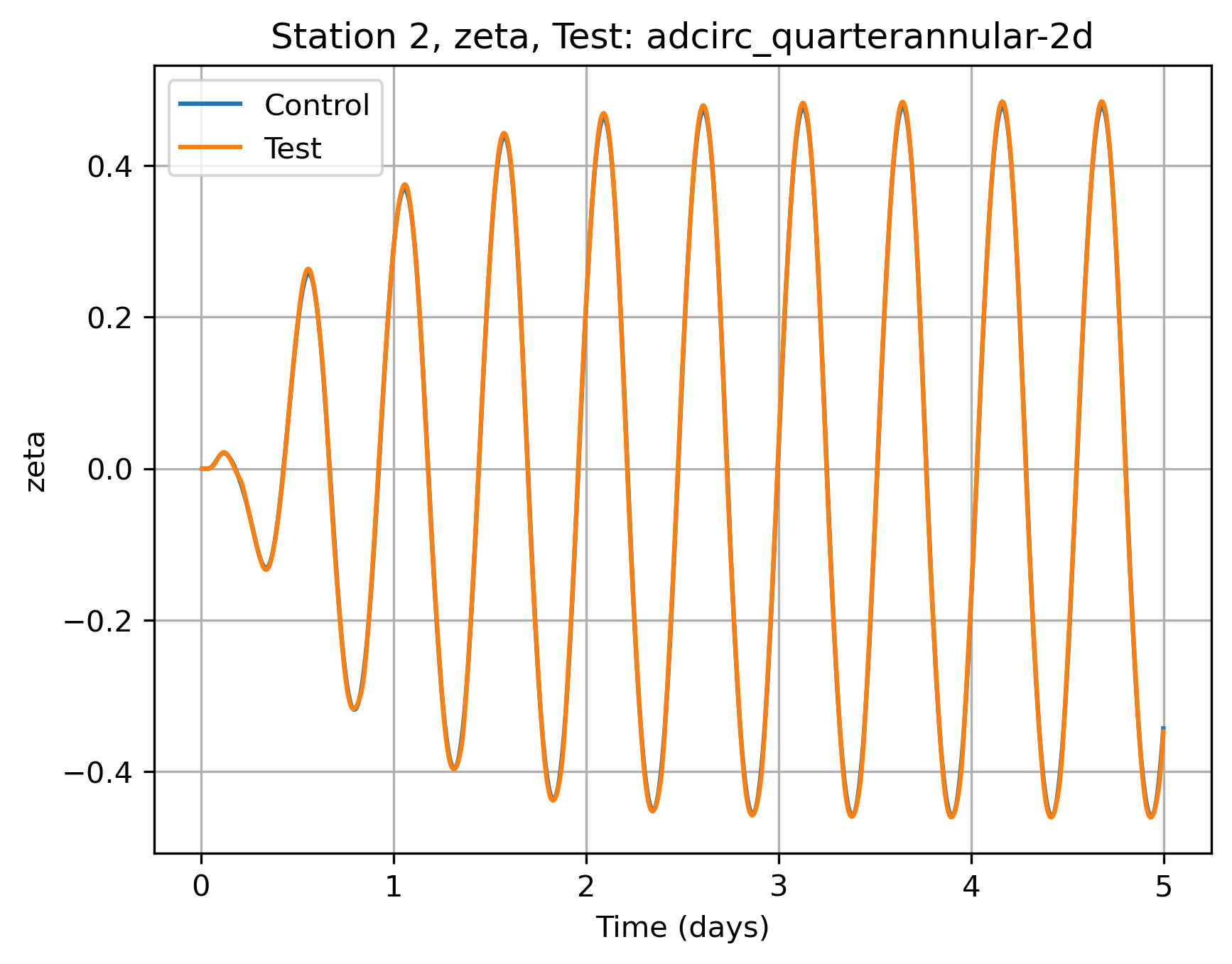}
    \includegraphics[width=0.40\linewidth]{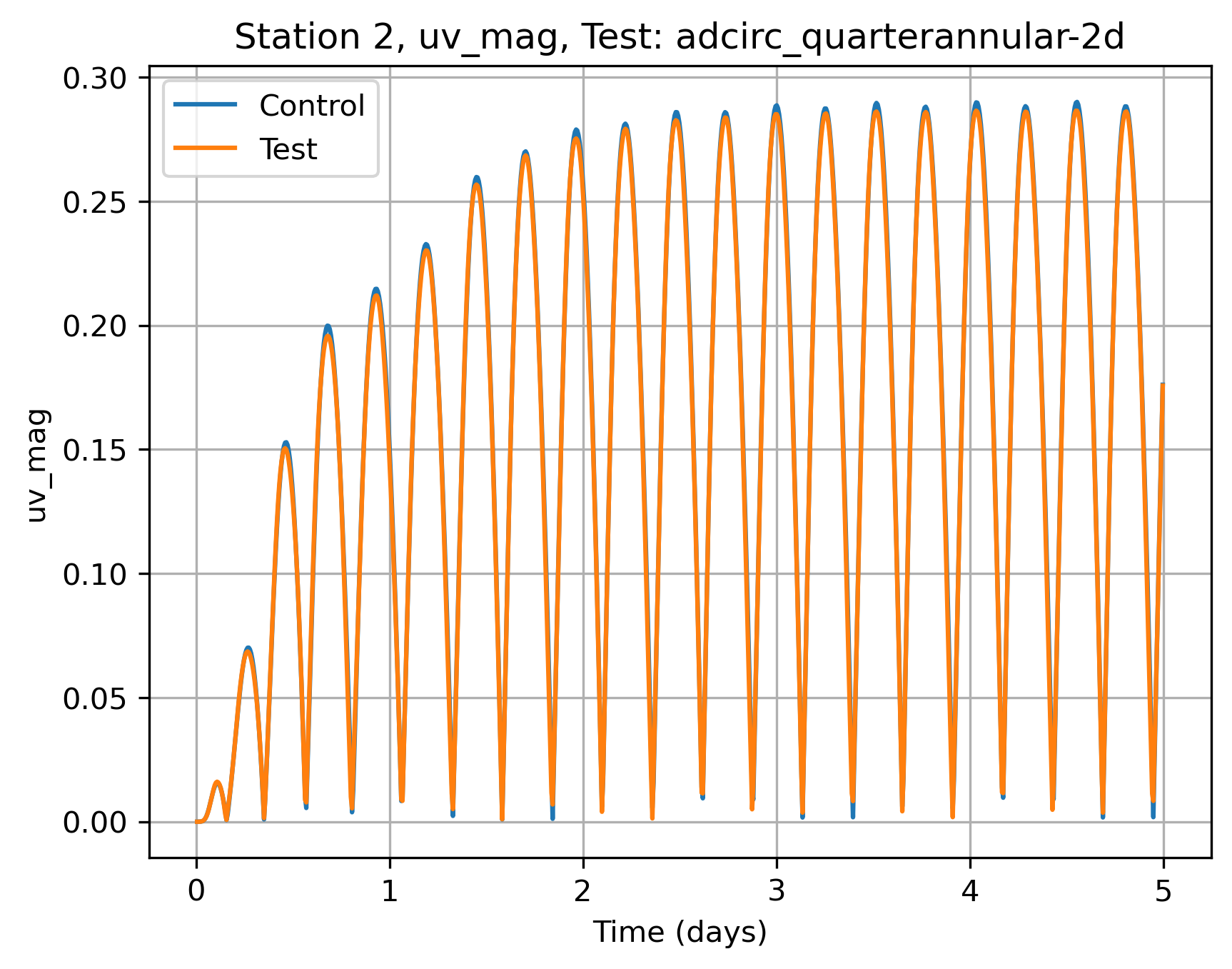}

    \caption{Time series output of CG ("Control") and DG-CG ("Test") for the Quarter Annular case at three locations, labeled in Figure \ref{fig:qa_domain}. The left column shows water elevation $\zeta$ and the right shows $||\uu||$.}
    \label{fig:qa_output}
\end{figure}

This test case is based on Lynch and Gray polar geometry domain with varying bathymetry; it is one of the standard test cases in the ADCIRC repository and is used to check for spurious oscillations and/or dissipation of a scheme. The grid and boundary conditions are shown in Figure \ref{fig:qa_domain}. The grid contains 96 elements and 63 nodes.
The domain ranges from $r_1=60,960$ m to $r_2=152,400$m, with bathymetry depth varying from $h_1=3.048$m to $h_2=19.05$m quadratically.  The radial spacing is 15,240 m. The elevation boundary is forced periodically with an amplitude of 0.3048 m and a period of 44,712 s. This case was run for 5 days with $\Delta t=174.656$s. Comparison of output elevation and velocities at the three stations is shown in Figure \ref{fig:qa_output} and indicates minuscule differences from the reference CG scheme.

\subsection{Shinnecock Inlet} \label{sec:shin}
\begin{figure}
    \centering
    \includegraphics[width=0.80\linewidth]{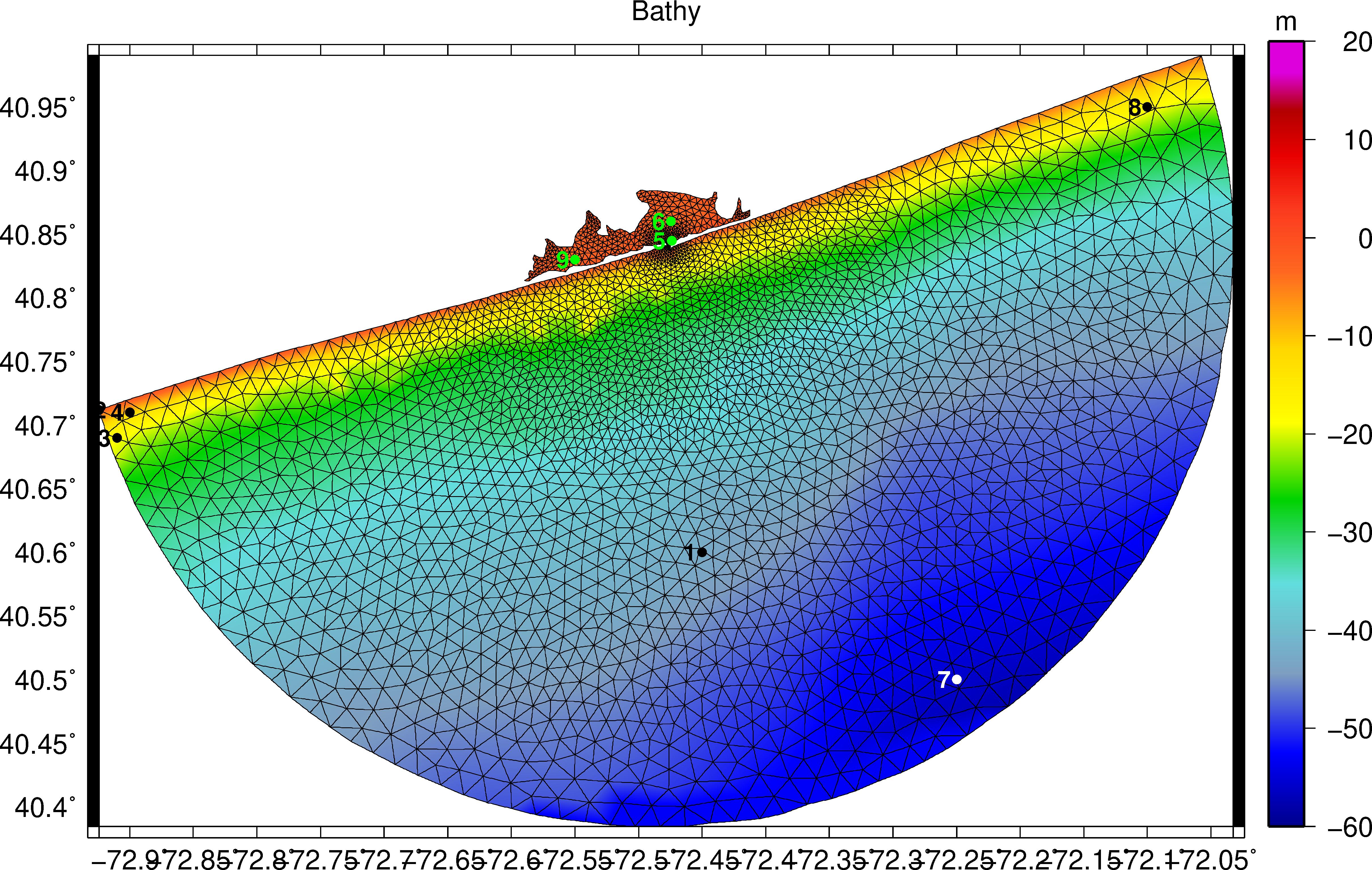}
    \caption{Bathymetry and discretization of the Shinnecock Inlet case. Labels 1-9 indicate the locations used to compare outputs.}
    \label{fig:shin-bathy}
\end{figure}

\begin{figure}
    \centering
    \includegraphics[width=0.80\linewidth]{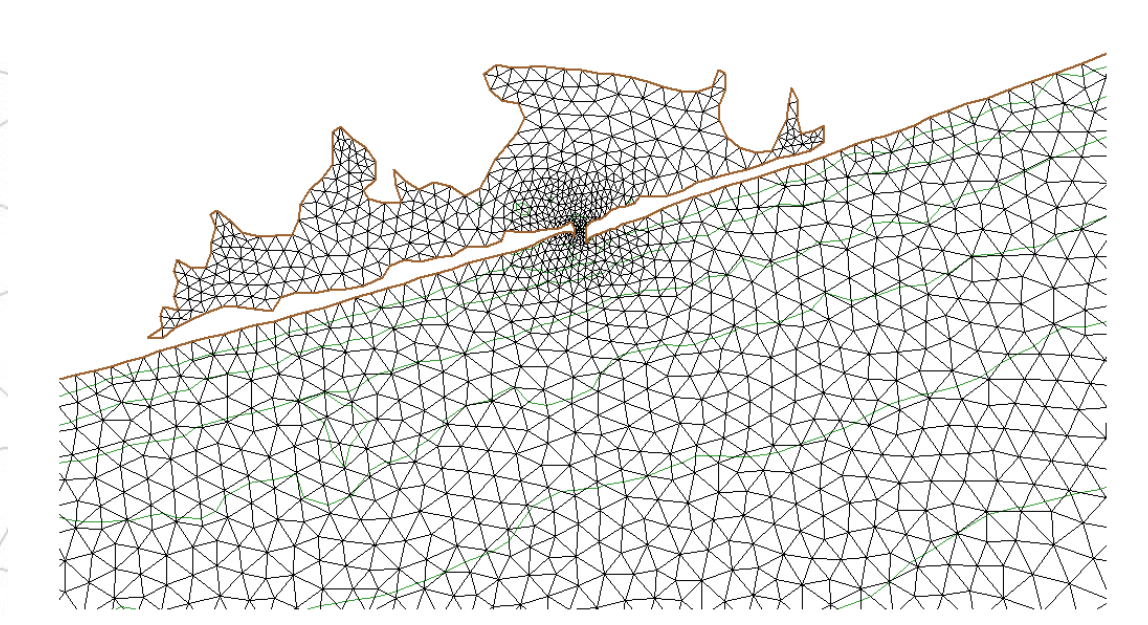}
    \caption{Zoom-in of the finite element mesh at the Shinnecock Inlet entrance.}
    \label{fig:shin_zoom}
\end{figure}

\begin{figure}
    \centering
    \includegraphics[width=0.99\linewidth]{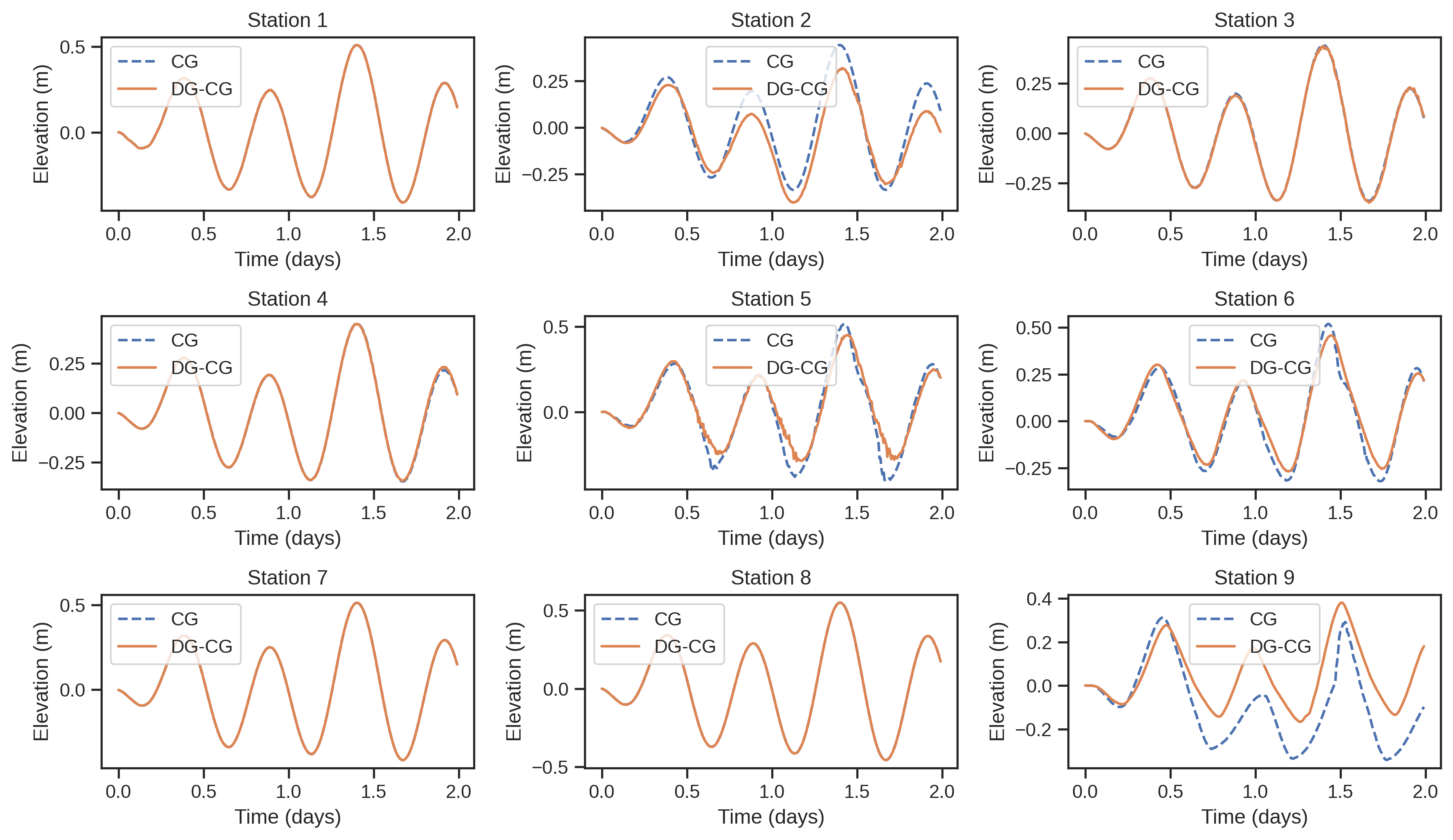}
    \caption{Water elevation outputs for the Shinnecock Inlet case at the 9 locations.}
    \label{fig:shin-stations}
\end{figure}

This test case is also a standard one in ADCIRC, and was developed in a study by the United States Army Corps of Engineers. It is an inlet located at the outer shore of Long Island. It contains 5,780 elements and 3,070 nodes, see Figure \ref{fig:shin-bathy}, and the resolution increases rapidly as we go closer to the inlet, this is shown in Figure \ref{fig:shin_zoom}. Beyond the barrier, there is also wetting and drying, making this a practical test case that is not too large. The outer segment is forced periodically, and  advection and hybrid bottom friction are enabled.
This case was run for 2 days with $\Delta t = 1$s.  Output comparison between CG and DG-CG are shown in Figure \ref{fig:shin-stations}. At stations away from the choke point, we observe almost identical results. At the choke point, we start to observe differences in the peak values of the time series which are likely caused by the different wetting and drying schemes. 
\subsection{Rain on a hill - modified lake-at-rest}
\begin{figure}[h!]
    \centering
    \includegraphics[width=0.80\linewidth]{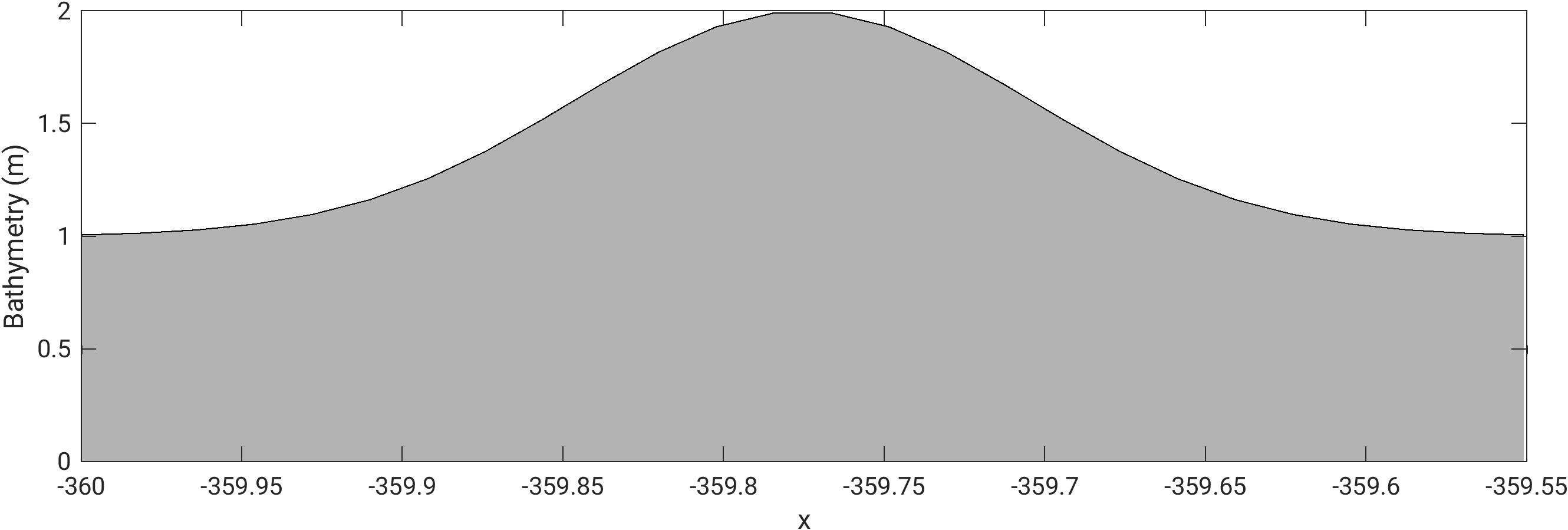}
    \caption{Bathymetry along the x-direction of the mass conservation test case. Dimension is in spherical coordinates.}
    \label{fig:tub}
\end{figure}

In this test case, we fill a small rectangular box with constant rain. The mesh contains 325 nodes and 576 elements, and spans 4,500 m in the y-direction and 9,000 m in the x-direction. 
%The coordinates are [-4.003e7, -3.998e7; 0, 7991]. 
All boundaries are walls, and the x-varying bathymetry is set to be: 
\begin{align}
    h_b(x) = e^{-w^2(x-c)^2} + 1,
\end{align}
where $w = 0.001$ and $c$ the midpoint to represent a slope, see Figure \ref{fig:tub}. The offset is chosen so that the simulation starts completely dry (since the geoid is zero by default). We choose a thin layer $H_0$ of $1 \times 10^{-4} \mm$. With this test case,  we seek to to illustrate the following three items:
\begin{itemize}
    \item Total mass is conserved from the addition of a source term in the continuity equation.
    \item The modified wetting/drying scheme allows a completely dry domain to become wet from a source term alone. 
    \item The velocity remains close to zero when the whole domain is submerged and the water levels out, i.e., this resembles a lake-at-rest test case at this steady state. 
\end{itemize}

To verify the correctness and stability of rain simulation, we compare the total volume of rain to the total volume of water in the box after 2 days, after which the rain is stopped. 
The volume of the hump is approximately $ 4.847 \times 10^7 \mm^3. $
After 2 days, the surface elevation is approximately 2.417 m, uniformly throughout the domain. Thus the net volume in the domain is: 
\begin{align*}
(2.417 \times 9000 \times 4500) \mm^3 - (4.847 \times 10^7) \mm^3 = 4.941 \times 10^7 \mm^3. 
\end{align*}
The rain intensity is $7.0556 \times 10^{-6}$ m/s (1 inch per hour). Thus the total volume of rain after 2 days is: 
\begin{align*}
(7 \times 10^{-6} \times 86400 \times 2 \times 9000 \times 4500) \mm = 4.937 \times 10^7 \mm^3. 
\end{align*}
The corresponding relative error is then $0.07 \%$. 
 Note that, unlike the original lake-at-rest case, where the bathymetry starts out fully submerged, the steady state velocity
retains some amount of the initial perturbation and decays slowly to around $10^{-3}$ to $10^{-5}$ m/s in magnitude.

\subsection{Neches River} \label{sec:neches}
\begin{figure}[h!]
    \centering
    \includegraphics[width=0.7\textwidth]{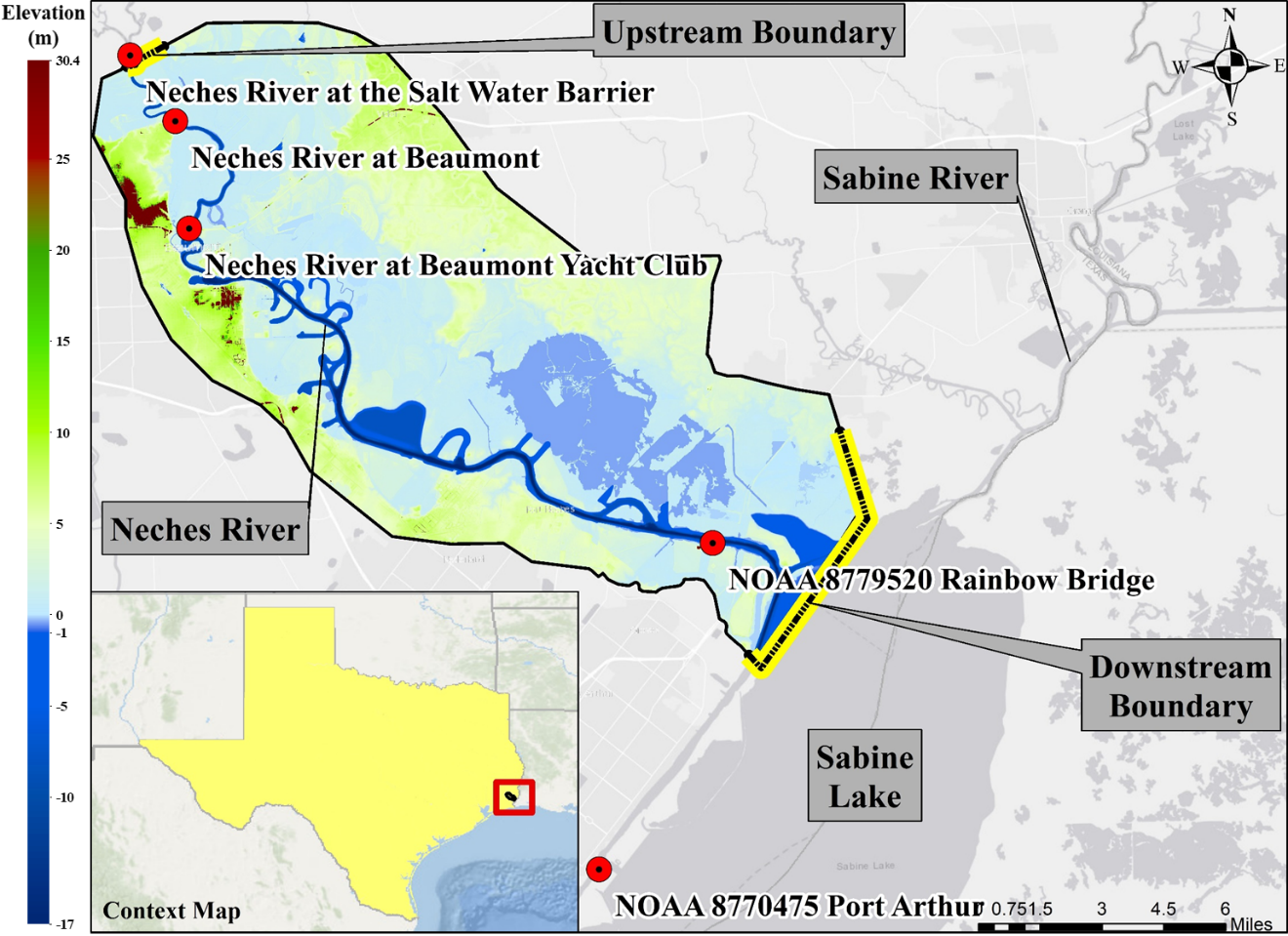}

    \caption{Computational domain of the Neches river case. Observation gauges are marked as red circles and boundaries are highlighted in yellow \citep{loveland2021developing}.}
    \label{fig:neches}
\end{figure}

\begin{figure}[h!]
    \centering

    \includegraphics[width=0.4\textwidth]{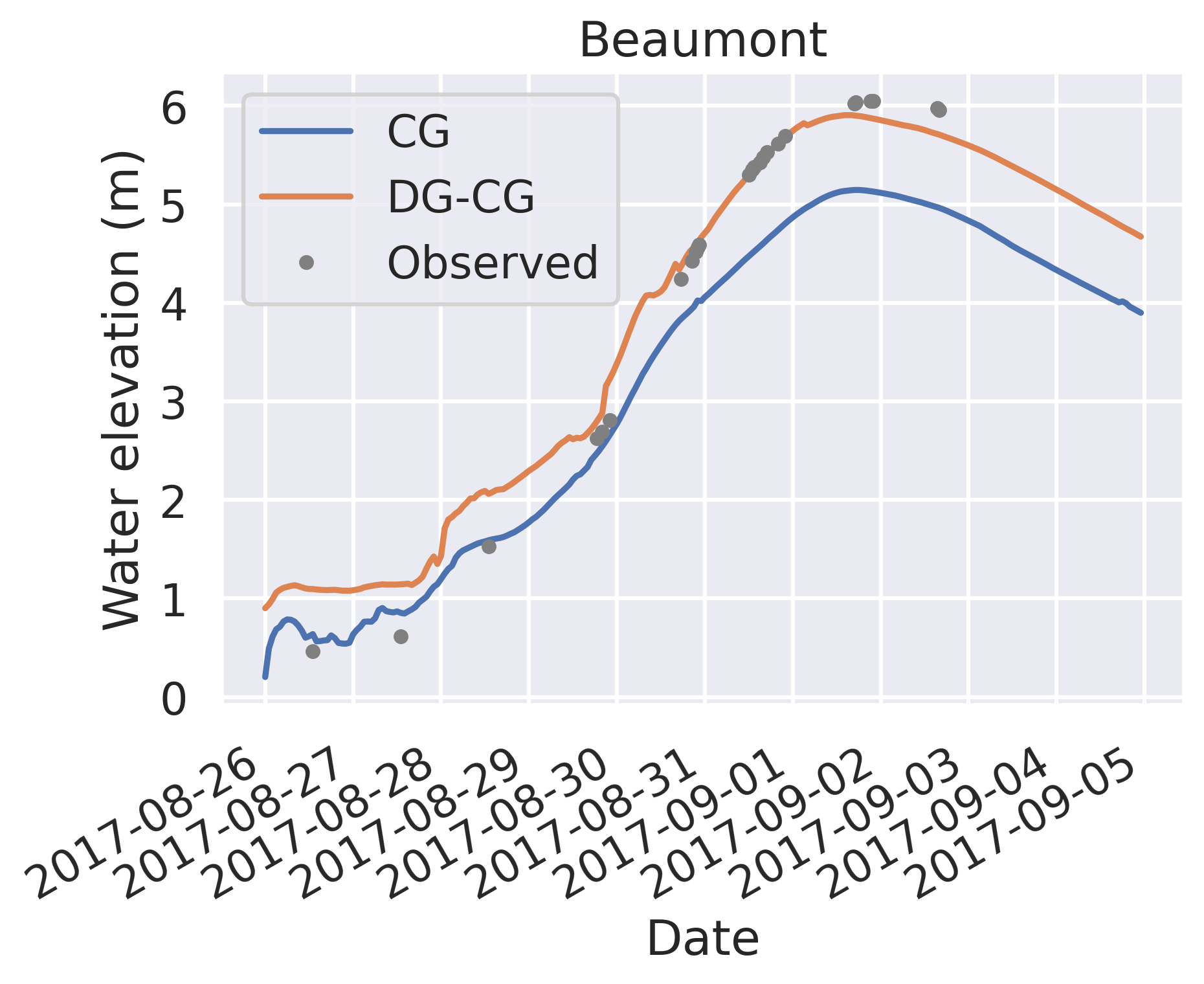}
  \includegraphics[width=0.4\textwidth]{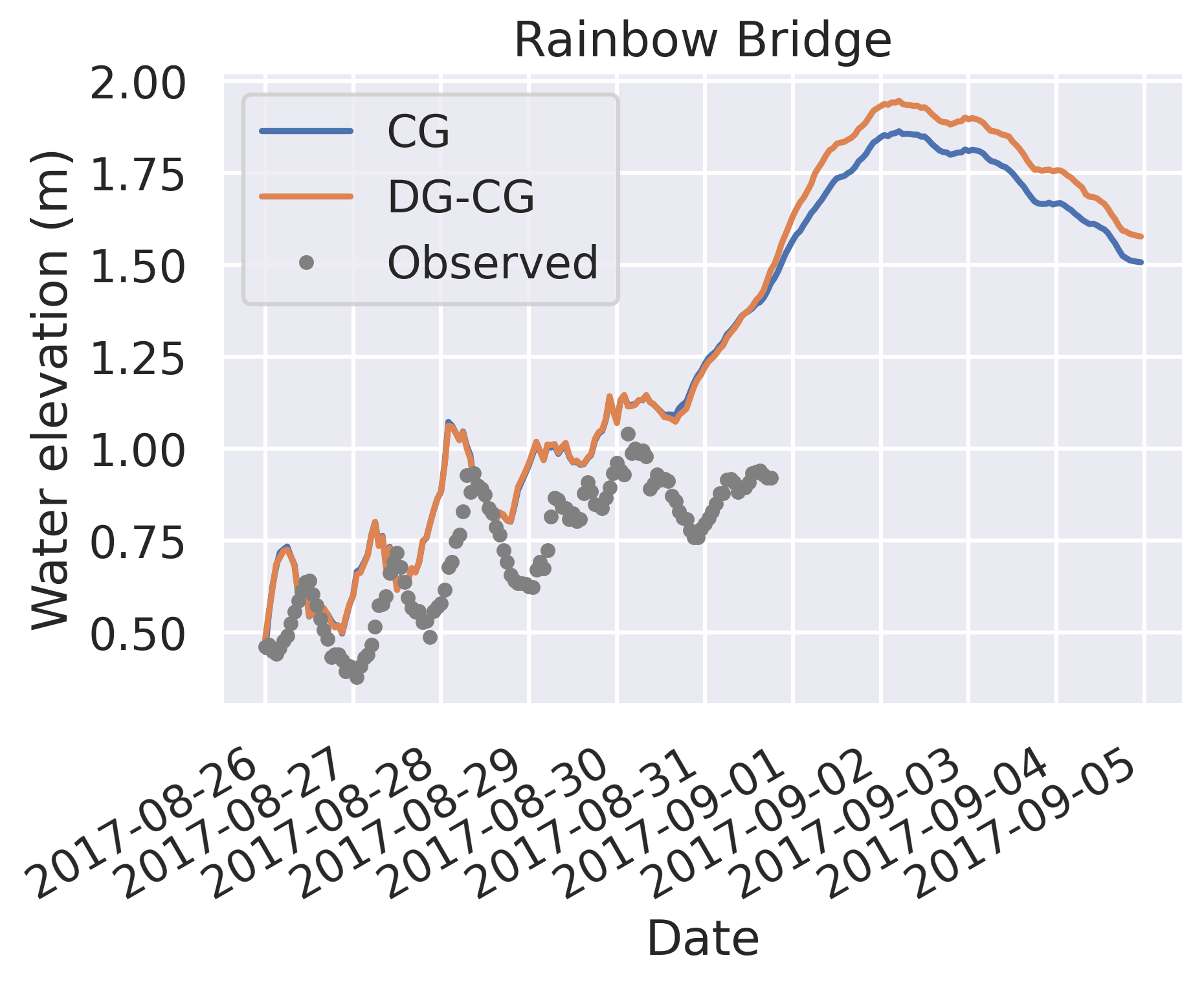}\\
      \includegraphics[width=0.4\textwidth]{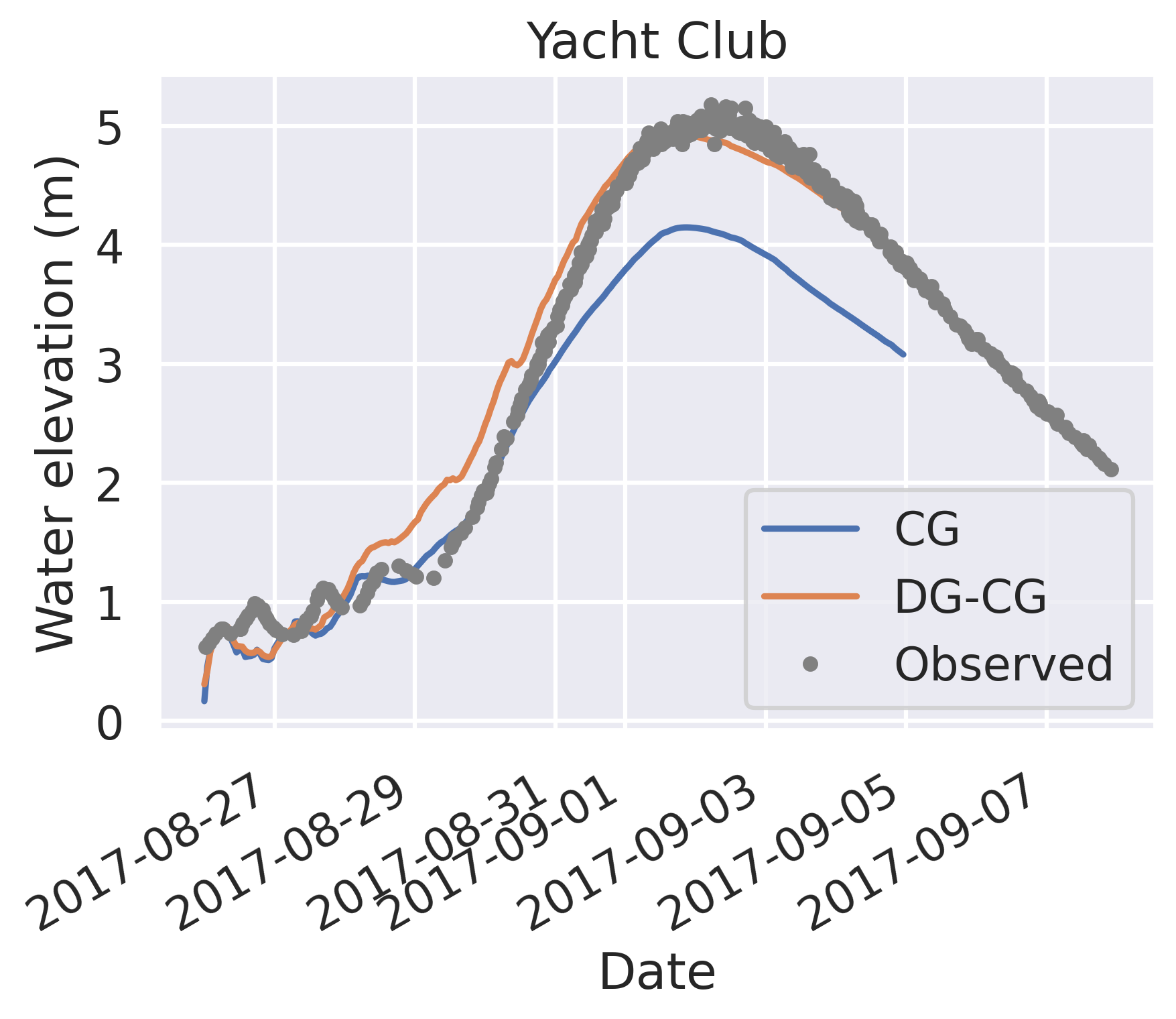}
    \includegraphics[width=0.4\textwidth]{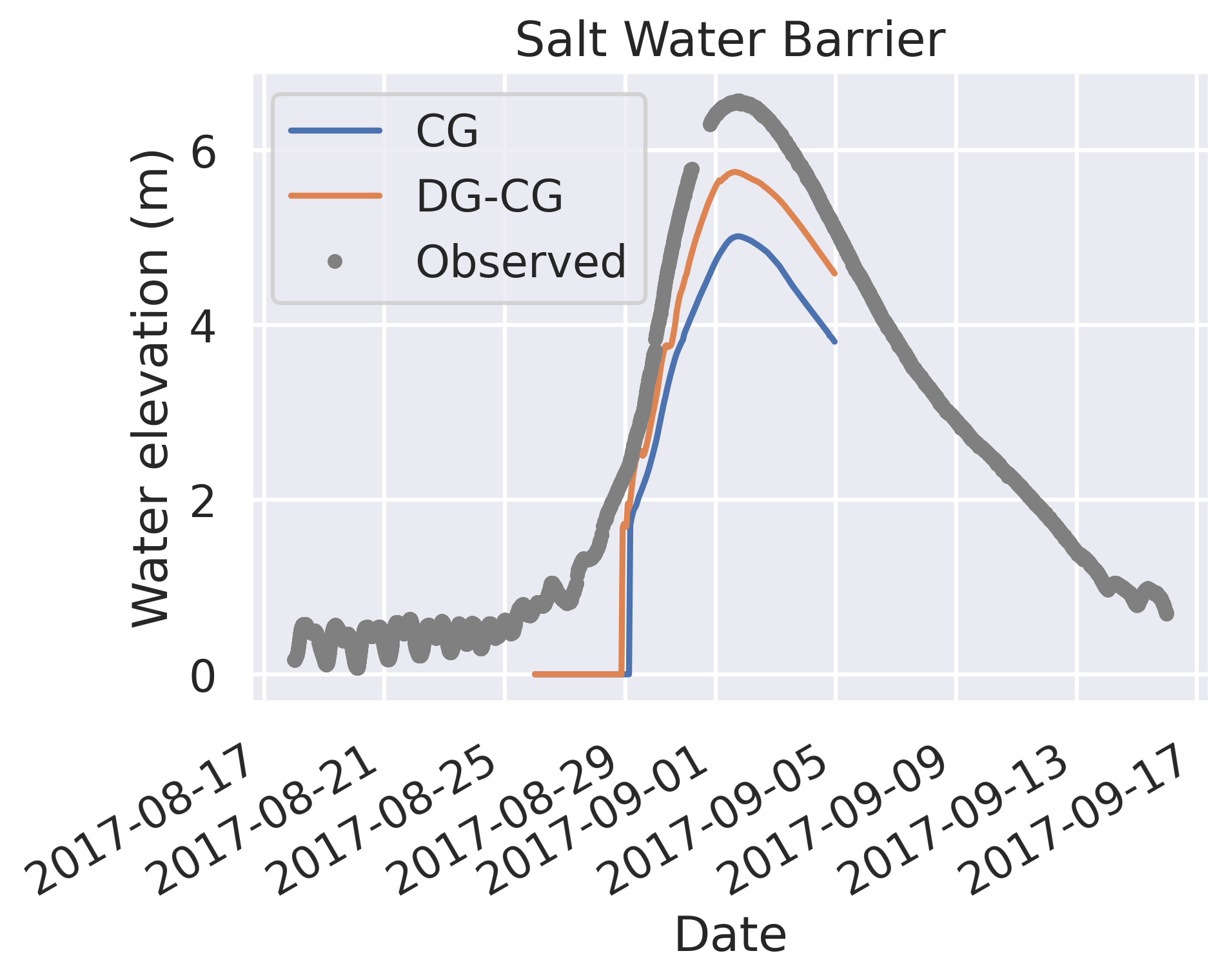}

    \caption{Simulated water elevation as well as observed values at various stations along the Neches River. Flat portions shown with zero elevation indicate dry nodes, not necessarily having zero elevation. Datum is NAVD88. Note that simulation output is limited by the range of input flow data.}
    \label{fig:neches_gauges}
\end{figure}
The next test case is based on an approach to model compound flooding through river-coastal ocean interaction. In this case, we consider the Lower Neches River watershed  near the Texas-Louisiana border as shown in Figure~\ref{fig:neches}. During Hurricane Harvey (2017), this watershed was subjected to extreme rainfall and large portions of its surrounding areas were inundated during and after the event, see e.g.,~\citep{watson2018characterization} for details on the flooding during this event. 
The model domain has been extracted from the  ADCIRC mesh used in the Hurricane Harvey case in Section~\ref{sec:harveyandike}~\cite{hope2013hindcast}, and contains 122,839 elements and 62,075 nodes and is shown in Figure~\ref{fig:neches}.
Hence, the highly detailed unstructured mesh, bathymetry and topography, as well as parameters such as Manning's $n$  are preserved. 
This event was also studied using ADCIRC and compared to HEC-RAS in~\citep{loveland2021developing} and the present study uses the same input data used in Loveland \emph{et al.}. We only present key features of this model and refer readers to the original publication and~\cite{wichitrnithed2024discontinuous} where we extended the DG-SWEM model for greater detail. 

\noindent As in \citep{loveland2021developing}, river flow data was extracted from a validated HEC-RAS model and a USGS gauge and implemented as flux boundary condition at the upstream. At the downstream end of the river which terminates into Sabine Lake, a time-varying elevation boundary condition was applied based on the same HEC-RAS model and the closest NOAA gauge. The locations of these inputs and the gauges used for validation are shown in Figure \ref{fig:neches}.

Comparisons of outputs from CG, DG-CG, as well as observed data from August 20 to September 1, 2017 are shown in Figure \ref{fig:neches_gauges}. We observe underprediction from the CG solver in all cases except at the Rainbow Bridge. We also observe that in those cases, the output from DG-CG is higher and closer to peak observations. At Rainbow Bridge, both solvers match closely (while slightly overpredicting) up to the end of observed data. 
That the DG scheme is able to better handle flows with high advection than ADCIRC has also been demonstrated in other studies~\citep{dawson2011discontinuous,wichitrnithed2024discontinuous}. 
The mismatch between wet/dry elevation values at the beginning of each solver can be attributed to differences in wetting-and-drying criteria as described in Section~\ref{sec:wetdry}.

\subsection{Hurricane Ike (2008) and Hurricane Harvey (2017)} \label{sec:harveyandike}
The final scenarios comprise of Hurricane Ike (2008) and Hurricane Harvey (2017). 
Both these hurricanes led to extensive flooding in the Houston-Galveston Bay area, though with different characteristics.  
Hurricane Ike represents a highly advective case with extremely high peak surge levels, and Harvey is representative of a compound flooding scenario, where most of the flood comes from rainfall and river discharge combined with the, relatively minor, surge that propagated up  Galveston Bay. 
\begin{figure}[h!]
    \centering
    \includegraphics[width=0.60\linewidth]{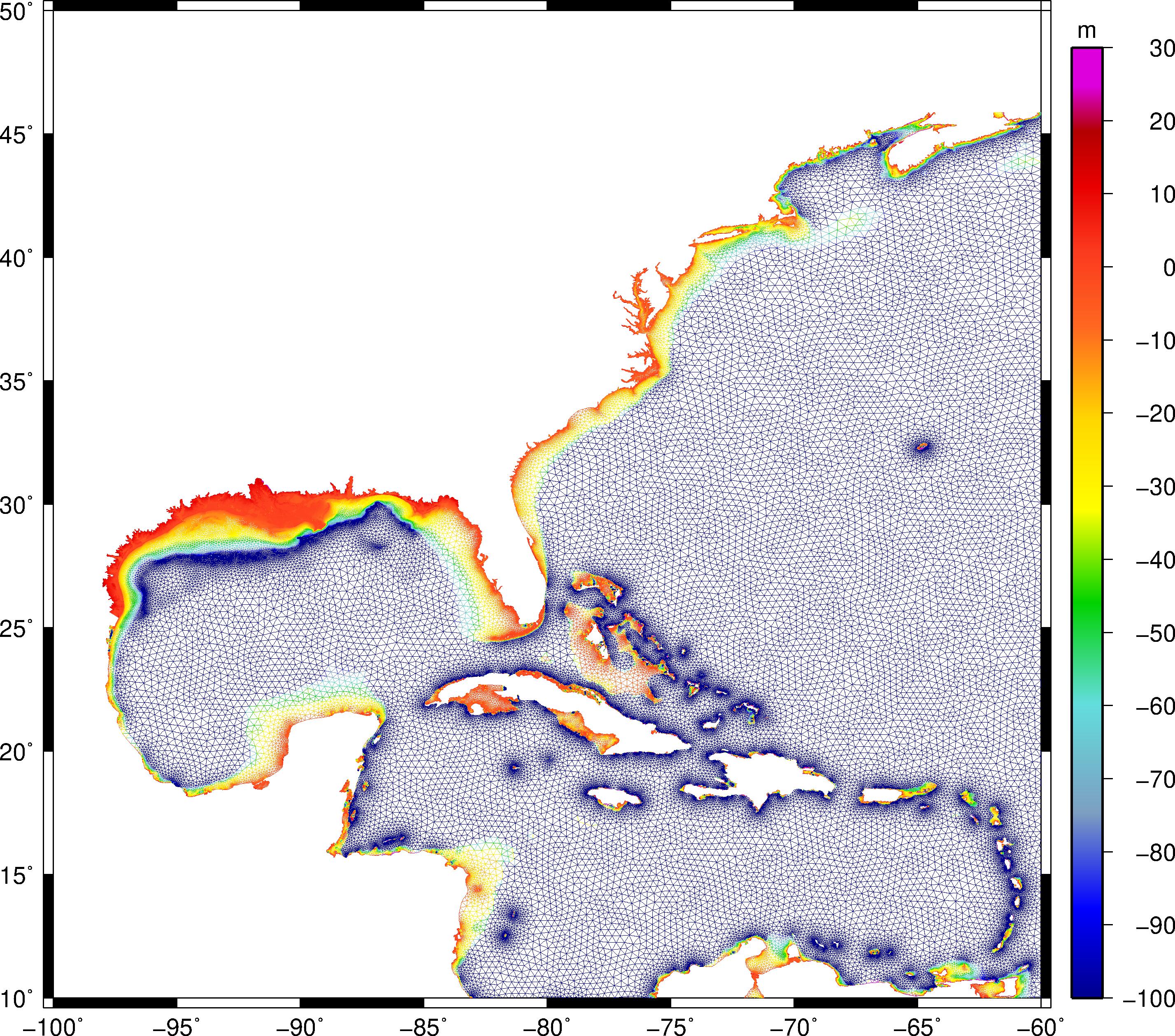}
    \caption{Bathymetry of the mesh used in the Hurricane Ike case. Note that the colorbar is capped for the sake of presentation. }
    \label{fig:120m}
\end{figure}

In Figure \ref{fig:120m},  the computational mesh used is shown. This mesh consists of 2 million nodes and contains a recently updated version of the detailed floodplain, with a smallest element size of 120 meters~\cite{Contreras2023-qi}. Manning's $n$ friction parameter is variable throughout the domain and was determined from observed datas~\cite{Contreras2023-qi}. As shown in Figure~\ref{fig:120m}, the mesh is highly unstructured with increased refinement in the Texas-Louisiana coastal region. For Hurricane Ike, we use a proprietary wind data in the  OWI format from OceanWeather Inc., see Section 4.1 in~\cite{hope2013hindcast} for further details on the wind data. 
For Hurricane Harvey, we use a more refined version of the mesh which contains around 8 million nodes; this is needed to cover the floodplain region where Harvey primarily affected. Wind data in this case is obtained from the NHC Hurricane Database (HURDAT) in the Best Track format.   Lastly, for both hurricanes, we obtain tides using OceanMesh2D~\cite{Pringle2018}, which in turn utilizes the TPXO9 tidal model~\cite{Egbert2002} to define a periodic forcing based on the Q1, O1, P1, K1, N2, M2, S2, and K2 tidal constituents. Additionally, we apply a elevation boundary condition corresponding to these tides along the 60$^\circ$ meridian, i.e., the eastern portion of the mesh shown in Figure~\ref{fig:120m}.

\begin{figure}[H]
    \centering
    \includegraphics[width=0.40\linewidth]{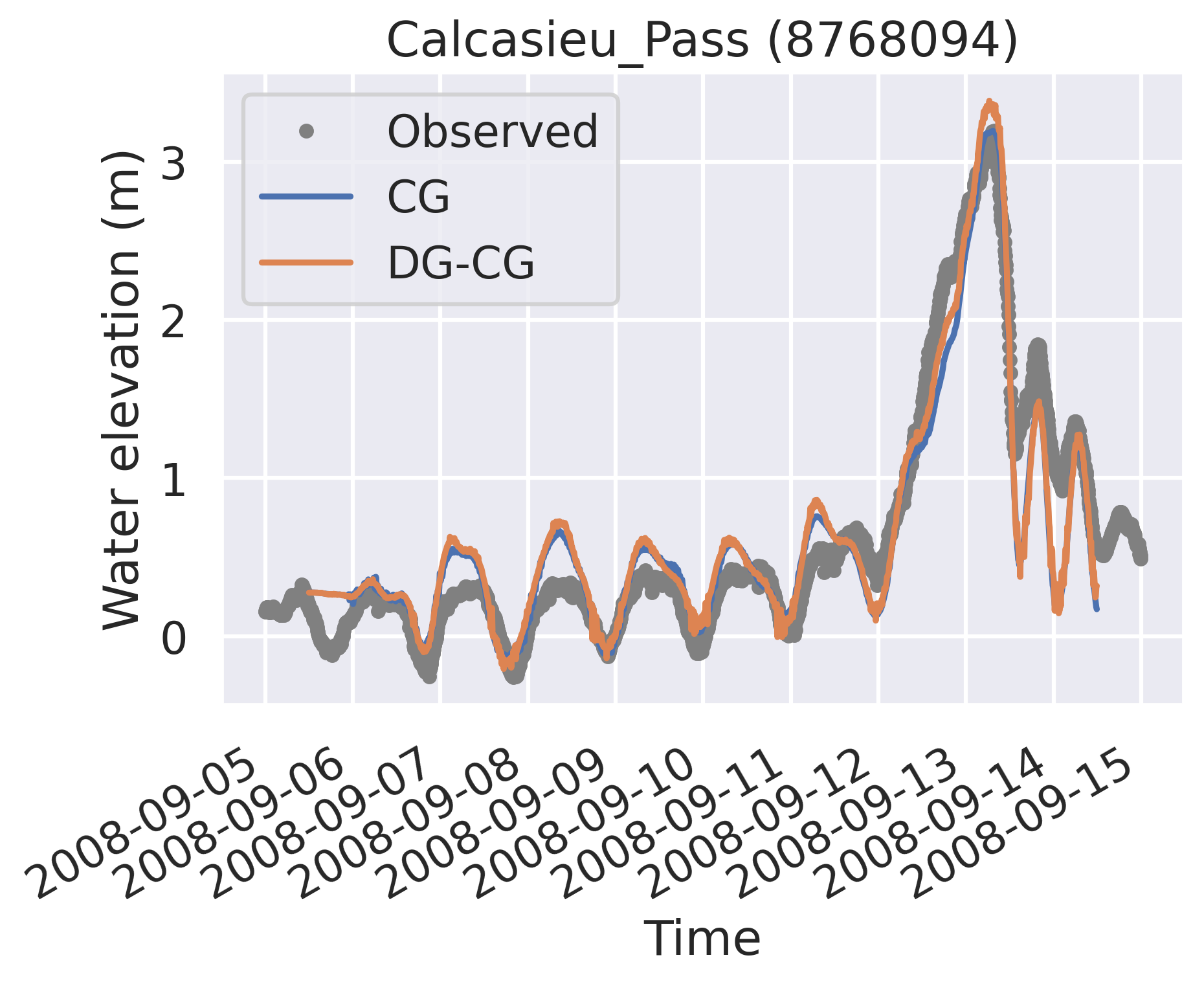}
    \includegraphics[width=0.40\linewidth]{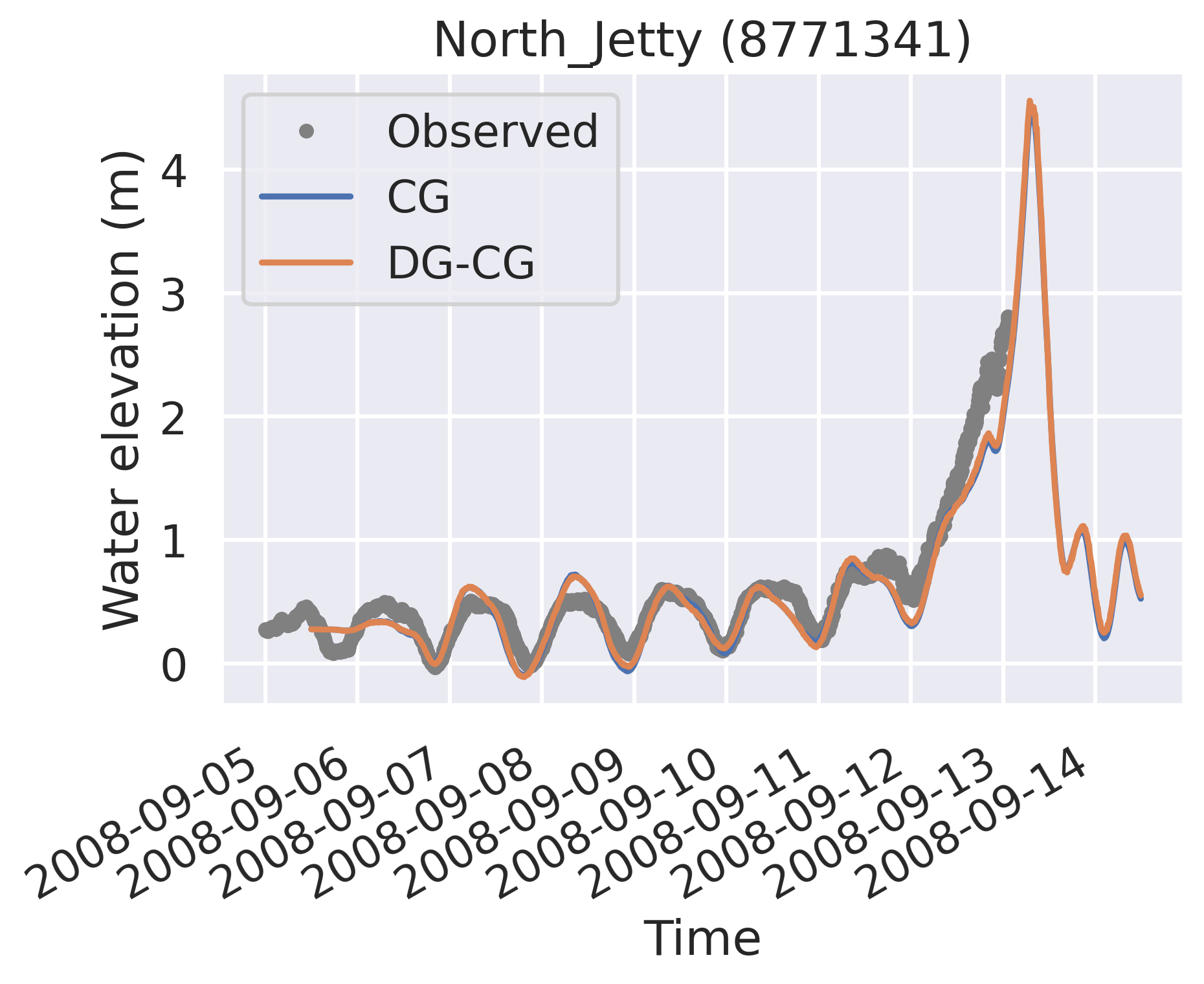} \\
\includegraphics[width=0.40\linewidth]{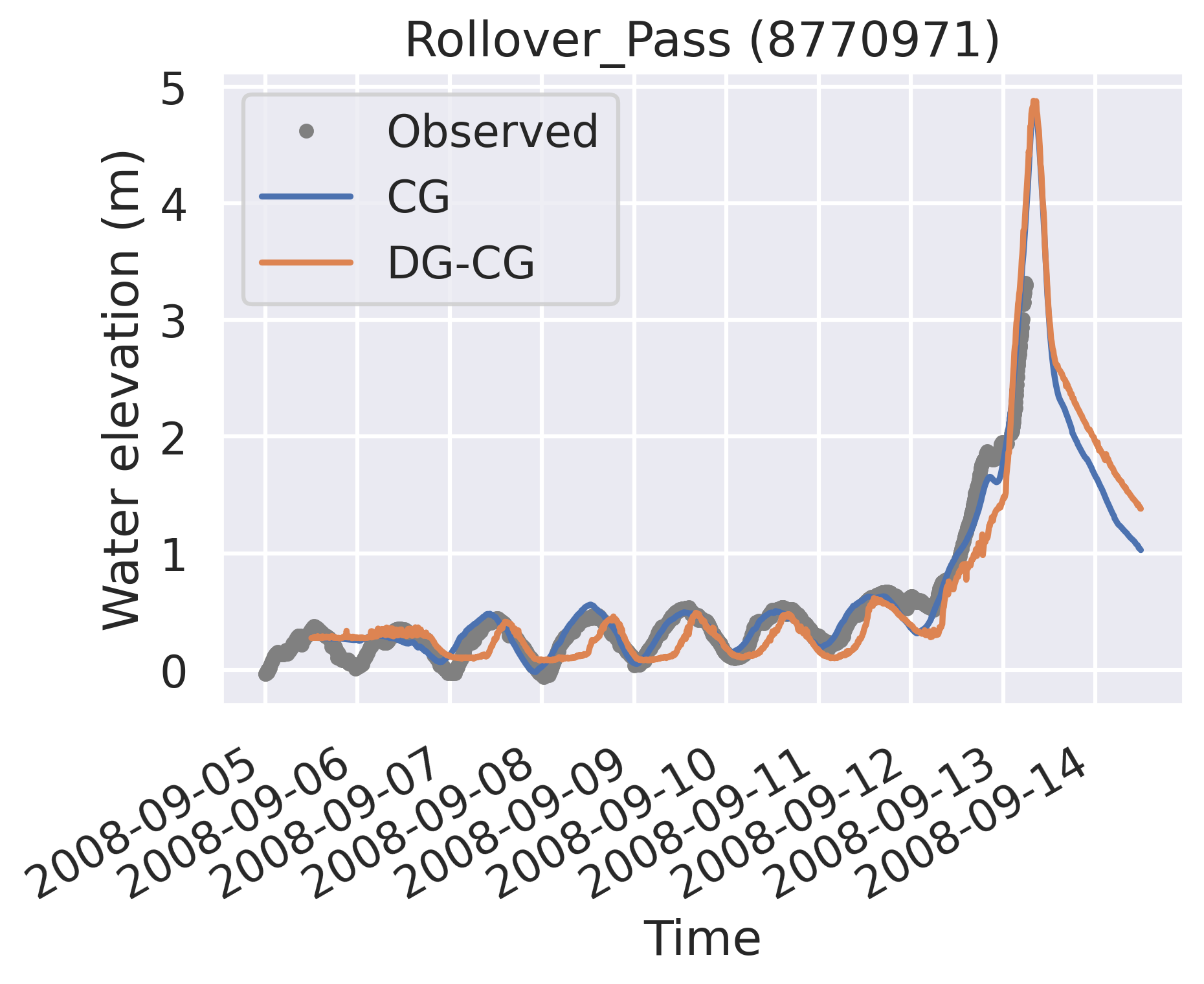}
    \includegraphics[width=0.40\linewidth]{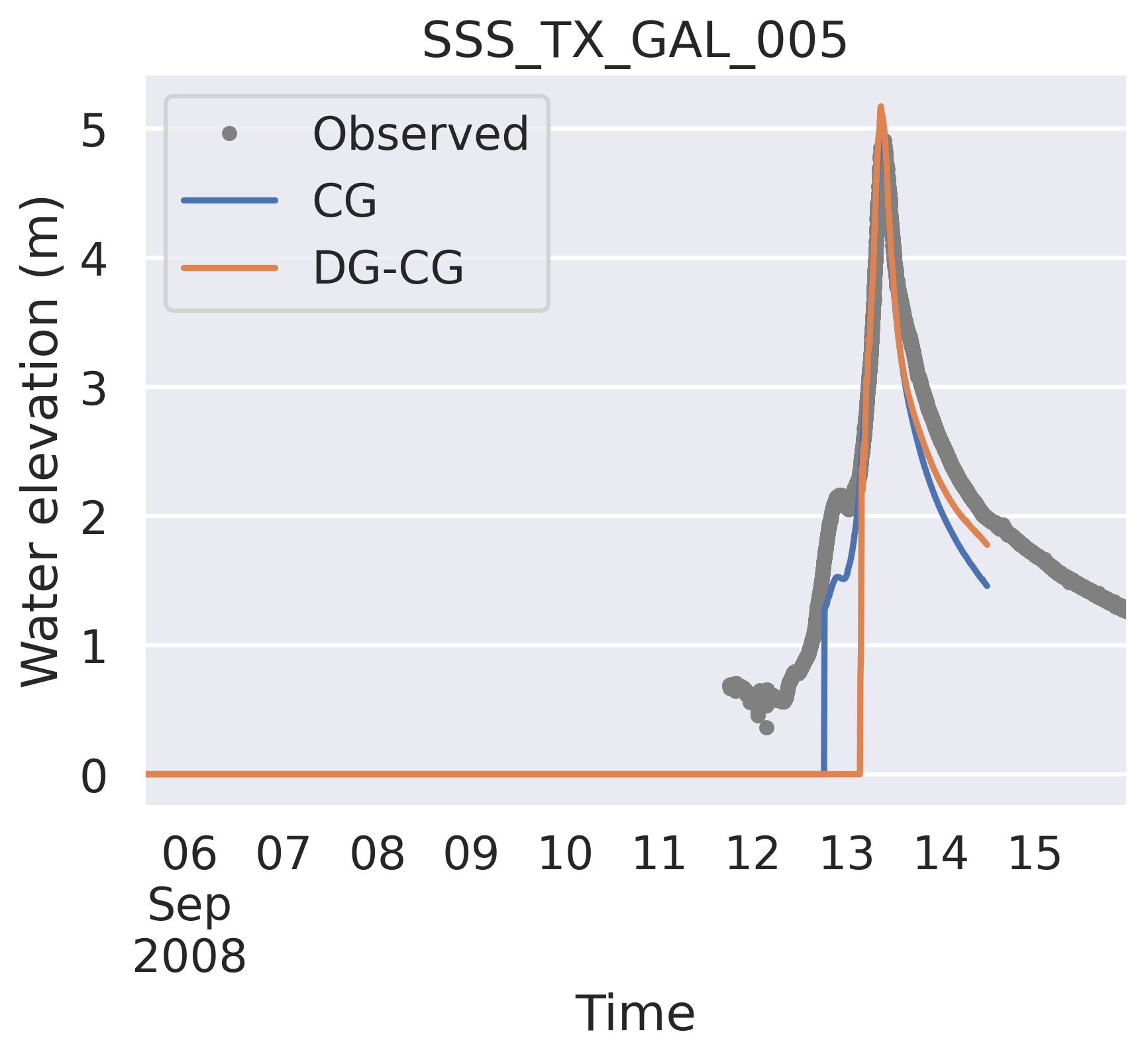} \\
\includegraphics[width=0.40\linewidth]{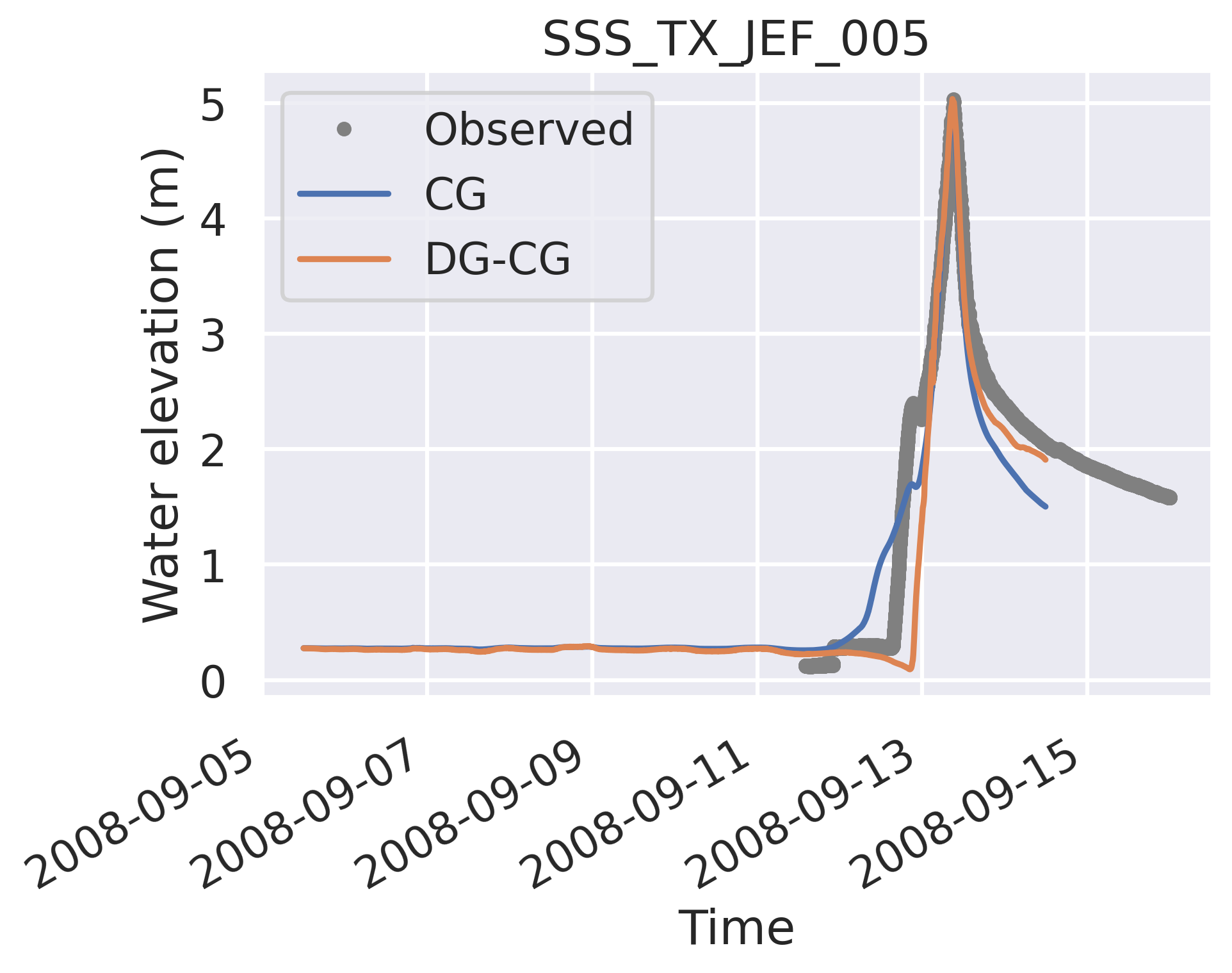}
    \includegraphics[width=0.40\linewidth]{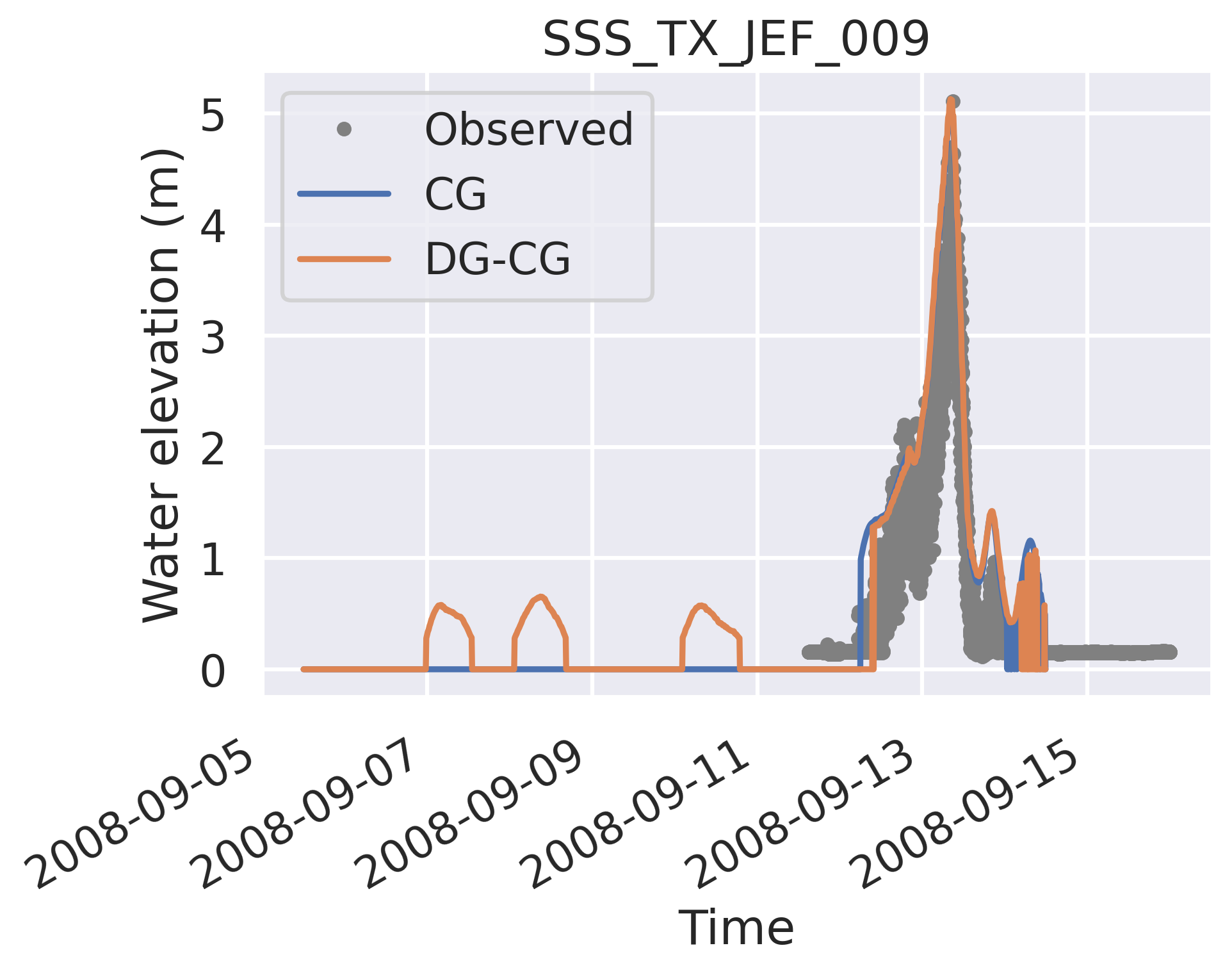}
    \caption{Comparison of simulated water levels and observed data for Hurricane Ike at nine NOAA stations.}
    \label{fig:elev_ike}
\end{figure}

To compare results, we plot both water elevation measurements at several NOAA gauges.
Results for Hurricane Ike are shown in Figures~\ref{fig:elev_ike}. At all stations, the DG/CG solver is almost identical to CG for peak surge. Small differences leading to the peak surge levels at some stations are likely due to the more aggressive wetting and drying algorithm in DG/CG (Section \ref{sec:wetdry}), which limits more flows across some wet/dry interfaces. 

\newpage 
\begin{figure}[H]
    \centering
    \includegraphics[width=0.40\linewidth]{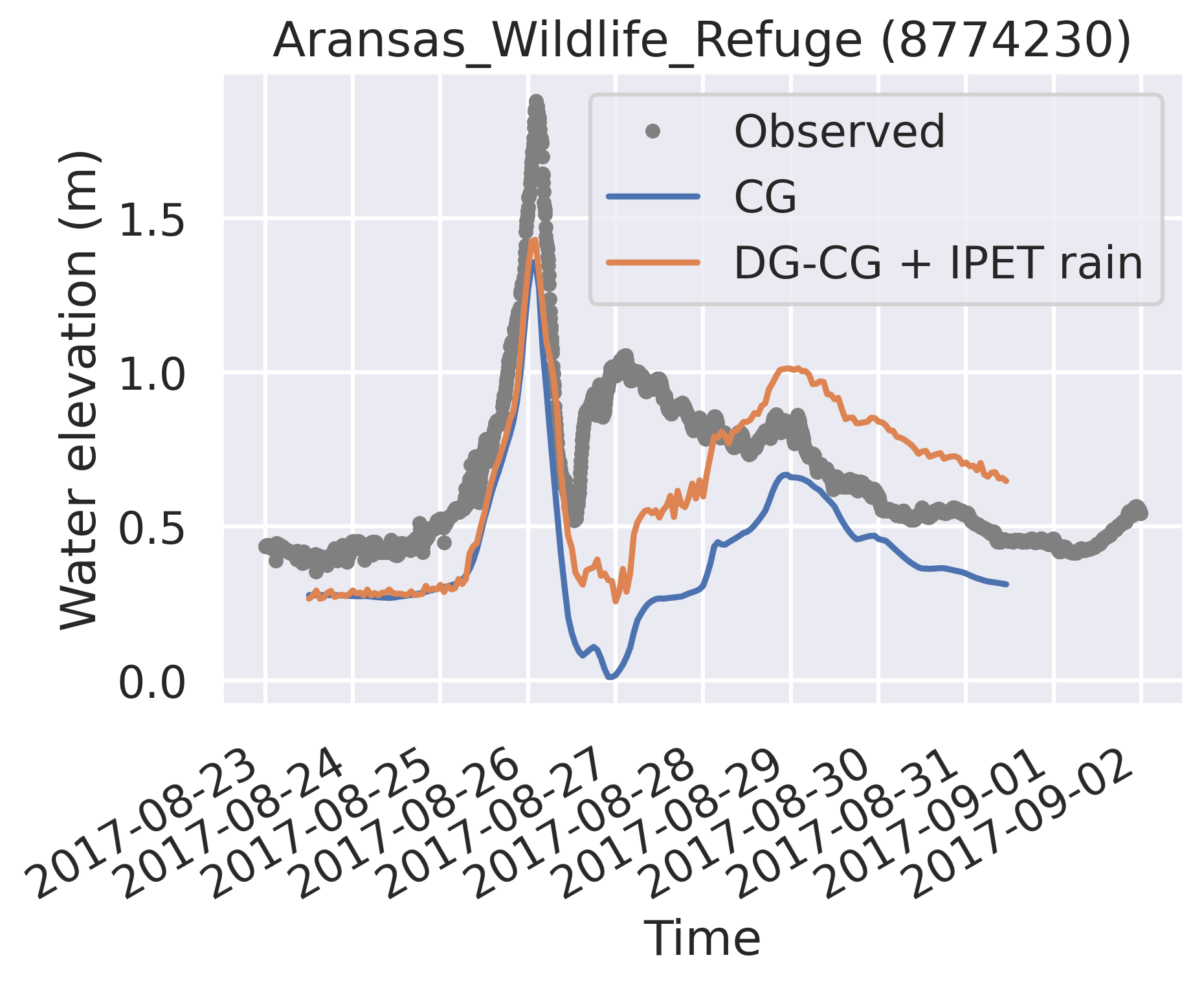}
    \includegraphics[width=0.40\linewidth]{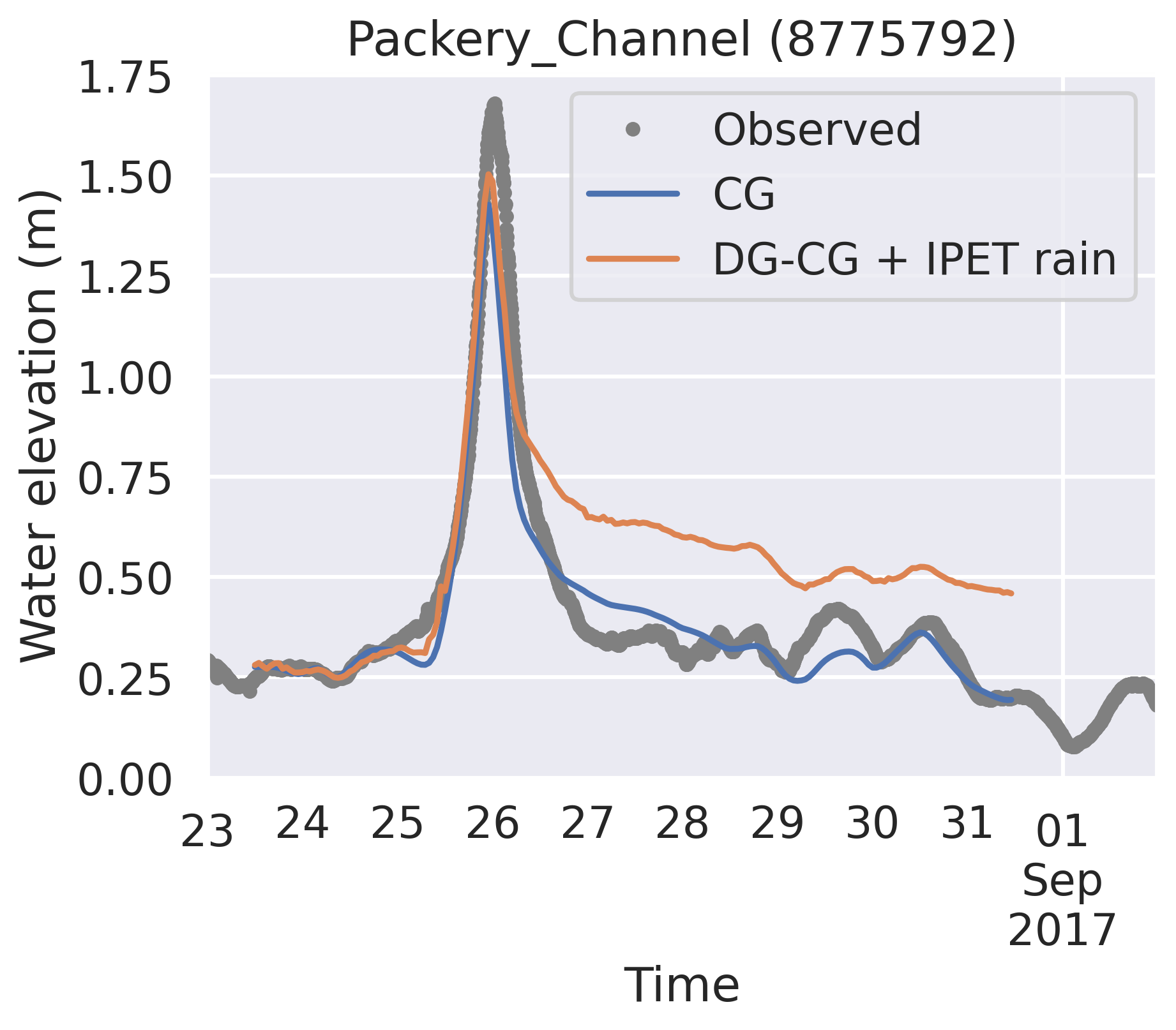} \\
\includegraphics[width=0.40\linewidth]{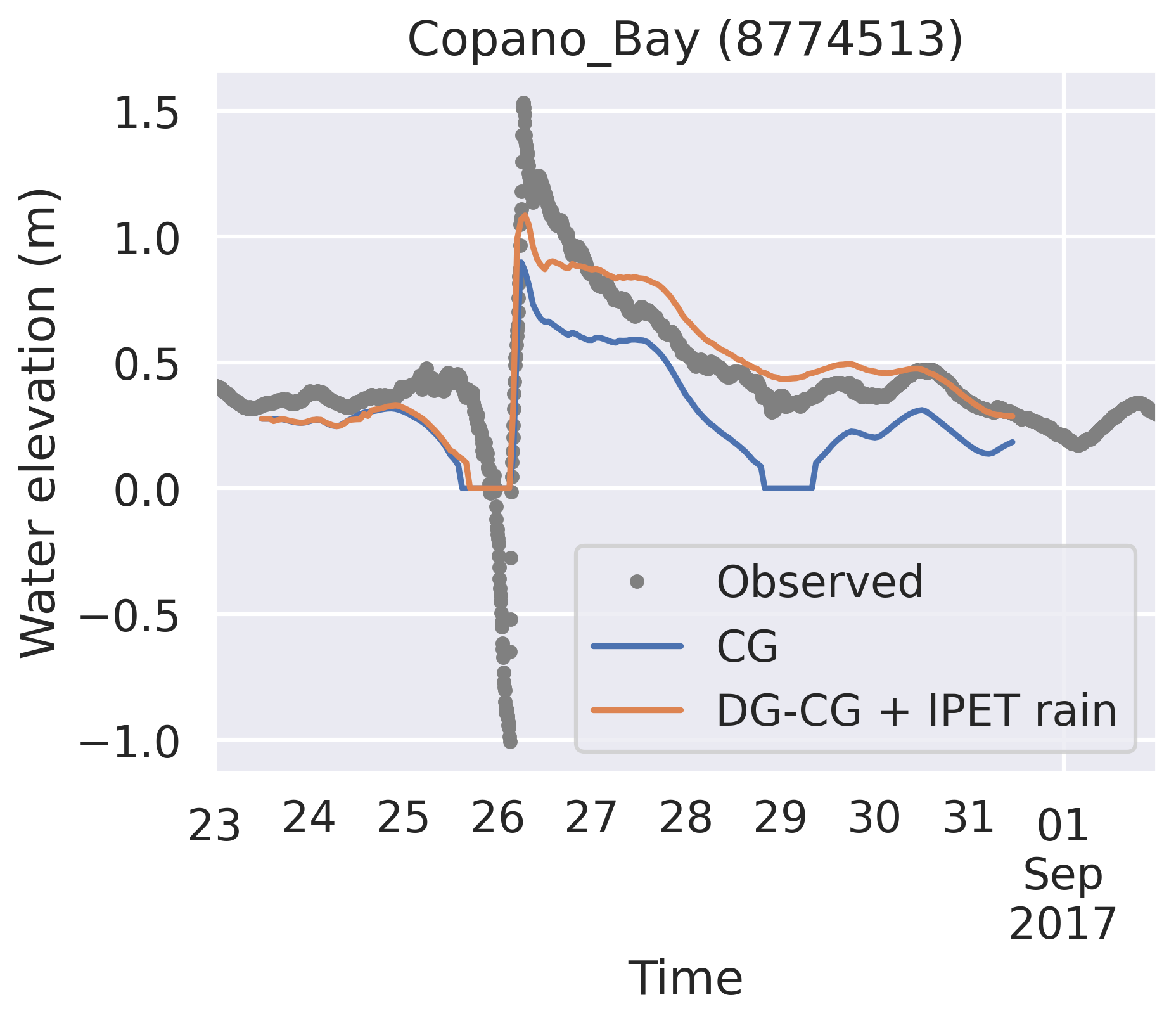}
    \includegraphics[width=0.40\linewidth]{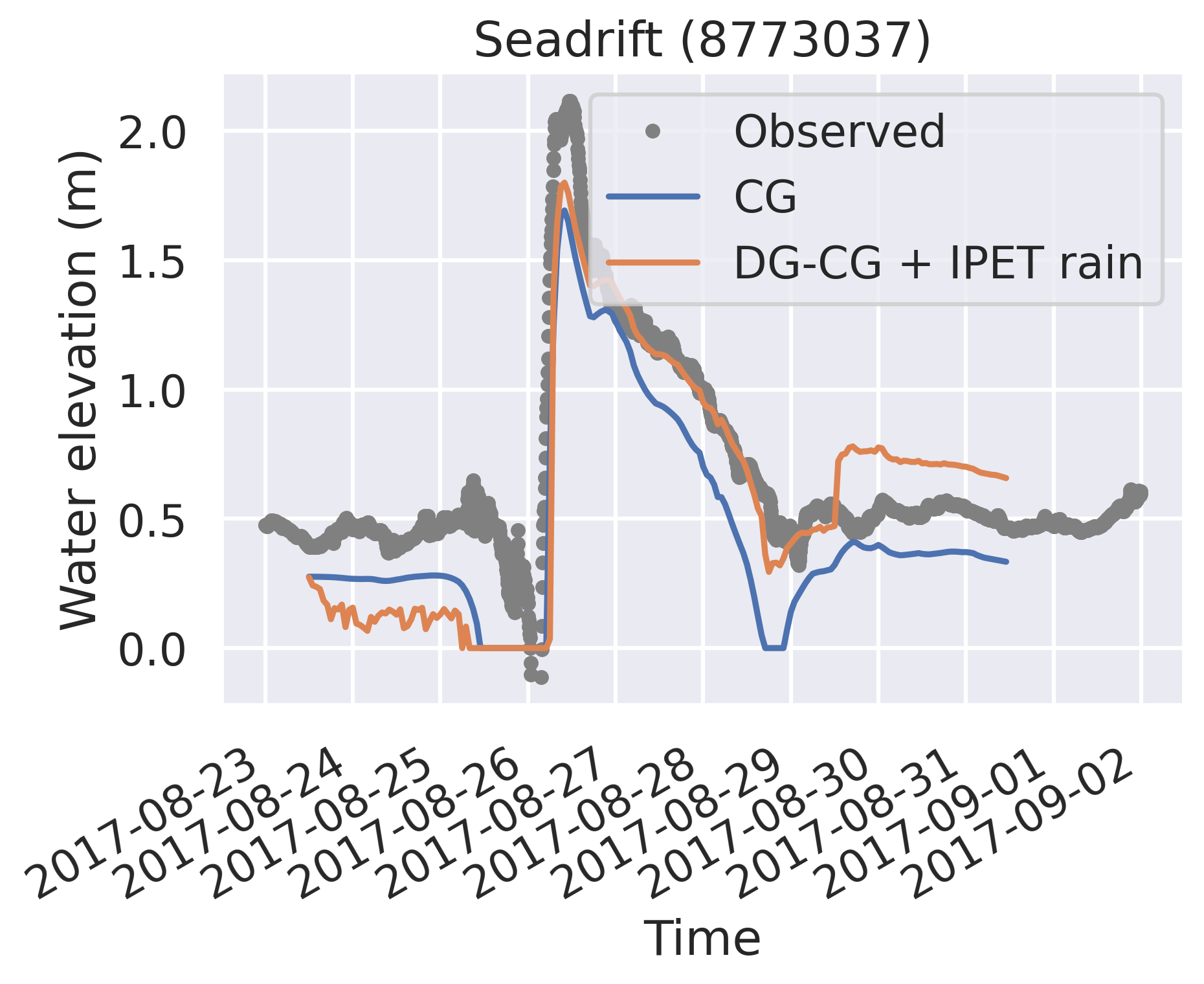} \\
\includegraphics[width=0.40\linewidth]{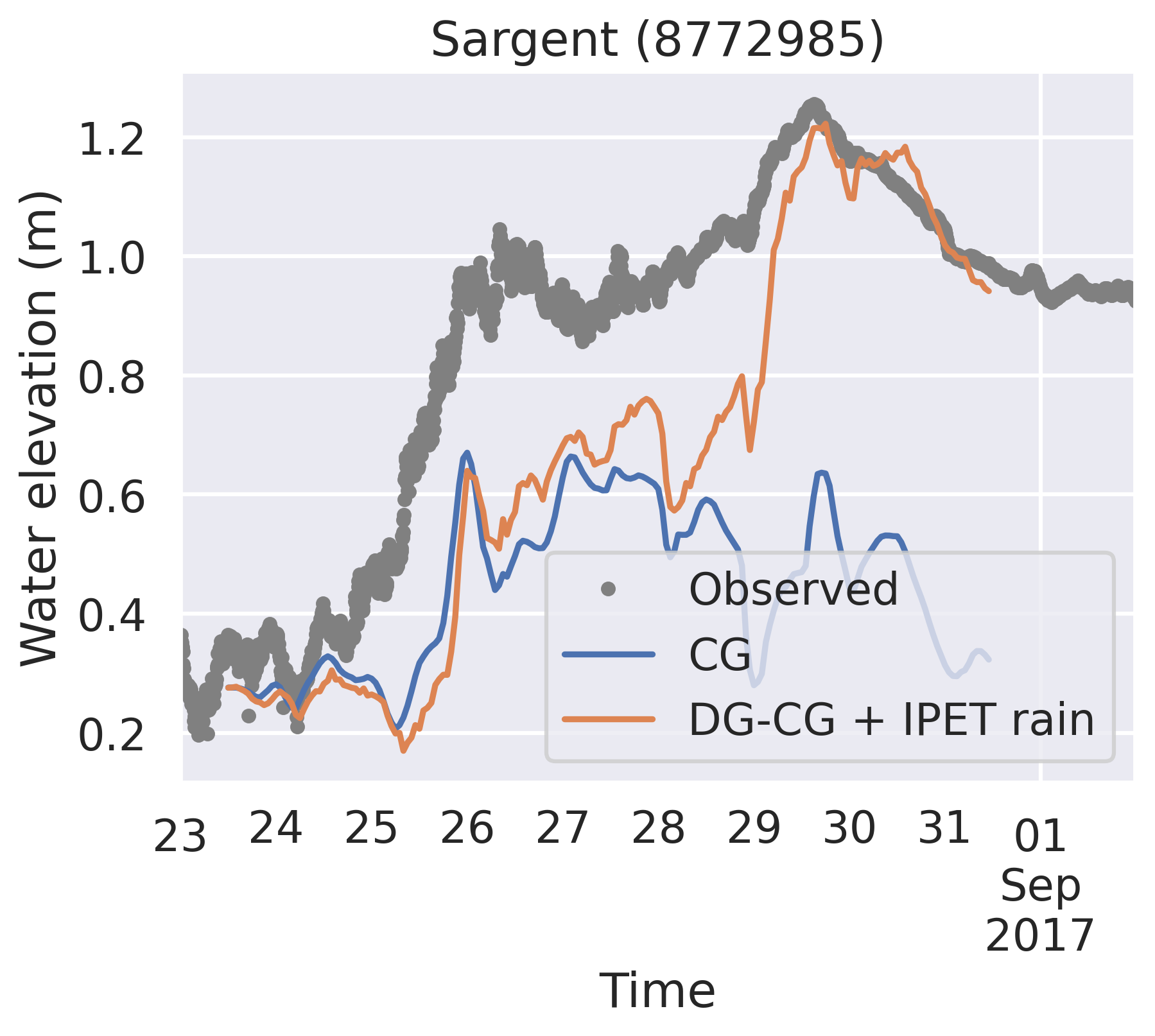}
    \includegraphics[width=0.40\linewidth]{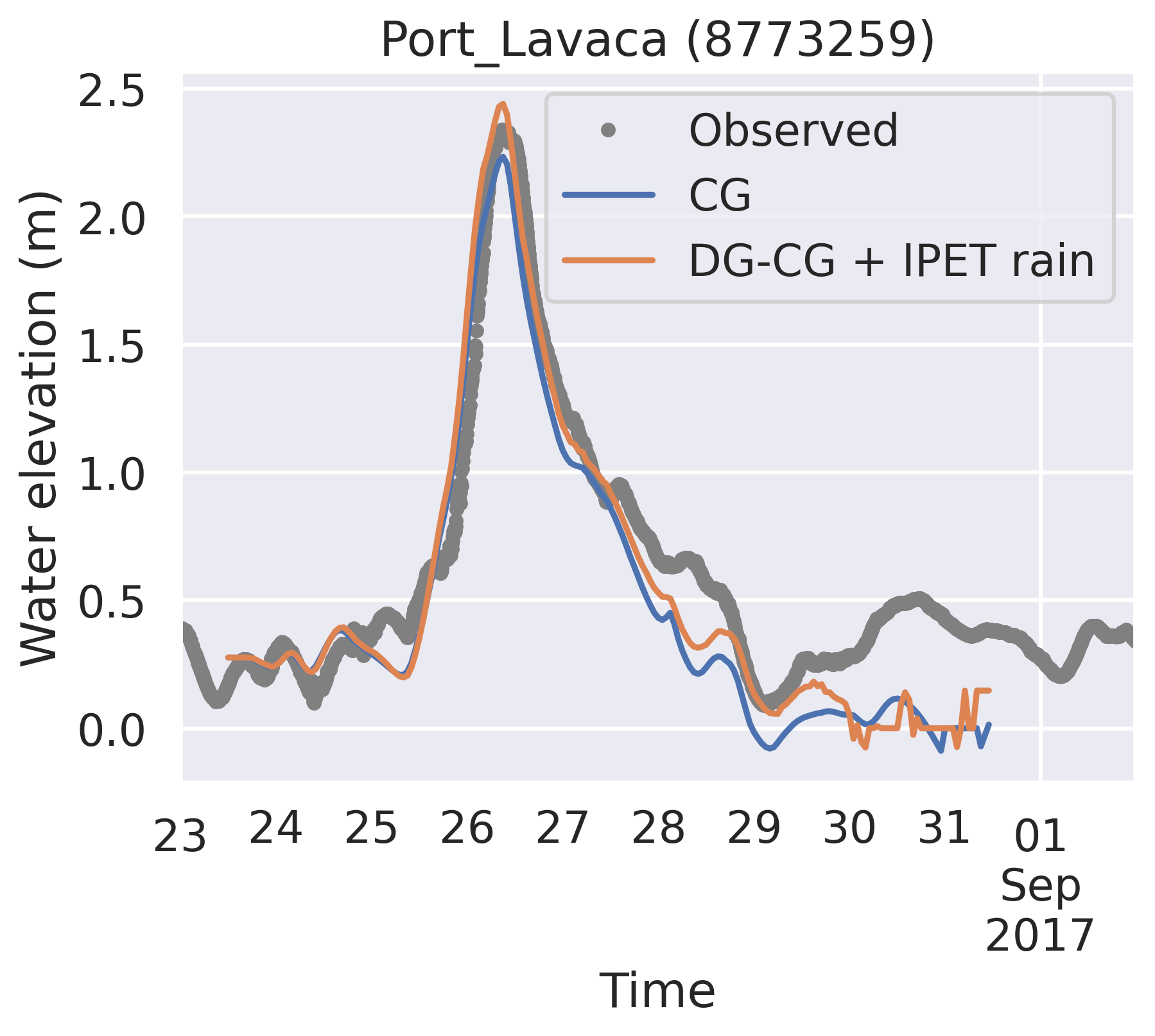}

    \caption{Comparison of water elevation at six NOAA stations during Hurricane Harvey. Flat regions indicate that nodes are considered dry in the simulation.}
    \label{fig:elev_harvey}
\end{figure}
\begin{figure}[H]
    \centering
    \includegraphics[width=0.6\linewidth]{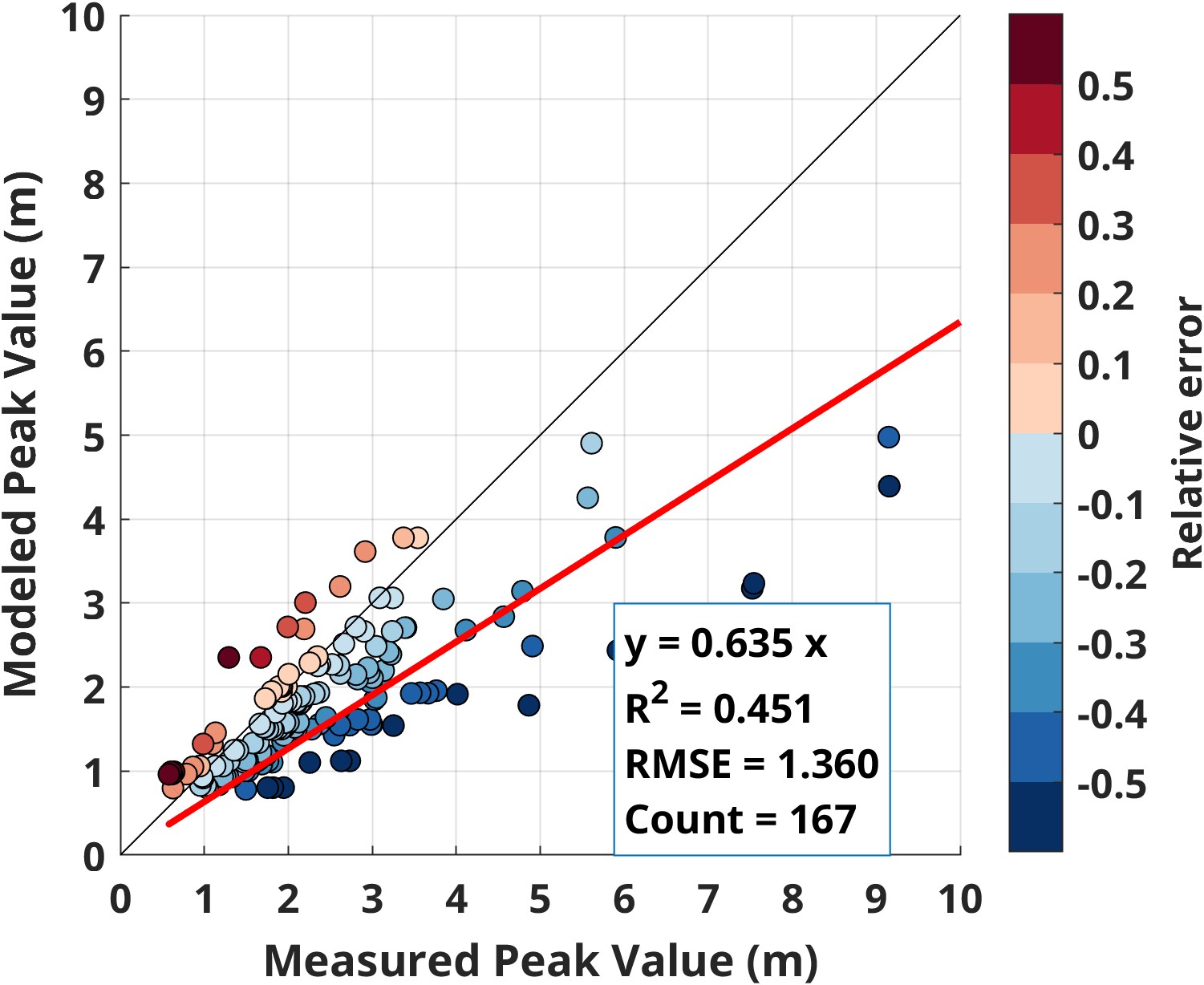} \\
    \includegraphics[width=0.60\linewidth]{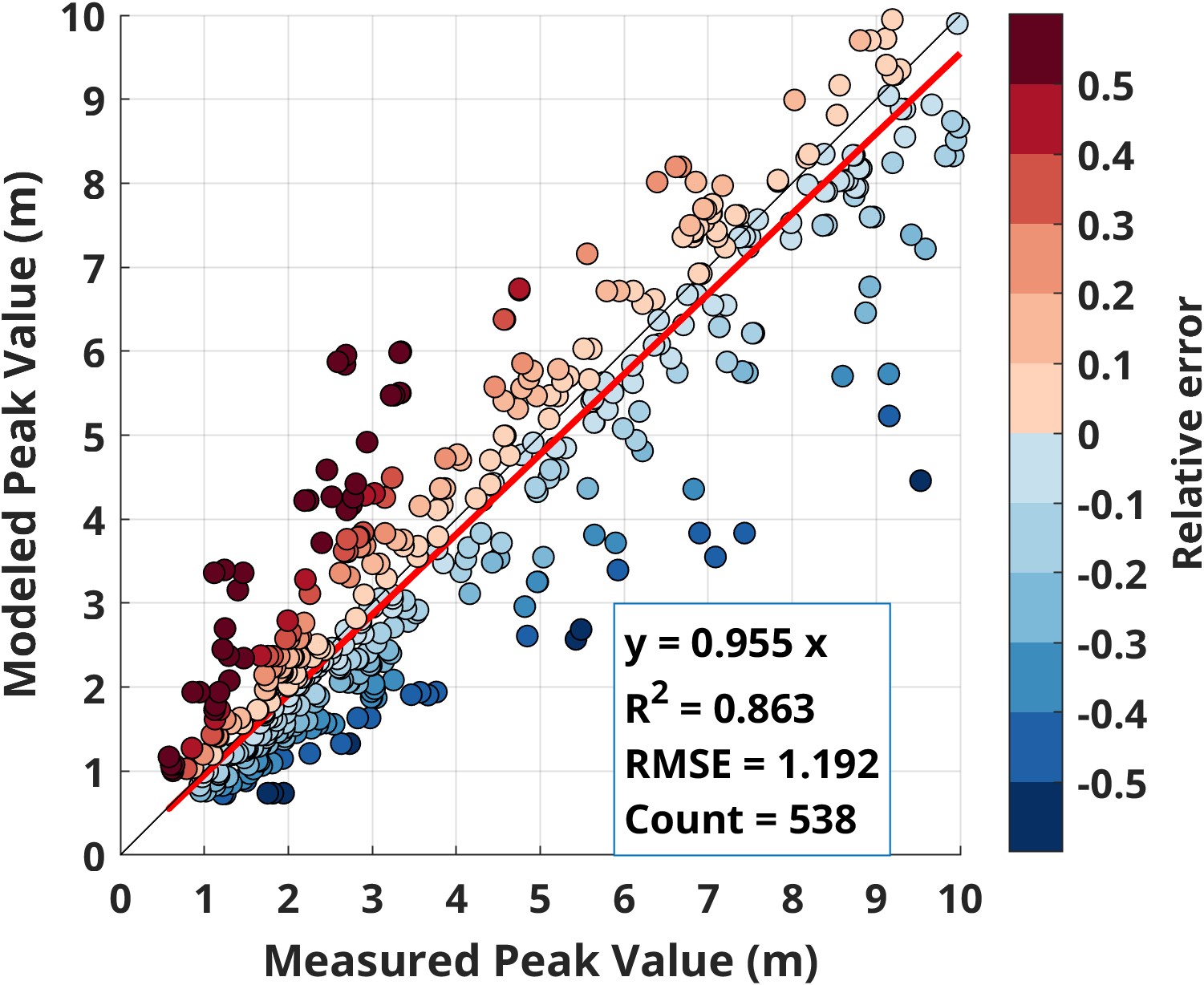}
    \caption{High water mark comparison for the Hurricane Harvey case. Horizontal axis indicates the observed USGS high water marks. Vertical axis indicates maximum water level from simulation, interpolated at the corresponding location. Top: CG output. Bottom: DG-CG with IPET rain. Only wet locations are counted.}
    \label{fig:HWMs_Harvey}
\end{figure}

Water elevation comparison for the Harvey case is shown in in Figure \ref{fig:elev_harvey}. We can clearly observe the effect of rain in the DG-CG case which raises the water level following the peak surge. This is most apparent at Aransas Wildlife Refuge (8774230) and Sargent (8772985). In the former, the secondary peak is delayed in DG-CG but matches the actual peak level better than CG without rain. In the latter, both solvers underpredict initial water levels, but DG-CG eventually matches the peak level. On the other hand, the parametric rain approach can also lead to overprediction as seen in the Packery Channel station. We also compare the simulated maximum water levels with the observed high water marks (HWM) \cite{hwm} located in the region around Harvey's landfall (Figure \ref{fig:HWMs_Harvey}). Note that data counts differ between CG and DG-CG because we only count nodes that are wet. We see a marked improvement in the DG-CG case, indicating the impact of precipitation in compound flooding. A limitation in DG-CG wetting and drying for this case is that it was necessary to turn off momentum advection terms to prevent instability. Although these terms do not make a significant impact in this case, we do aim to make the DG-CG wetting and drying scheme more robust so as to handle highly refined meshes such as this one.

\section{Computational Performance} \label{sec:performance}
This section analyzes the performance differences between CG and DG/CG coupling. Essentially, we are investigating the performance penalty we incur when solving the continuity equation with DG. We start with the serial case to see the detailed breakdown of contributions, and then move on to parallel scaling which is affected by the addition of halo exchanges. In all cases, the program is compiled using Intel compilers, and the setup is shown in Table \ref{tab:setup}.

\begin{table}[h]
    \centering
    \begin{tabular}{l r}
    \hline
        Processors & Intel 8280 "Cascade Lake" \\
        Cores/Node & 56 \\
        Frequency & 2.7 GHz \\ 
        Memory/Node & 192 GB DDR4 \\ 
        Interconnect & Mellanox Infiniband HDR-100 \\
        Compiler & Intel Compiler Family 19.1.1 (icc, icpc, ifx) \\
        MPI Implementation & Intel MPI 19.0.9 \\
        Optimization flags & -O3 -xHost -qopenmp \\
        \hline
    \end{tabular}
    \caption{Experimental configuration on the Frontera Supercomputer.}
    \label{tab:setup}
\end{table}

\subsection{Serial performance}
\subsubsection{Optimization}
As mentioned, the higher degrees of freedom in the DG method unavoidably requires more time to compute than the CG scheme. On the other hand, the higher degree of data parallelism from having most computations be element-local makes it attractive to use vector instructions. The general idea and motivation is outlined in \citep{Brus2017-ve} which involves significant loop reordering and blocking. To keep the code maintainable as part of ADCIRC, we opt to use a simpler approach and try to rely on automatic vectorization and basic optimizations as much as possible \citep{Unknown2010-wl}. Wherever possible, loops with independent iterations and large trip count (e.g. number of elements) are prepended with the OpenMP \verb|!$omp simd| pragma, and inner loops are unrolled. The compiled code and assembly is checked using Intel Advisor to verify if the vectorization took place. Vectorizing the slope limiter  requires a little more effort, as the automatic vectorizations were only partial. We manually performed  explicit blocking on the number of elements and created vector versions of the local arrays private to each element.
Without significant restructuring, the interior edge integration routine is not very suitable to vectorization since multiple edges can modify the same element. Nevertheless, we still observe significant improvement from inlining the numerical flux function by moving it into the same module.

\subsubsection{Test case}
We again use the Shinnecock inlet test case from Section \ref{sec:shin}. This case is not too large to be run serially and includes wetting and drying. 
The case is run and timing is reported using Intel Advisor in Table \ref{tab:serial}. We show both the total run time, and the breakdown of continuity solving time in the DG/CG case. In the original DG/CG code, the DG solver takes around 75\% of the total time, mostly due to edge integration and slope limiting. After the aforementioned optimizations, the DG solver's runtime is cut by half and now takes around 60\% of the total time. This results in DG/CG being around 18\% slower than CG for this serial case.

\begin{table}[H]
    \centering
    \begin{tabular}{l r r r r}
    \hline
        & DG/CG & Optimized DG/CG & CG &  Speedup over CG \\
         \hline
         \textbf{Total} & \textbf{19.9} & \textbf{11.8} &  \textbf{10.03} &  \textbf{-1.18}$\times$\\ \hline \hline
        {Continuity Eq} & 15.1 & 7.03 &  5.6 &  -1.24$\times$ \\ 
        $\rightarrow$ Edge integration & 6.14 & 2.90  & - & -\\
        $\rightarrow$ Slopelimiter & 4.06 & 1.94  & - & - \\ 
        $\rightarrow$ Area integration & 1.71 & 1.25  & - & -\\ 
            $\rightarrow$ Modal2Nodal & 1.69 & 0.16  & - & -\\ \hline\hline
        Momentum Eqs & 2.29 & 2.34 &  2.14 &  -1.09$\times$ \\
    \hline
    \end{tabular}
    \caption{Run time breakdown (s) for the serial Shinnecock inlet case.}
    \label{tab:serial}
\end{table}
\subsection{Parallel scalability}
\begin{figure}[h!]
    \centering
    \includegraphics[width=0.8\linewidth]{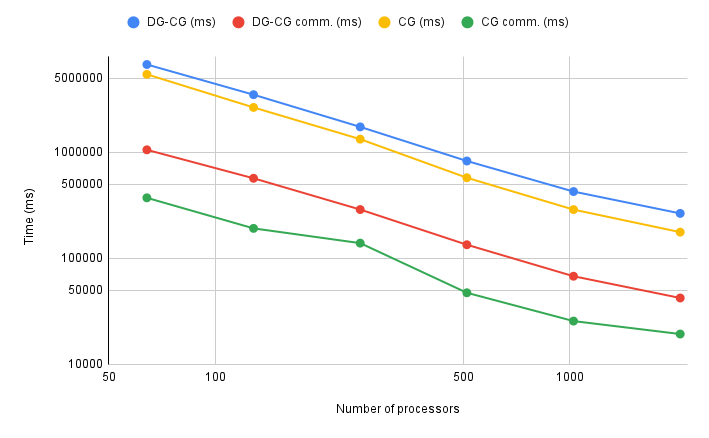}
    \caption{Strong scaling of CG and DG-CG on the 120m grid. Blue and yellow lines indicate the total time. Red and green indicate MPI communication time (including data packing).}
    \label{fig:strong}
\end{figure}

To evaluate the parallel performance of the code, we use the 120m-spaced grid and measure strong scaling from 64 to 2,048 processors. This mesh consists of a periodic tidal boundary and forcing from Hurricane Ike. A simulation duration of 1 day and $\Delta t = 1$ s were used. We run both CG and DG-CG; for CG we use the explicit solver option and therefore no iterative matrix solve is performed. In Table~\ref{tab:parallel_runtime}, we  show the runtime breakdown and speedup. This data is also plotted in Figure \ref{fig:strong}. We use the Tuning and Analysis Utilities (TAU) \cite{Shende2006-mq} to measure the time. Most of the communication takes place during neighbor exchange of updated variables, and scale quite well with more processors. However, the overhead is still significant compared to the CG code because we now update additional variables like wet/dry flags and elevation before/after slope limiting. 
\begin{table}[H]
    \centering
    \begin{tabular}{l r r r r}
    \hline
       $P$ & $N/P$ & Time (s) & Neighbor comm. (s) & Speedup  \\
        \hline
        \multicolumn{5}{c}{\textbf{DG-CG}} \\ 
         64 & 30,253 & 6,776 & 1514 & - \\
         128 & 15,126 & 3,513 & 824 & 1.93 \\
         256 & 7,563 & 1,743 & 409 & 2.01 \\
         512 & 3,781 & 830 & 191 & 2.10 \\
         1024 & 1,890 & 427 & 97.2 & 1.94 \\
         2048 & 945 & 244 & 58 & 1.61 \\
         \hline
         \multicolumn{5}{c}{\textbf{CG}} \\ 
         64 & 30,253 & 5471 & 372 & - \\
         128 & 15,126 & 2663 & 191.6 & 2.05 \\
         256 & 7,563 & 1335 & 139 & 1.99 \\
         512 & 3,781 & 576.7 & 47.2 & 2.31 \\
         1024 & 1,890 & 288 & 25.49 & 2.00 \\
         2048 & 945 & 176 & 19.23 & 1.63 \\
         \hline
    \end{tabular}
    \caption{Average runtime (s) across all CPU cores. Speedup is relative to the previous runtime, i.e. ideally 2$\times$ for each configuration.}
    \label{tab:parallel_runtime}
\end{table}
\section{Concluding Remarks} \label{sec:conclusions}
We have presented recent developments to compound flood modeling by combining DG and CG schemes to solve the SWE. In particular, we exploit the conservation and stability properties of the DG method to add rainfall to the simulation while maintaining the computational performance of the CG method for momentum. To ascertain spatially and temporally varying rainfall intensity we use parametric rainfall models from literature as well as interpolated rain data from past events.

We have shown results from extensive numerical experiments which highlight the capabilities and properties of our methodology, including conservation properties and compound flooding during a hurricane with significant rainfall.  In particular, we note the enhancements due to the addition of rainfall in the results for Hurricane Harvey (2017) in the areas close to the hurricane track, indicating the potential of using such parametric rainfall models in compound flood simulations. Comparisons to results from ADCIRC for river runoff in the Neches river further highlights  the capabilities of our DG-CG methodology.
On the practical side, we have also created a template for incorporating this solver into the ADCIRC codebase in a modular way, with coupling minimized.

We note that while the addition of the rainfall and resulting runoff to the solver is a significant step towards modeling compound floods, there are a plethora of other hydraulic and hydrological processes that may also impact compound flood events  not explicitly accounted for here. These include, e.g., evapotranspiration, infiltration, and interception. Inclusion of these will be the focus of future works on the further extension of our model. Further improvements to the wetting and drying scheme are also needed to ensure stability in the general case without overly limiting the fluxes.

\section*{Acknowledgements}
This work has been supported by the United States National Science Foundation NSF PREEVENTS Track 2 Program, under NSF Grant Numbers 1855047 
This material is also based on work supported by the US Department of Homeland Security under Grant No.2015-ST-061-ND0001-01. The views and conclusions contained in this document are those of the authors and should not be interpreted as necessarily representing the official policies, either expressed or implied, of the US Department of Homeland Security.

The authors also would like to gratefully acknowledge the use of the ``ADCIRC", ``DMS23001", and ``DMS21031" allocations on the Frontera supercomputer at the Texas Advanced Computing Center at the University of Texas at Austin. 

\pagebreak

\bibliographystyle{elsarticle-num}
\bibliography{references}

@ARTICLE{Shende2006-mq,
  title     = "The tau parallel performance system",
  author    = "Shende, Sameer S and Malony, Allen D",
  journal   = "Int. J. High Perform. Comput. Appl.",
  publisher = "SAGE Publications",
  volume    =  20,
  number    =  2,
  pages     = "287--311",
  month     =  may,
  year      =  2006,
  url       = "http://dx.doi.org/10.1177/1094342006064482",
  language  = "en"
}

@ARTICLE{Dawson2006-ve,
  title     = "Continuous, discontinuous and coupled discontinuous–continuous
               {Galerkin} finite element methods for the shallow water equations",
  author    = "Dawson, Clint and Westerink, Joannes J and Feyen, Jesse C and
               Pothina, Dharhas",
  journal   = "Int. J. Numer. Methods Fluids",
  publisher = "Wiley",
  volume    =  52,
  number    =  1,
  pages     = "63--88",
  month     =  sep,
  year      =  2006,
  url       = "https://onlinelibrary.wiley.com/doi/10.1002/fld.1156",
  language  = "en"
}

@ARTICLE{Brackins2020-xp,
  title     = "Evaluation of parametric precipitation models in reproducing
               tropical cyclone rainfall patterns",
  author    = "Brackins, John T and Kalyanapu, Alfred J",
  journal   = "J. Hydrol. (Amst.)",
  publisher = "Elsevier BV",
  volume    =  580,
  number    =  124255,
  pages     =  124255,
  month     =  jan,
  year      =  2020,
  url       = "http://dx.doi.org/10.1016/j.jhydrol.2019.124255",
  language  = "en"
}

@ARTICLE{Lynch1978-ih,
  title     = "Analytic solutions for computer flow model testing",
  author    = "Lynch, Daniel R and Gray, William G",
  journal   = "J. Hydraul. Div.",
  publisher = "American Society of Civil Engineers (ASCE)",
  volume    =  104,
  number    =  10,
  pages     = "1409--1428",
  month     =  oct,
  year      =  1978,
  url       = "https://ascelibrary.org/doi/epdf/10.1061/JYCEAJ.0005086",
  language  = "en"
}

@MANUAL{Unknown2010-wl,
  title = "A Guide to Vectorization with Intel® C++ Compilers",
  year  =  2010,
  url   = "https://www.intel.com/content/dam/develop/external/us/en/documents/31848-compilerautovectorizationguide.pdf"
}

@ARTICLE{Brus2017-ve,
  title   = "Performance and Scalability Improvements for Discontinuous {Galerkin}
             Solutions to Conservation Laws on Unstructured Grids",
  author  = "Brus, S R and Wirasaet, D and Westerink, J J and Dawson, C",
  journal = "J. Sci. Comput.",
  volume  =  70,
  number  =  1,
  pages   = "210--242",
  month   =  jan,
  year    =  2017,
  url     = "https://doi.org/10.1007/s10915-016-0249-y"
}

@TECHREPORT{Contreras2023-qi,
  title       = "A {Channel-to-Basin} Scale {ADCIRC} Based Hydrodynamic
                 Unstructured Mesh Model for the {US} {East and Gulf of Mexico}
                 Coasts",
  author      = "Contreras, Maria Teresa and Woods, Brendan and Blakely,
                 Coleman and Wirasaet, Damrongsak and Westerink, Joannes and
                 Cobell, Zach and Pringle, William and Moghimi, Saeed and
                 Vinogradov, Sergey and Myers, Edward and Seroka, Greg and
                 Lalime, Michael and Funakoshi, Yuji and Van der Westhuysen,
                 Andre and Abdolali, Ali and Valseth, Eirik and Dawson, Clint",
  number      = "NOS CS 54",
  institution = "National Oceanic and Atmospheric Administration",
  month       =  jan,
  year        =  2023,
  url         = "https://repository.library.noaa.gov/view/noaa/48079/noaa_48079_DS1.pdf"
}

@ARTICLE{Dubiner1991-ts,
  title   = "Spectral methods on triangles and other domains",
  author  = "Dubiner, Moshe",
  journal = "J. Sci. Comput.",
  volume  =  6,
  number  =  4,
  pages   = "345--390",
  month   =  dec,
  year    =  1991,
  url     = "https://doi.org/10.1007/BF01060030"
}

@ARTICLE{Lynch1979-ux,
  title   = "A wave equation model for finite element tidal computations",
  author  = "Lynch, Daniel R and Gray, William G",
  journal = "Comput. Fluids",
  volume  =  7,
  number  =  3,
  pages   = "207--228",
  month   =  sep,
  year    =  1979,
  url     = "https://www.sciencedirect.com/science/article/pii/0045793079900379"
}

@misc{grib2,
title={{NCEP WMO GRIB2} Documentation},
url={https://www.nco.ncep.noaa.gov/pmb/docs/grib2/grib2_doc/}
}

@article{wahl2015increasing,
  title={Increasing risk of compound flooding from storm surge and rainfall for major {US} cities},
  author={Wahl, Thomas and Jain, Shaleen and Bender, Jens and Meyers, Steven D and Luther, Mark E},
  journal={Nature Climate Change},
  volume={5},
  number={12},
  pages={1093--1097},
  year={2015},
  publisher={Nature Publishing Group}
}

@misc{hwm,
author = {{US Geological Survey}},
title = "Flood Event Viewer",
url = {https://stn.wim.usgs.gov/FEV/#2017Harvey},
note = {Accessed: June 2022}
}

@ARTICLE{Shu1987-ws,
   title     = "{TVB} uniformly high-order schemes for conservation laws",
   author    = "Shu, Chi-Wang",
   journal   = "Math. Comput.",
   publisher = "American Mathematical Society (AMS)",
   volume    =  49,
   number    =  179,
   pages     = "105--121",
   year      =  1987,
   language  = "en"
 }

@article{loveland2021developing,
  title={Developing a Modeling Framework to Simulate Compound Flooding: When Storm Surge Interacts With Riverine Flow},
  author={Loveland, Mark and Kiaghadi, Amin and Dawson, Clint N and Rifai, Hanadi S and Misra, Shubhra and Mosser, Helena and Parola, Alessandro},
  journal={Frontiers in Climate},
  volume={2},
  pages={609610},
  year={2021},
  publisher={Frontiers Media SA}
}

@article{santiago2019comprehensive,
  title={A comprehensive review of compound inundation models in low-gradient coastal watersheds},
  author={Santiago-Collazo, F{\'e}lix L and Bilskie, Matthew V and Hagen, Scott C},
  journal={Environmental Modelling \& Software},
  volume={119},
  pages={166--181},
  year={2019},
  publisher={Elsevier}
}

@article{orton2020flood,
  title={Flood hazard assessment from storm tides, rain and sea level rise for a tidal river estuary},
  author={Orton, PM and Conticello, FR and Cioffi, F and Hall, TM and Georgas, N and Lall, U and Blumberg, AF and MacManus, K},
  journal={Natural hazards},
  volume={102},
  number={2},
  pages={729--757},
  year={2020},
  publisher={Springer}
}

@article{kumbier2018investigating,
  title={Investigating compound flooding in an estuary using hydrodynamic modelling: a case study from the {Shoalhaven River}, {Australia}},
  author={Kumbier, Kristian and Carvalho, Rafael C and Vafeidis, Athanasios T and Woodroffe, Colin D},
  journal={Natural Hazards and Earth System Sciences},
  volume={18},
  number={2},
  pages={463--477},
  year={2018},
  publisher={Copernicus GmbH}
}

@article{luettich1992adcirc,
  title={{ADCIRC}: an advanced three-dimensional circulation model for shelves, coasts, and estuaries. {R}eport 1, Theory and methodology of {ADCIRC-2DD1} and {ADCIRC-3DL}},
  author={Luettich, Richard Albert and Westerink, Joannes J and Scheffner, Norman W and others},
  journal={ },
  year={1992},
  publisher={Coastal Engineering Research Center (US)}
}

@article{Pringle2020,
  AUTHOR = {Pringle, W. J. and Wirasaet, D. and Roberts, K. J. and Westerink, J. J.},
  TITLE = {Global Storm Tide Modeling with ADCIRC v55: Unstructured Mesh Design and Performance},
  JOURNAL = {Geoscientific Model Development Discussions},
  VOLUME = {2020},
  YEAR = {2020},
  PAGES = {1--30},
  URL = {https://gmd.copernicus.org/preprints/gmd-2020-123/},
  DOI = {10.5194/gmd-2020-123}
}

@article{kubatko2006hp,
  title={hp discontinuous {Galerkin} methods for advection dominated problems in shallow water flow},
  author={Kubatko, Ethan J and Westerink, Joannes J and Dawson, Clint},
  journal={Computer Methods in Applied Mechanics and Engineering},
  volume={196},
  number={1-3},
  pages={437--451},
  year={2006},
  publisher={Elsevier}
}

@article{dawson2011discontinuous,
  title={Discontinuous {Galerkin} methods for modeling hurricane storm surge},
  author={Dawson, Clint and Kubatko, Ethan J and Westerink, Joannes J and Trahan, Corey and Mirabito, Christopher and Michoski, Craig and Panda, Nishant},
  journal={Advances in Water Resources},
  volume={34},
  number={9},
  pages={1165--1176},
  year={2011},
  publisher={Elsevier}
}

@article{dawson2002discontinuous,
  title={Discontinuous and coupled continuous/discontinuous {Galerkin} methods for the shallow water equations},
  author={Dawson, Clint and Proft, Jennifer},
  journal={Computer Methods in Applied Mechanics and Engineering},
  volume={191},
  number={41-42},
  pages={4721--4746},
  year={2002},
  publisher={Elsevier}
}

@article{bunya2009wetting,
  title={A wetting and drying treatment for the {Runge--Kutta discontinuous Galerkin} solution to the shallow water equations},
  author={Bunya, Shintaro and Kubatko, Ethan J and Westerink, Joannes J and Dawson, Clint},
  journal={Computer Methods in Applied Mechanics and Engineering},
  volume={198},
  number={17-20},
  pages={1548--1562},
  year={2009},
  publisher={Elsevier}
}

@book{tan1992shallow,
  title={Shallow water hydrodynamics: Mathematical theory and numerical solution for a two-dimensional system of shallow-water equations},
  author={Tan, {Wei-Yan}},
  year={1992},
  publisher={Elsevier}
}

@article {Tuleya:2007,
      author  = {Robert E Tuleya and Mark DeMaria and Robert J Kuligowski},
      title   = {Evaluation of {GFDL} and Simple Statistical Model Rainfall Forecasts for {U.S.} Landfalling Tropical Storms},
      journal = {Weather and Forecasting},
      year    = {2007},
      volume  = {22},
      number  = {1},
      pages   = {56--70},
      month   = {},
      note    = {},
      key     = {}}

@article{leveque1998balancing,
  title={Balancing source terms and flux gradients in high-resolution {Godunov} methods: the quasi-steady wave-propagation algorithm},
  author={LeVeque, Randall J},
  journal={Journal of computational physics},
  volume={146},
  number={1},
  pages={346--365},
  year={1998},
  publisher={Elsevier}
}

@misc{Pringle2018,
author = {Pringle, WIlliam},
title = {{ OceanMesh2D: User guide - Precise distance-based two-dimensional automated mesh generation toolbox intended for coastal ocean/shallow water}},
year = {2018}
}

@article{Egbert2002,
author = {Egbert, G.D. and Erofeeva, S.Y.},
issn = {07390572},
journal = {Journal of Atmospheric and Oceanic Technology},
number = {2},
pages = {183--204},
publisher = {American Meteorological Society},
title = {{Efficient inverse modeling of barotropic ocean tides}},
volume = {19},
year = {2002}
}

@article{hope2013hindcast,
  title={Hindcast and validation of {Hurricane Ike (2008)} waves, forerunner, and storm surge},
  author={Hope, Mark E and Westerink, Joannes J and Kennedy, Andrew B and Kerr, PC and Dietrich, J Casey and Dawson, C and Bender, Christopher J and Smith, JM and Jensen, Robert E and Zijlema, Marcel and others},
  journal={Journal of Geophysical Research: Oceans},
  volume={118},
  number={9},
  pages={4424--4460},
  year={2013},
  publisher={Wiley Online Library}
}

@techreport{watson2018characterization,
  title={Characterization of peak streamflows and flood inundation of selected areas in southeastern {Texas} and southwestern {Louisiana from the August and September} 2017 flood resulting from {Hurricane Harvey}},
  author={Watson, Kara M and Harwell, Glenn R and Wallace, David S and Welborn, Toby L and Stengel, Victoria G and McDowell, Jeremy S},
  year={2018},
  institution={US Geological Survey}
}

@article{wichitrnithed2024discontinuous,
  title={A discontinuous {Galerkin} finite element model for compound flood simulations},
  author={Wichitrnithed, Chayanon and Valseth, Eirik and Kubatko, Ethan J and Kang, Younghun and Hudson, Mackenzie and Dawson, Clint},
  journal={Computer Methods in Applied Mechanics and Engineering},
  volume={420},
  pages={116707},
  year={2024},
  publisher={Elsevier}
}

@book{vreugdenhil1994numerical,
  title={Numerical methods for shallow-water flow},
  author={Vreugdenhil, Cornelis Boudewijn},
  volume={13},
  year={1994},
  publisher={Springer Science \& Business Media}
}

@article{bunya2010high,
  title={A high-resolution coupled riverine flow, tide, wind, wind wave, and storm surge model for southern {Louisiana and Mississippi}. Part I: Model development and validation},
  author={Bunya, S and Dietrich, J Cl and Westerink, JJ and Ebersole, BA and Smith, JM and Atkinson, JH and Jensen, Resio and Resio, DT and Luettich, RA and Dawson, C and others},
  journal={Monthly weather review},
  volume={138},
  number={2},
  pages={345--377},
  year={2010},
  publisher={American Meteorological Society}
}

@article{blanton2012urgent,
  title={Urgent computing of storm surge for {North Carolina's} coast},
  author={Blanton, Brian and McGee, John and Fleming, Jason and Kaiser, Carola and Kaiser, Hartmut and Lander, Howard and Luettich, Rick and Dresback, Kendra and Kolar, Randy},
  journal={Procedia Computer Science},
  volume={9},
  pages={1677--1686},
  year={2012},
  publisher={Elsevier}
}

@incollection{funakoshi2012development,
  title={Development of extratropical surge and tide operational forecast system {(ESTOFS)}},
  author={Funakoshi, Yuji and Feyen, Jesse and Aikman, Frank and Tolman, Hendrik and van der Westhuysen, Andre and Chawla, Arun and Rivin, Ilya and Taylor, Arthur},
  booktitle={Estuarine and Coastal Modeling (2011)},
  pages={201--212},
  year={2012}
}

@article{carey1983finite,
  title={Finite elements: a second course},
  author={Carey, Graham F and Oden, J Tinsley},
  year={1983},
  publisher={Englewood Cliffs: Prentice-Hall,}
}

@book{ern2004theory,
  title={Theory and practice of finite elements},
  author={Ern, Alexandre and Guermond, Jean-Luc},
  volume={159},
  year={2004},
  publisher={Springer}
}

@article{cangialosi2018tropical,
  title={Tropical cyclone report: {Hurricane Irma (AL112017)}},
  author={Cangialosi, John P and Latto, Andrew S and Berg, Robbie},
  journal={National Hurricane Center},
  volume={28},
  pages={2020},
  year={2018}
}

@article{callaghan2020extreme,
  title={Extreme rainfall and flooding from {Hurricane Florence}},
  author={Callaghan, Jeff},
  journal={Tropical Cyclone Research and Review},
  volume={9},
  number={3},
  pages={172--177},
  year={2020},
  publisher={Elsevier}
}

@article{pachev2023one,
  title={One-way coupling of {E3SM} with {ADCIRC} demonstrated on {Hurricane Harvey}},
  author={Pachev, Benjamin and Leung, L Ruby and Zhou, Tian and Dawson, Clint},
  journal={Natural Hazards},
  pages={1--25},
  year={2023},
  publisher={Springer}
}

@article{savant2020theory,
  title={Theory, formulation, and implementation of the Cartesian and spherical coordinate two-dimensional depth-averaged module of the {Adaptive Hydraulics (AdH)} finite element numerical code},
  author={Savant, Gaurav and Berger, Rutherford C and Trahan, Corey J and Brown, Gary L},
  year={2020},
  publisher={Coastal and Hydraulics Laboratory (US)}
}

@article{dawson2024swemnics,
  title={{SWEMniCS}: a software toolbox for modeling coastal ocean circulation, storm surges, inland, and compound flooding},
  author={Dawson, Clint and Loveland, Mark and Pachev, Benjamin and Proft, Jennifer and Valseth, Eirik},
  journal={npj Natural Hazards},
  volume={1},
  number={1},
  pages={44},
  year={2024},
  publisher={Nature Publishing Group UK London}
}

@article{zhang2014schism,
  title={{SCHISM} theory manual},
  author={Zhang, YJ},
  journal={On the Internet at: http://ccrm. vims. edu/schism/combined\_theory\_manual. pdf},
  year={2014}
}

@article{zhang2020simulating,
  title={Simulating compound flooding events in a hurricane},
  author={Zhang, Yinglong J and Ye, Fei and Yu, Haocheng and Sun, Weiling and Moghimi, Saeed and Myers, Edward and Nunez, Karinna and Zhang, Ruoyin and Wang, Harry and Roland, Aron and others},
  journal={Ocean Dynamics},
  volume={70},
  pages={621--640},
  year={2020},
  publisher={Springer}
}

@book{luettich2004formulation,
  title={Formulation and numerical implementation of the {2D/3D ADCIRC} finite element model version 44. XX},
  author={Luettich, Richard Albert and Westerink, Joannes J},
  volume={20},
  year={2004},
  publisher={R. Luettich Chapel Hill, NC, USA}
}

@book{Prony1790,
  author    = {Prony, Gaspard de},
  title     = {Nouvelle Architecture Hydraulique},
  publisher = {Firmin-Didot},
  year      = {1790},
  address   = {Paris}
}

@article{Manning1891,
  author  = {Manning, Robert},
  title   = {On the flow of water in open channels and pipes},
  journal = {Transactions of the Institution of Civil Engineers of Ireland},
  year    = {1891},
  volume  = {20},
  pages   = {161--207}
}

\end{document}